\documentclass[useAMS,usenatbib]{mn2e}

 \usepackage{times}

\usepackage{graphicx}
\usepackage{supertabular}
\usepackage{subfigure}
\usepackage{rotating}
\usepackage{lscape}
\usepackage{epsfig}

\def\squig{$\sim\!\!$}
\def\lesssim{\mathrel{\hbox{\rlap{\hbox{\lower4pt\hbox{$\sim$}}}\hbox{$<$}}}}
\def\gtrsim{\mathrel{\hbox{\rlap{\hbox{\lower4pt\hbox{$\sim$}}}\hbox{$>$}}}}

\def\arcsec{\hbox{$^{\prime\prime}$}}
\def\arcmin{$^{\prime}$}
\def\deg{\hbox{$^\circ$}}
\def\power{WHz$^{-1}$sr$^{-1}$}

\title[A sample of mJy radio sources at 1.4 GHz in the Lynx and
  Hercules fields]{A sample of mJy radio sources at 1.4 GHz in the Lynx
  and Hercules fields - I. Radio imaging, multicolour photometry and spectroscopy} 
\author[E. E. Rigby, I. A. Snellen and
  P. N. Best]{E. E. Rigby$^{1}$\thanks{E-mail:eer@roe.ac.uk},
  I. A. G. Snellen$^{2}$ and P. N. Best$^{1}$ \\ 
$^1$SUPA\thanks{Scottish Universities Physics Alliance}, University of
  Edinburgh, Institute for Astronomy, Royal Observatory, Edinburgh EH9
  3HJ, UK\\ 
$^{2}$Leiden Observatory, Leiden University, Niels Bohrweg 2, NL-2300RA Leiden, The Netherlands}

\begin{document}


\pagerange{\pageref{firstpage}--\pageref{lastpage}} \pubyear{2002}

\maketitle

\label{firstpage}

\begin{abstract}
With the goal of identifying high redshift radio galaxies with FRI
classification, here are presented high resolution, 
wide--field radio observations, near infra--red and optical imaging
and multi--object spectroscopy of two fields of the Leiden--Berkeley Deep
Survey. These fields, Hercules.1 and Lynx.2, contain a complete sample
of 81 radio sources with S$_{\rm 1.4 GHz}>0.5$ mJy within 0.6 square
degrees. This sample will form the basis for a study of the population
and cosmic evolution of high redshift, low power, FRI radio sources which will be presented
in Paper II. Currently, the host galaxy identification fraction is 86\% with 11
sources remaining unidentified at a level of \emph{r$^{\prime}$}$\geqslant$25.2 mag
(Hercules; 4 sources) or \emph{r$^{\prime}$}$\geqslant$24.4 mag (Lynx; 7 sources) or
\emph{K}$\gtrsim$20 mag. Spectroscopic redshifts have been determined for
49\% of the sample and photometric redshift estimates are presented for the 
remainder of the sample.  
\end{abstract}


\section{Introduction}

Radio--loud active galaxies that display extended jet emission from
their central cores can be divided into two main
types; Fanaroff \& Riley class I and II (FRI and FRII; Fanaroff \& Riley
1974). The galaxies of FRI type are `edge--darkened' with the majority of
their emission confined to their central regions and jets that flare out
close to the nucleus. On the other hand, the FRII
galaxies are `edge--brightened' meaning the bulk of their emission
originates from the hotspots at the ends of their highly collimated
jets. The FRII galaxies are the more luminous of the two classes and typically have
$P_{\rm 178MHz}$ $>10^{24-25}$ \power~ but there is significant
overlap at the break luminosity. The FRIs and FRIIs have also been
suggested as the unbeamed parent populations of BL Lac objects and
flat spectrum quasars respectively (Jackson \& Wall, 1999).   

The differences between the two FR classes are not confined to the
lobe morphology. For instance, Zirbel \& Baum (1995) found that the
FRIIs produce 10--50 times more emission line luminosity than the FRIs
at a particular radio core power. Additionally, optical observations
by Owen \& Laing (1989) found that the host galaxies of FRIs tended to
be larger and more luminous than those of FRIIs, though later work by
Ledlow \& Owen (1996) suggested that this result was caused by a
combination of sample selection effect, and only observing a small
range in radio power.  

It is not yet clear whether the observed morphological differences 
between the two FR classes are the result of fundamental differences 
in the properties of the central engine (e.g. lower accretion
rates in FRIs leading to advection  dominated accretion flow, or a slower
FRI black hole spin) as advocated by e.g. Baum et. al. (1995),  
or differences in the interactions of the jets with their environments as suggested by the
work of Gopal--Krishna \& Wiita (2000) and Gawro\'{n}ski et al. (2006). This intrinsic/extrinsic
question is of vital importance for understanding the relationship between these objects; if the
intrinsic difference model is correct then the FRIs and FRIIs are two
discrete classes of object, however if the evidence suggests the other
model is correct, the underlying properties of the classes would be the
same. In the latter case the two classes may simply represent 
different stages in the evolution of a radio galaxy, i.e. it starts
out as a powerful, high--luminosity FRII and as it ages its jets
become less powerful and it becomes an FRI (e.g. Willott et al. 2001). 

One of the key ways in which the differences between the two classes
can be investigated is through their evolution with cosmic
epoch since, if the extrinsic model is correct, then FRIs and FRIIs of
the same luminosity should evolve in  
similar ways. FRIIs are known to undergo strong cosmic evolution, with density
enhancements for the most  
luminous of a factor of 100-1000 out to redshifts of 1--2, compared to
a factor of \squig~10 for the less luminous, sources (Wall, 1980);
the behaviour of the FRIs is less clear. Low--redshift studies initially  
suggested that they had a constant space density (e.g. Jackson \& Wall
1999; Willott et al. 2001) and this appeared to be confirmed
at higher redshift by Clewley \& Jarvis (2004). However, they selected
the FRIs in their sample using a luminosity cut; this could lead to
FRIs being missed (particularly the more luminous FRIs which may
evolve the most) since the FRI/FRII break luminosity is not fixed, 
but is a function of host galaxy magnitude (Ledlow \& Owen
1996). It is clear therefore, that to define a robust sample of
distant FRIs, in order to obtain an accurate picture of their high
redshift behaviour, radio morphological classification is a
necessity. 

Determining the cosmic evolution of FRI radio galaxies is also important
because of the impact that they may have on galaxy formation and 
evolution. Models of galaxy formation are increasingly turning to
these objects to solve the problem of  
massive galaxy over--growth (e.g. Bower et al., 2006). It is predominantly
the lower luminosity sources that provide the necessary feedback for
this, (Best et al., 2006), and may possibly be limited to the FRI population  
alone. As such understanding the little studied FRI sources and their
evolution could be critical to deciphering this mechanism. 

The first significant attempt at determining the FRI high--redshift space
density was carried out by Snellen \& Best (2001) using the
Hubble Deep Field and Flanking Fields (HDF+FF). Two z$>$1 FRI galaxies were detected in this
area, which calculations showed to be broadly consistent with an FRI space
density enhancement comparable to that of less luminous FRII
galaxies at that redshift, and inconsistent (a probability of $<$1\%) with a non-evolving FRI
population. However, with only two detected FRIs the uncertainties in
this result are clearly large.

The area of sky used in the analysis of Snellen \& Best (2001)
was only large enough to give a first estimate of the high redshift space
density of FRIs. This work, therefore, uses a deep, wide--field, 
Very Large Array (VLA) A--array survey an order of magnitude larger
than the HDF+HFF which will enable the space density to be directly
measured for the first time. Here we present the initial radio,
optical and infra--red imaging, along with the spectroscopic
observations which form the basis of the work. The layout of the
paper is as follows: in Sections \ref{sample} and \ref{new_rad} the sample is defined and the new
radio observations taken are described; in Sections \ref{opt_chap} and \ref{id_chap}, the optical and
infra--red imaging are presented and the host galaxies identified;
Sections \ref{spec_chap} and \ref{sb_chap} describe the spectroscopic observations of a subset of the
sample. Finally, Section \ref{z_est_1} outlines the redshift estimation methods
used for the remainder of the sample and the conclusions of the paper
can be found in Section \ref{concl}.   

\section{\protect\label{sample} The radio sample}
\protect\label{sample}

The survey was split over two fields -- one in the constellation of Lynx at
right ascension, $\alpha=8h45$, declination, $\delta=+44.6\deg$ (J2000) 
and one in Hercules at $\alpha=17h20$, $\delta=+49.9\deg$
(J2000). These fields were
chosen because of the existence of previous low resolution
radio observations by Windhorst et al. (1984), Oort \& Windhorst
(1985) and Oort \& van Langevelde (1987). Additionally, the Hercules 
field has some previous optical and spectroscopic observations by 
Waddington et al. (2000). Alongside this, the Lynx field is also covered
by the Sloan Digital Sky Survey (SDSS; York et al. 2000; Stoughton et al. 2002).

The two fields were originally observed as part of the
Leiden--Berkeley Deep Survey (LBDS) in which they were referred to as
Lynx.2 and Hercules.1. For the remainder of this paper they will be
referred to as Lynx and Hercules respectively. It should be noted that
the Lynx.2 field is unusual in that it does not contain any radio
sources brighter than 6 mJy over nearly a square degree. However, the
resulting underrepresentation in the radio source counts above this
level should not affect the conclusions of this paper, as it is the
faint end of the RLF that is being investigated here. This section
outlines the previous work in the two fields. 

\subsection{Sample definition and previous radio work}

The LBDS survey was constructed to provide photometry for faint galaxies and quasars, via
multicolour plates obtained with the 4m Mayall Telescope at Kitt Peak
(Kron, 1980; Koo \& Kron 1982). Radio follow--up of nine of the LBDS fields (including
Hercules and Lynx that are used here)
was performed subsequently using the 3km Westerbork Synthesis Radio Telescope 
(WSRT) at 1.4 GHz, at a resolution of 12.5\arcsec\ (Windhorst et
al. 1984), reaching a rms noise level, at the field centre, of 0.12--0.28 mJy. Their radio
sample consists of 306 sources which satisfied the sample 
selection criteria of peak signal to noise ($S_{P}/N$) $ \geq 5\sigma$ out to an attenuation
factor, $A(r), \leq 5$ (for WSRT this corresponds to a radius of $\leq
0.464\deg$ or $\leq$ 28\arcmin). 

The Hercules and Lynx fields were reobserved by Oort \& van
Langevelde (1987) and Oort \& Windhorst (1985)
respectively, again using the 3 km WSRT at 1.4 GHz with a 12.5\arcsec\ beam. These two sets
of observations were a factor 2--3 deeper than the original Windhorst et
al. (1984) ones, reaching a 5$\sigma$ flux limit of 0.45 mJy
for Hercules and 0.30 mJy for Lynx at the pointing
centre. 

\begin{figure*}
\centering
\begin{minipage}{150mm}
\includegraphics[scale=0.5]{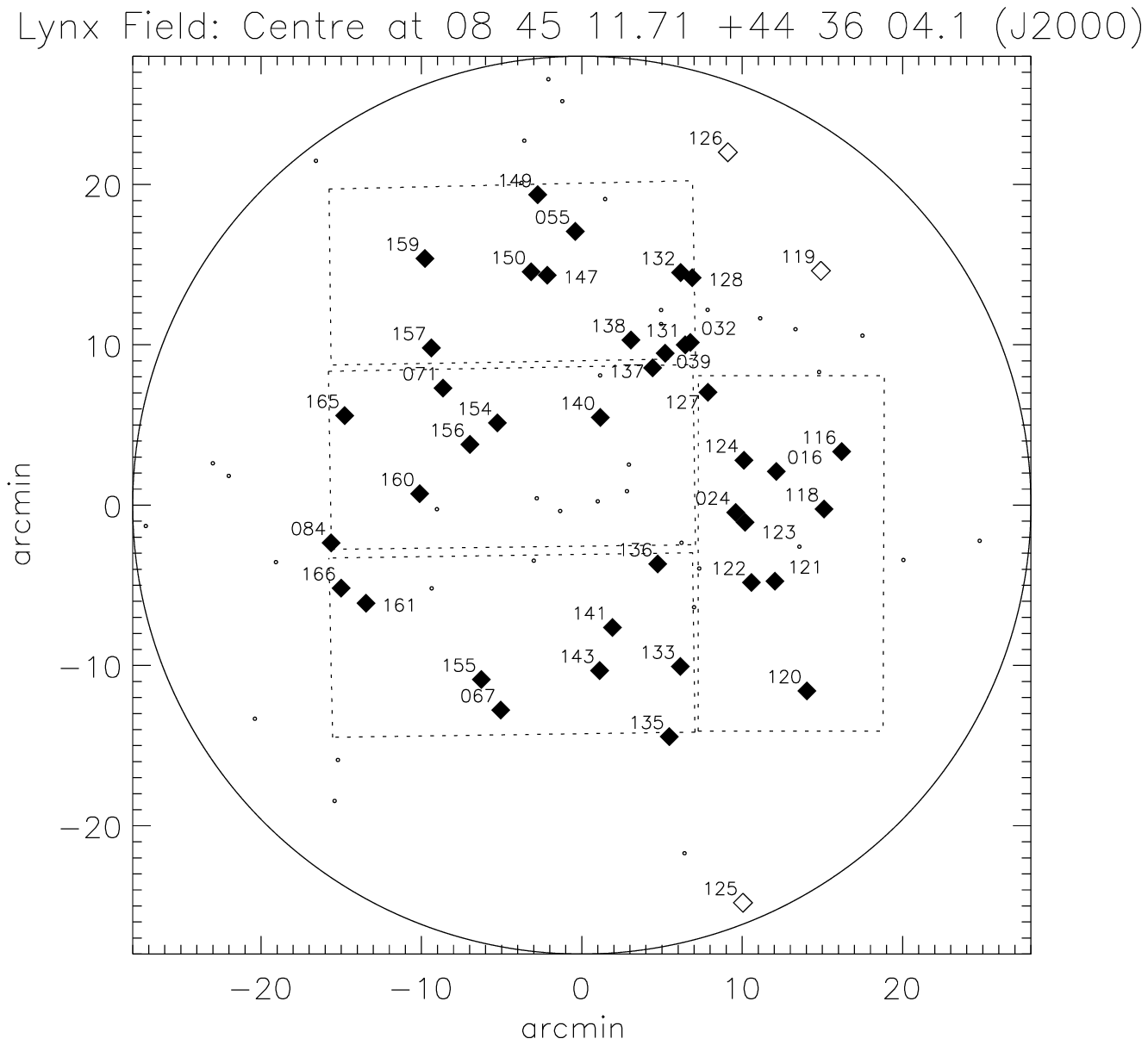}
\includegraphics[scale=0.5]{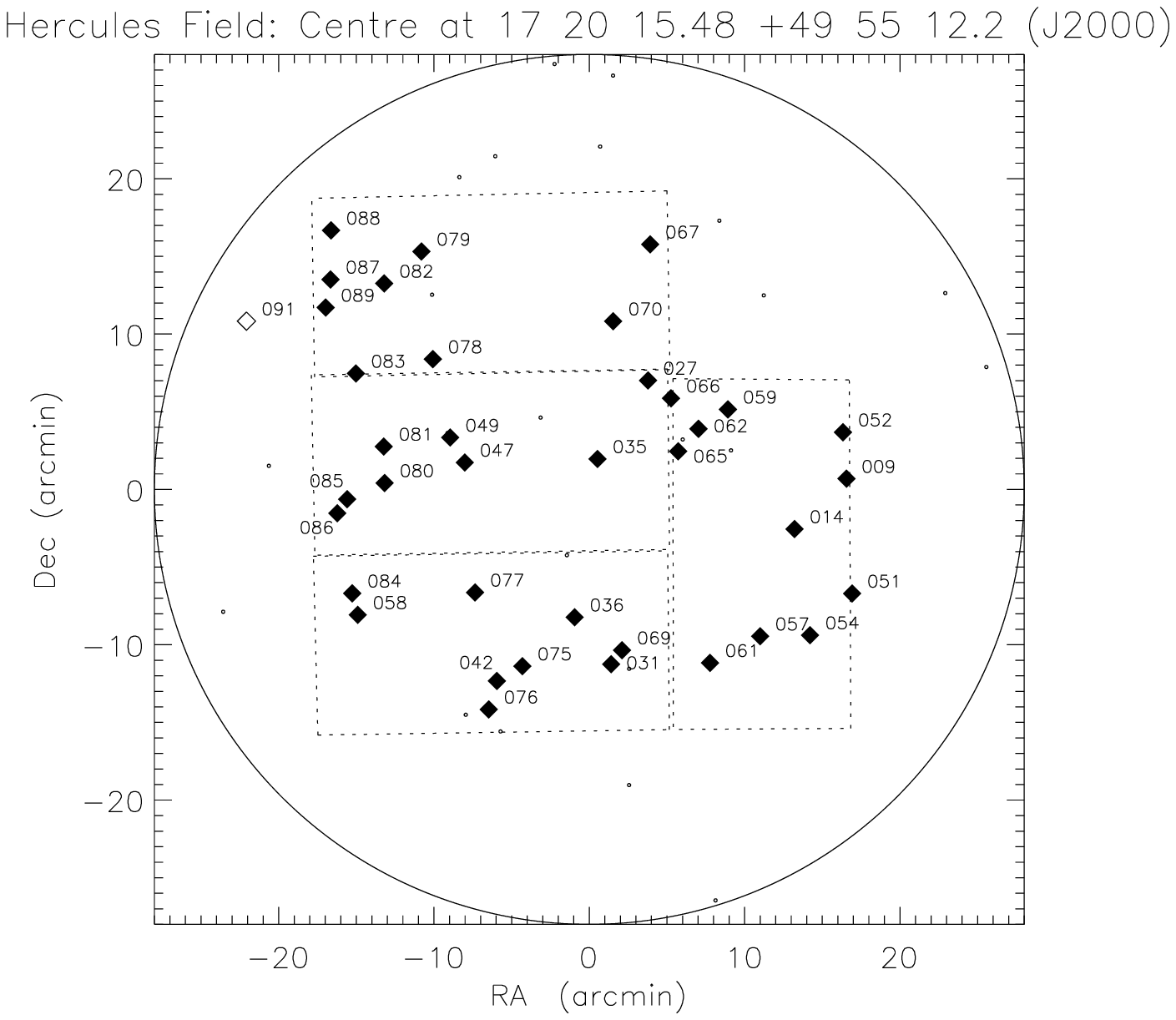}       
\caption{\protect\label{CCD_plot} The distribution of the radio
sources in the Lynx and Hercules fields; the labels correspond to the last
3 characters of the source names. The large circle in both plots
represents the 0.464$\deg$ radius of the previous WSRT observations of
e.g. Windhorst et al. (1984); this is also approximately the 30\arcmin primary beam of the VLA at
1.4 GHz. The dotted line represents the area of the sample which is
defined by the field of view of the optical imaging. Dots indicate the
positions of the additional sources which were not included in the sample
because they either fell below the flux density limit of 0.5 mJy, or
they are not covered by the sample area. Open diamonds indicate
sources not included in the complete sample, but which were included in
the infra--red observations.} 
\end{minipage}
\end{figure*}

The sample used in this work is a subset of the combined Hercules
and Lynx sources, as it is limited by the field of view size of the
optical imaging described in \S\ref{opt_chap}; this is illustrated in
Figure \ref{CCD_plot}. A flux limit of 0.5 mJy was also imposed to
remove the faintest, most poorly detected sources and provide a more
uniform limiting flux density across the two fields.  
Table \ref{a_table_herc} gives the flux densities for the Hercules field
(Oort \& van Langevelde 1987) and Table \ref{a_table_lynx} gives the
same for the Lynx field (Oort \& Windhorst 1985) for the sources
included in this work. Table \ref{a_table_no} gives the 1.4 GHz flux densities
of sources which were not covered by the optical observations, but were
included in the subsequent infra--red imaging; consequently they are
not part of the complete sample, but are included here for
completeness. These, and other source parameters, were measured by
Oort and collaborators using an elliptical Gaussian fitting method in two dimensions. 

\subsection{Sample completeness}

The selection criteria of the LBDS meant that source weighting was
necessary to account for incompleteness in the starting sample. This incompleteness arises from
two factors: the attenuation of the WSRT primary beam and the
resolution bias. The effect of the first of these factors, the decreasing
sensitivity at increasing radial distance from the pointing centre, is
to make the probability of detecting a source depend on where it is 
located in the map; a source which just satisfies the selection
criteria at the centre would have been missed if it was located at the
map edge. To correct for this each source was assigned a weight that
was inversely proportional to the area over which it would have met
the selection criteria and hence, have been included in the sample
(Windhorst et al. 1984). The total sample area for the current work is
limited by the size of the optical imaging, so the attenuation weights
for the sources considered here were recalculated to account for this.

The corrections for the incompleteness due to the resolution bias (the
fact that a resolved source will be more 
difficult to detect than a point source of the same total flux) were
found, by Windhorst et al. (1984), through
detailed modelling of the source detection algorithms. It was
subsequently found that sources with higher flux densities were also
those with a large angular size (e.g. Windhorst et al. 1993) which, as Windhorst,
Mathis \& Neuschaefer (1990) showed, meant that the
original weights were overestimated. Waddington et al. (2000) derived
a new expression for the resolution bias that took this into account;
it is that method which is used in the current calculations.  

These two re--derived correction factors, the attenuation weight and resolution
weight, were then multiplied together to give the final weights. For further details of
the source weighting methods used in the LBDS see Windhorst et
al. (1984) and Waddington et al. (2000). See Tables \ref{a_table_herc}
and \ref{a_table_lynx} for the weight calculated for each source.

\section{\protect\label{new_rad} New radio observations}
\protect\label{new_rad}
\subsection{The observations}
The VLA data were taken with the array in A configuration on 22nd April 2002
for the Hercules field, and 15th February 2002 for the Lynx field.
Both sets of observations were taken at a frequency of 1.4 GHz (L--band), in
four IF spectral line mode to enable wide field imaging, using 16 channels
of width 781.25 kHz. The IFs were centred on 1391.3 and
1471.1 MHz observing dual polarization. Both the Hercules and the Lynx
fields were observed for 8.5 hours each.

The data were calibrated using the NRAO AIPS package. Because of the
non-coplanar geometry of the telescope array, in order to image the entire
area of the primary beam, a three--dimensional Fourier
transform  is required. As this would be computationally impractical, the standard
pseudo--three--dimensional Fourier transform technique, as
incorporated in the task IMAGR, (Perley 1999) was
adopted: this technique divides the field of view into numerous smaller
facets, within each of which a two--dimensional Fourier transform provides a
sufficiently good approximation. For each of the Hercules and Lynx fields,
256 by 256 pixel facets (with 0.35\arcsec\ per pixel) were centred on all the sources that were already
known from the imaging of Windhorst, Oort and collaborators, discussed above. These sub-fields were
cleaned and self-calibrated using the AIPS tasks IMAGR and CALIB. The
self calibration consisted of multiple phase--only cycles followed by
one final amplitude and phase calibration. The resulting maps have a
resolution of 1.6\arcsec\, and reach a noise limit of 15 $\mu$Jy.

\subsection{Source detection and flux density measurement}

The Hercules sources of Oort \& van Langevelde (1987) and Lynx sources of
Oort \& Windhorst (1985) were all detected in our VLA 
observations. This gives a sample of 81 sources, evenly spread over
the two fields; the distribution on the sky can be seen in Figure
\ref{CCD_plot}. Flux densities were measured for these sources using
the AIPS task \emph{imfit} to fit a Gaussian, if they were pointlike, or the task
\emph{tvstat} to sum within a defined area, if they showed significant
extension; the method used for each individual source is indicated in
Tables \ref{a_table_herc} and \ref{a_table_lynx}. The values were then
corrected for the attenuation of the VLA primary beam. For some objects, where
previously one source was detected, these higher resolution   
observations have resolved it into more than one component each
associated with a different host galaxy. In these cases, the
sub--components are labelled \emph{a}, \emph{b}, etc. and 
the low resolution flux density has been assigned to each new component
according to the A--array flux density ratio. These components are
only retained in the sample if they remain above the 0.5 mJy flux density
limit. 

\begin{figure}
\centering
\includegraphics[scale=0.3, angle=90]{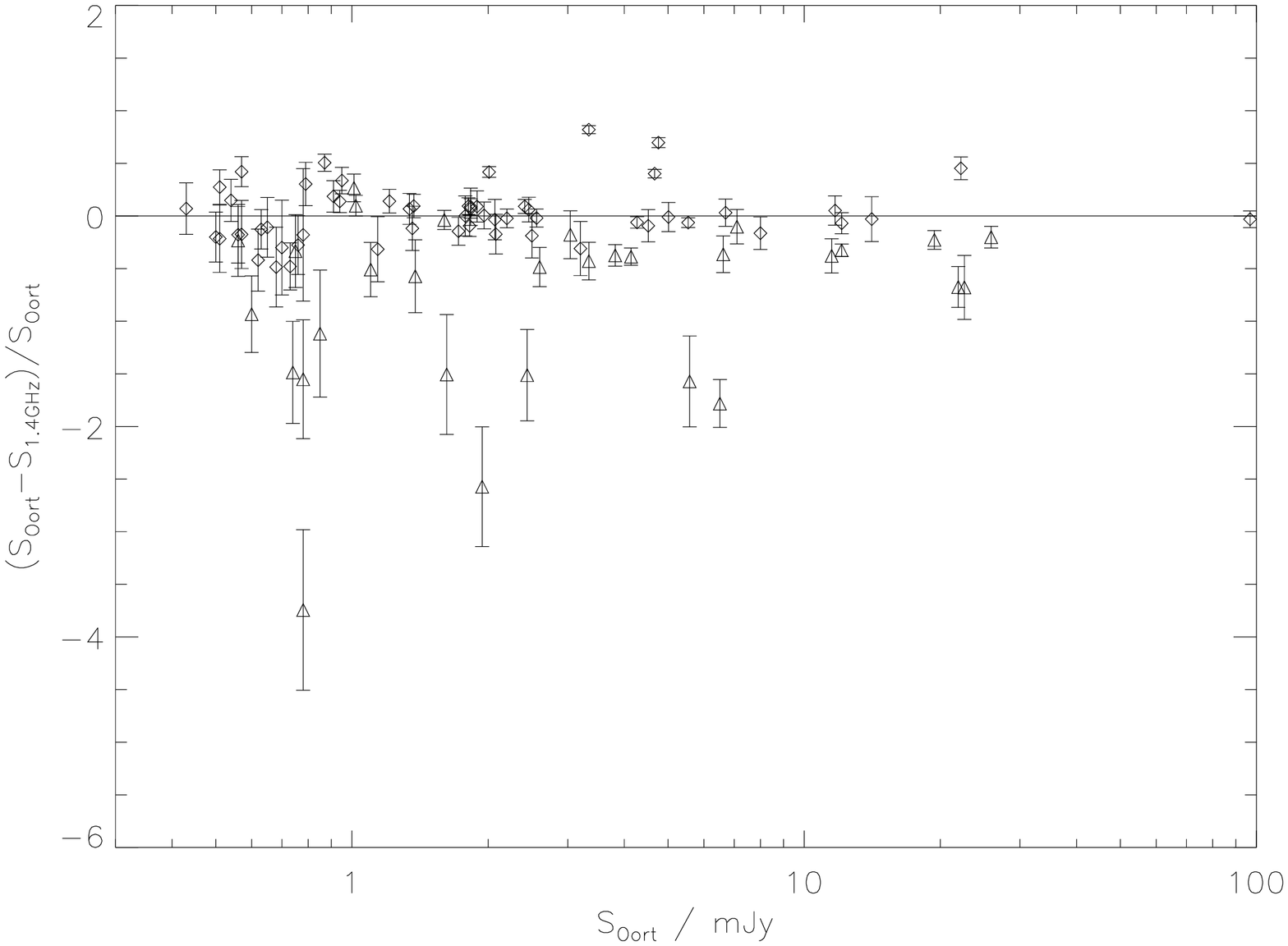}
\caption{\label{oort_comp} A comparison of the VLA A--array ($S_{\rm 1.4 GHz}$) and Oort
  et al. (1985) or Oort \& van Langevelde (1987) ($S_{\rm Oort}$) flux densities for both fields. Extended,
  \emph{tvstat} measured sources are shown as triangles and 
  compact, \emph{imfit} measured sources are shown as diamonds.}
\end{figure}

Figure \ref{oort_comp} shows a comparison between the WSRT flux densities
measured by Oort et al. (1985; Lynx field) or Oort \& van Langevelde
(1987; Hercules field) and those measured for the new VLA A--array
observations. The values are generally in good agreement for 
the compact sources; those that are different have $S_{\rm
  Oort}>S_{\rm 1.4 GHz}$ which suggests that flux has been lost at the
higher VLA resolution, and possibly indicates a resolved out FRI--type
structure. The extended sources, however, are mainly underestimated by
Oort et al. ($S_{\rm Oort}<S_{\rm 1.4 GHz}$) as a result of their elliptical Gaussian measuring method
which misses any extended flux. It should be noted that a subset of
the two fields (36\% of the sample) were also observed with
the VLA by Oort et al. (1987). These measurements have not been used
for flux density comparisons though, due to the small number of
sources included. 

The positions of the detected sources are given in Tables
\ref{a_table_herc} and \ref{a_table_lynx} along 
with the measured flux densities and primary beam correction factors,
$C_{\rm PB}$,  (i.e. $S_{\rm Cor}=C_{\rm PB}~S_{\rm Meas}$) used. The
corresponding radio contour images can be found in Appendix
\ref{rad_im}.  

\begin{table*}
\centering
\begin{minipage}{13.0cm}
\caption{\protect\label{a_table_herc}The Hercules radio source positions from the
VLA A--array observations along with the A--array and Oort et al. 1987 1.4 GHz
primary beam corrected flux densities, source weights and primary beam
correction factor, $C_{\rm PB}$ (see text 
for full details). An `I' in the final column indicates an
\emph{imfit} measured flux density; a `T' indicates a \emph{tvstat} measurement.}
\begin{tabular}{l|cc|c|c|c|c|c}
\hline
\multicolumn{6}{l}{Hercules}  \\
\hline
Name    &\multicolumn{2}{c|}{RA/DEC (J2000)} & $S_{\rm Oort}$ (mJy) &
$S_{\rm 1.4 GHz}$ (mJy) & Weight & $C_{\rm PB}$ & Measure\\
\hline	               
53w052  & 17 18 34.14 & 49 58 53.0  &  8.00   $\pm$  0.34 & 9.31   $\pm$   1.17 & 1.00 & 2.26 &I\\
53w054a & 17 18 47.30 & 49 45 49.0  &  2.07   $\pm$  0.19 & 2.14   $\pm$   0.35 & 1.00 & 2.34 &I\\
53w054b & 17 18 49.97 & 49 46 12.2  &  2.08   $\pm$  0.19 & 2.44   $\pm$   0.32 & 1.00 & 2.20 &I\\
53w057  & 17 19 07.29 & 49 45 44.8  &  1.96   $\pm$  0.14 & 1.95   $\pm$   0.21 & 1.00 & 1.82 &I\\
53w059  & 17 19 20.18 & 50 00 21.2  &  19.40  $\pm$  1.0  & 23.81  $\pm$   1.21 & 1.00 & 1.34 &T\\
53w061  & 17 19 27.34 & 49 44 01.9  &  4.76   $\pm$  0.43 & 1.44   $\pm$   0.18 & 1.02 & 1.69 &I\\
53w062  & 17 19 31.93 & 49 59 06.2  &  0.73   $\pm$  0.10 & 1.08   $\pm$   0.07 & 1.42 & 1.19 &I\\
53w065  & 17 19 40.05 & 49 57 39.2  &  5.54   $\pm$  0.20 & 5.89   $\pm$   0.14 & 1.00 & 1.11 &I\\
53w066  & 17 19 42.96 & 50 01 03.9  &  4.27   $\pm$  0.17 & 4.53   $\pm$   0.15 & 1.00 & 1.18 &I\\
53w067  & 17 19 51.27 & 50 10 58.7  &  21.9   $\pm$  0.90 & 36.68  $\pm$   3.97 & 1.00 & 2.15 &T\\
53w069  & 17 20 02.52 & 49 44 51.0  &  3.82   $\pm$  0.17 & 5.25   $\pm$   0.31 & 1.00 & 1.36 &T\\
53w070  & 17 20 06.07 & 50 06 01.7  &  2.56   $\pm$  0.14 & 2.61   $\pm$   0.17 & 1.00 & 1.39 &I\\
53w075  & 17 20 42.37 & 49 43 49.1  &  96.8   $\pm$  3.3  & 99.82  $\pm$   6.83 & 1.00 & 1.51 &I\\
53w076  & 17 20 55.82 & 49 41 02.2  &  1.94   $\pm$  0.17 & 6.93   $\pm$   0.92 & 1.00 & 2.21 &T\\
53w077  & 17 21 01.32 & 49 48 34.0  &  6.51   $\pm$  0.39 & 18.11  $\pm$   1.01 & 1.00 & 1.31 &T\\
53w078  & 17 21 18.17 & 50 03 35.2  &  0.74   $\pm$  0.12 & 1.84   $\pm$   0.20 & 1.40 & 1.62 &T\\
53w079  & 17 21 22.75 & 50 10 31.0  &  11.7   $\pm$  0.5  & 11.1   $\pm$   1.68 & 1.00 & 2.87 &I\\
53w080  & 17 21 37.48 & 49 55 36.8  &  25.9   $\pm$  0.9  & 31.11  $\pm$   2.42 & 1.00 & 1.63 &T\\
53w081  & 17 21 37.86 & 49 57 57.6  &  12.1   $\pm$  0.5  & 12.93  $\pm$   1.08 & 1.00 & 1.68 &I\\
53w082  & 17 21 37.64 & 50 08 27.4  &  2.50   $\pm$  0.19 & 2.97   $\pm$   0.47 & 1.00 & 2.86 &I\\
53w083  & 17 21 48.95 & 50 02 39.7  &  5.01   $\pm$  0.25 & 5.06   $\pm$   0.64 & 1.00 & 2.28 &I\\
53w084  & 17 21 50.43 & 49 48 30.5  &  0.68   $\pm$  0.12 & 1.01   $\pm$   0.19 & 1.51 & 2.53 &I\\
53w085  & 17 21 52.48 & 49 54 34.1  &  4.52   $\pm$  0.22 & 4.94   $\pm$   0.66 & 1.00 & 2.02 &I\\
53w086a & 17 21 56.42 & 49 53 39.8  &  1.62   $\pm$  0.30 & 4.06   $\pm$   0.54 & 1.09 & 2.17 &T\\
53w086b & 17 21 57.65 & 49 53 33.8  &  2.44   $\pm$  0.30 & 6.13   $\pm$   0.74 & 1.00 & 2.22 &T\\
53w087  & 17 21 59.10 & 50 08 42.9  &  5.58   $\pm$  0.35 & 14.35  $\pm$   2.23 & 1.00 & 4.23 &T\\
53w088  & 17 21 58.90 & 50 11 52.7  &  14.1   $\pm$  0.7  & 14.52  $\pm$   2.92 & 1.00 & 6.07 &I\\
53w089  & 17 22 01.05 & 50 06 54.7  &  3.04   $\pm$  0.26 & 3.58   $\pm$   0.62 & 1.00 & 3.71 &T\\
66w009a & 17 18 32.76 & 49 55 53.4  &  1.14   $\pm$  0.21 & 1.50   $\pm$   0.22 & 1.23 & 2.22 &I\\
66w009b & 17 18 33.73 & 49 56 03.2  &  0.70   $\pm$  0.21 & 0.91   $\pm$   0.16 & 5.69 & 2.19 &I\\
66w014  & 17 18 53.51 & 49 52 39.1  &  3.34   $\pm$  0.51 & 0.60   $\pm$   0.09 & 1.18 & 1.66 &I\\
66w027  & 17 19 52.11 & 50 02 12.7  &  0.57   $\pm$  0.11 & 0.67   $\pm$   0.13 & 3.75 & 1.19 &I\\
66w031  & 17 20 06.87 & 49 43 57.0  &  0.76   $\pm$  0.14 & 0.97   $\pm$   0.12 & 2.32 & 1.43 &I\\
66w035  & 17 20 12.32 & 49 57 09.7  &  0.63   $\pm$  0.09 & 0.71   $\pm$   0.06 & 1.77 & 1.01 &I\\
66w036  & 17 20 21.46 & 49 46 58.3  &  0.78   $\pm$  0.11 & 3.70   $\pm$   0.29 & 1.57 & 1.20 &T\\
66w042  & 17 20 52.59 & 49 42 52.4  &  0.78   $\pm$  0.14 & 1.99   $\pm$   0.26 & 1.61 & 1.70 &T\\
66w047  & 17 21 05.43 & 49 56 56.0  &  0.60   $\pm$  0.10 & 1.16   $\pm$   0.10 & 2.32 & 1.20 &T\\
66w049  & 17 21 11.25 & 49 58 32.4  &  1.38   $\pm$  0.27 & 2.17   $\pm$   0.22 & 3.09 & 1.28 &I\\
66w058  & 17 21 48.23 & 49 47 07.3  &  1.89   $\pm$  0.16 & 1.72   $\pm$   0.24 & 1.01 & 2.33 &I\\
\hline
\end{tabular}
\end{minipage}
\end{table*}

\begin{table*}
\centering
\begin{minipage}{130mm}
\caption{\protect\label{a_table_lynx}The Lynx radio source positions from the
VLA A--array observations along with the A--array and Oort et al. 1985 1.4 GHz
primary beam corrected flux densities, source weights and primary beam
correction factor, $C_{\rm PB}$ (see text 
for full details). An `I' in the final column indicates an
\emph{imfit} measured flux density; a `T' indicates a \emph{tvstat} measurement. }
\begin{tabular}{l|cc|c|c|c|c|c}
\hline
\multicolumn{6}{l}{Lynx}  \\
\hline
Name    &\multicolumn{2}{c|}{RA/DEC (J2000)} & $S_{\rm Oort}$ (mJy) &
$S_{\rm 1.4 GHz}$ (mJy) & Weight & $C_{\rm PB}$ & Measure \\
\hline	               
55w116  & 08 43 40.72 & 44 39 24.7  & 1.36 $\pm$  0.12 & 1.52  $\pm$  0.25 & 1.00 & 2.22 &I\\
55w118  & 08 43 46.86 & 44 35 49.7  & 0.91 $\pm$  0.09 & 0.74  $\pm$  0.11 & 1.03 & 1.92 &I\\
55w120  & 08 43 52.89 & 44 24 29.0  & 1.83 $\pm$  0.16 & 1.67  $\pm$  0.29 & 1.00 & 2.68 &I\\
55w121  & 08 44 04.06 & 44 31 19.4  & 1.21 $\pm$  0.09 & 1.04  $\pm$  0.11 & 1.00 & 1.60 &I\\
55w122  & 08 44 12.33 & 44 31 14.9  & 0.56 $\pm$  0.08 & 0.66  $\pm$  0.12 & 1.27 & 1.45 &I\\
55w123  & 08 44 14.54 & 44 35 00.2  & 2.01 $\pm$  0.10 & 1.17  $\pm$  0.08 & 1.00 & 1.33 &I\\
55w124  & 08 44 14.93 & 44 38 52.2  & 4.67 $\pm$  0.17 & 2.79  $\pm$  0.16 & 1.00 & 1.35 &I\\
55w127  & 08 44 27.55 & 44 43 07.4  & 1.81 $\pm$  0.10 & 1.64  $\pm$  0.11 & 1.00 & 1.36 &I\\
55w128  & 08 44 33.05 & 44 50 15.3  & 3.34 $\pm$  0.18 & 4.77  $\pm$  0.54 & 1.00 & 2.05 &T\\
55w131  & 08 44 35.51 & 44 46 04.1  & 1.01 $\pm$  0.10 & 0.74  $\pm$  0.11 & 1.01 & 1.48 &T\\
55w132  & 08 44 37.12 & 44 50 34.7  & 1.10 $\pm$  0.11 & 1.66  $\pm$  0.23 & 1.01 & 2.05 &T\\
55w133  & 08 44 37.24 & 44 26 00.4  & 2.20 $\pm$  0.11 & 2.25  $\pm$  0.16 & 1.00 & 1.47 &I\\
55w135  & 08 44 41.10 & 44 21 37.7  & 2.60 $\pm$  0.14 & 3.86  $\pm$  0.44 & 1.00 & 1.98 &T\\
55w136  & 08 44 45.14 & 44 32 23.9  & 1.02 $\pm$  0.07 & 0.92  $\pm$  0.08 & 1.00 & 1.10 &T\\
55w137  & 08 44 46.90 & 44 44 37.9  & 1.60 $\pm$  0.09 & 1.66  $\pm$  0.11 & 1.00 & 1.29 &T\\
55w138  & 08 44 54.51 & 44 46 22.0  & 1.82 $\pm$  0.10 & 1.99  $\pm$  0.15 & 1.00 & 1.37 &I\\
55w140  & 08 45 06.06 & 44 40 41.2  & 0.79 $\pm$  0.08 & 0.55  $\pm$  0.06 & 1.25 & 1.06 &I\\
55w141  & 08 45 03.29 & 44 28 15.1  & 0.87 $\pm$  0.07 & 0.43  $\pm$  0.06 & 1.01 & 1.19 &I\\
55w143a & 08 45 05.49 & 44 25 45.0  & 2.41 $\pm$  0.11 & 2.19  $\pm$  0.13 & 1.00 & 1.34 &I\\
55w143b & 08 45 04.25 & 44 25 53.3  & 0.57 $\pm$  0.09 & 0.33  $\pm$  0.06 & 1.58 & 1.33 &I\\
55w147  & 08 45 23.83 & 44 50 24.6  & 1.72 $\pm$  0.11 & 1.97  $\pm$  0.19 & 1.00 & 1.82 &I\\
55w149  & 08 45 27.17 & 44 55 25.9  & 7.10 $\pm$  0.32 & 7.82  $\pm$  1.11 & 1.00 & 3.20 &T\\
55w150  & 08 45 29.47 & 44 50 37.4  & 0.95 $\pm$  0.10 & 0.63  $\pm$  0.10 & 1.02 & 1.88 &I\\
55w154  & 08 45 41.30 & 44 40 11.9  & 12.1 $\pm$  0.40 & 13.71 $\pm$  0.40 & 1.00 & 1.13 &T\\
55w155  & 08 45 46.89 & 44 25 11.6  & 1.83 $\pm$  0.10 & 1.70  $\pm$  0.14 & 1.00 & 1.55 &I\\
55w156  & 08 45 50.92 & 44 39 51.5  & 4.14 $\pm$  0.16 & 4.78  $\pm$  0.21 & 1.00 & 1.19 &T\\
55w157  & 08 46 04.44 & 44 45 52.7  & 1.37 $\pm$  0.10 & 1.24  $\pm$  0.12 & 1.00 & 1.68 &I\\
55w159a & 08 46 06.67 & 44 51 27.5  & 6.70 $\pm$  0.29 & 6.49  $\pm$  0.82 & 1.00 & 2.69 &I\\
55w159b & 08 46 06.82 & 44 50 54.1  & 0.75 $\pm$  0.13 & 1.00  $\pm$  0.19 & 1.00 & 2.55 &T\\
55w160  & 08 46 08.50 & 44 36 47.1  & 0.94 $\pm$  0.08 & 0.81  $\pm$  0.07 & 1.01 & 1.32 &I\\
55w161  & 08 46 27.32 & 44 29 56.9  & 1.34 $\pm$  0.14 & 1.25  $\pm$  0.15 & 1.02 & 1.87 &I\\
55w165a & 08 46 34.76 & 44 41 39.2  & 18.12$\pm$  0.54 & 18.88 $\pm$  1.54 & 1.00 & 2.06 &T\\
55w165b & 08 46 33.37 & 44 41 24.4  & 0.78 $\pm$  0.40 & 0.92  $\pm$  0.14 & 1.47 & 1.99 &I\\
55w166  & 08 46 36.02 & 44 30 53.5  & 2.46 $\pm$  0.14 & 2.31  $\pm$  0.26 & 1.00 & 2.07 &I\\
60w016  & 08 44 03.58 & 44 38 10.2  & 0.62 $\pm$  0.08 & 0.88  $\pm$  0.13 & 1.14 & 1.52 &I\\
60w024  & 08 44 17.83 & 44 35 36.9  & 0.51 $\pm$  0.09 & 0.37  $\pm$  0.05 & 2.06 & 1.29 &I\\
60w032  & 08 44 33.69 & 44 46 13.0  & 0.54 $\pm$  0.09 & 0.46  $\pm$  0.08 & 1.58 & 1.51 &I\\
60w039  & 08 44 42.50 & 44 45 32.5  & 0.65 $\pm$  0.09 & 0.72  $\pm$  0.16 & 1.26 & 1.38 &I\\
60w055  & 08 45 14.00 & 44 53 08.7  & 0.51 $\pm$  0.08 & 0.62  $\pm$  0.13 & 1.42 & 2.34 &I\\
60w067  & 08 45 40.47 & 44 23 20.1  & 0.56 $\pm$  0.09 & 0.69  $\pm$  0.15 & 1.31 & 1.70 &T\\
60w071  & 08 46 00.34 & 44 43 22.1  & 0.50 $\pm$  0.08 & 0.60  $\pm$  0.07 & 1.44 & 1.42 &I\\
60w084  & 08 46 39.86 & 44 33 44.5  & 0.85 $\pm$  0.17 & 1.80  $\pm$  0.37 & 1.38 & 2.08 &T\\
\hline
\end{tabular}
\end{minipage}
\end{table*}

\begin{table*}
\centering
\begin{minipage}{13.0cm}
\caption{\protect\label{a_table_no} The Hercules and Lynx field
  sources which did not fall within the optical field but were
  included in the infra--red observations. They are given a weight of
  0.00 as they are not part of the complete sample. They are included here
  for completeness. For full details see text.  An `I' in the final column indicates an
\emph{imfit} measured flux density; a `T' indicates a \emph{tvstat} measurement.} 
\begin{tabular}{l|cc|c|c|c|c|c}
\hline
Name    &\multicolumn{2}{c|}{RA/DEC (J2000)} & $S_{\rm Oort}$ (mJy) &
$S_{\rm 1.4 GHz}$ (mJy) & Weight & $C_{\rm PB}$ & Measure\\
\hline	               
53w091  & 17 22 32.73 & 50 06 01.9  & 22.6 $\pm$  1.1  & 37.93 $\pm$  6.62 & 0.00 & 7.43  &T\\
55w119  & 08 43 47.98 & 44 50 41.4  & 1.78 $\pm$  0.16 & 1.78  $\pm$  0.30 & 0.00 & 3.86  &I\\
55w125  & 08 44 15.28 & 44 11 16.2  & 22.2 $\pm$  1.20 & 12.15 $\pm$  2.30 & 0.00 & 12.01 &I\\
55w126  & 08 44 20.56 & 44 58 05.0  & 3.20 $\pm$  0.24 & 4.19  $\pm$  0.76 & 0.00 & 6.37  &I\\
\hline
\end{tabular}
\end{minipage}
\end{table*}

\section{\protect\label{opt_chap} Optical and infra--red imaging}          
\protect\label{opt_chap}
The FRI space density calculation depends on determining the redshifts
of the radio galaxies in the survey. To obtain all of these 
spectroscopically would have been very time consuming, so the aim was
to combine photometric redshift estimates with spectroscopic
follow--up of the best high redshift FRI candidates.  
Optical observations were carried out using the Wide Field Camera
(WFC) on the 2.5 m Isaac Newton Telescope (INT) in La Palma, and a
subsample of sources were also observed using the UKIRT Fast Track
Imager (UFTI) on UKIRT, the 3.8 m UK Infra--red Telescope located in
Hawaii. These two sets of observations are described in this section.  

\subsection{INT observations and data reduction}
The WFC consists of 4
thinned EEV 2kx4k CCDs with a pixel size of 13.5 $\mu$m, resulting in
a scale of $0.33\arcsec$/pixel and a combined field of view of
$\sim 34 \times 34~$arcmin$^2$. This large field of view makes the
WFC an ideal instrument for observing the two fields which are
of comparable size. 

The main WFC observations were split over two separate runs in April
2003 and April 2004. Unfortunately these were both largely weathered
out, so observations through two filters only were obtained - sloan 
\emph{r$^{\prime}$} and \emph{i$^{\prime}$}. Several exposures of 300s or 600s were taken, 
and the telescope was offset by 30\arcsec\ after every third
observation to fill the gaps between the CCDs and cover the whole
field. Full details of the observations can be found in Table
\ref{obs}. 

The April 2003 run took data on one night only; the 7th. Conditions were 
non-photometric, so these data have only been used to determine optical
identifications for the radio objects. For the April 2004 run useful
observations were only taken on the 15th. Standard star fields were
observed throughout the photometric night.

\begin{table*}
\centering
\begin{minipage}{126mm}
\caption{\label{obs} Summary of INT observations}
\begin{tabular}{c|c|c|c|c|c}
\hline
Field      & Band & Observation Date & Exposure Time & Photometric? & Seeing (\arcsec)      \\ 
\hline
Hercules   & r    & 07/04/03         & 24x300s       & No            &  2.5 \\
           & i    &   ''             & 10x300s       & No            &  1.9 \\
           & r    & 15/04/04         & 15x600s       & Yes           &  1.5 \\
           & i    &   ''             & 15x300s       & Yes           &  1.5 \\
Lynx       & i    & 07/04/03         & 9x300s        & No            &  2.1 \\
           & r    & 15/04/04         & 6x300s        & Yes           &  1.5 \\
           & i    & 06/01/05         & 1x300s        & Yes           &  3.0 \\
\hline
\end{tabular}
\end{minipage}
\end{table*}

One further observation of the Lynx field of 300s, in \emph{i$^{\prime}$}, was
obtained on 6th January 2005. This was needed because of the lack of
photometric \emph{i}--band data in this field. Although the night was photometric
the seeing was very poor ($\sim3\arcsec$). Standard star fields were again
observed throughout the night. This image was only used to
photometrically calibrate the Lynx field; the identifications were done with
the previous \emph{r} and \emph{i}--band images.  

All the images were processed using the IRAF software package. Bias
frames, taken at the beginning of each night, were averaged together to
make a master--bias for each detector which was then subtracted from the remaining
data. For the majority of the observations, a flat--field was made for
the four detectors, in each field and filter, by median combining the
separate science frames and rejecting 
pixels according to the readnoise and gain of the CCD. The exception
to this was the 15th April observations where twilight flats were
taken. Next the individual science frames were divided by the corresponding sky--flat,
which had been normalised using 
its pixel mean. The frames were registered using $\sim$10 stars and
shifted and the final images were then median combined and clipped
using the CCD noise properties as before. The offsets between frames
were sufficiently small that it was not necessary to account for 
distortions in this process.   

The astrometric calibration for the INT data was complicated by the
distortion of the WFC across the 4 CCDs (Taylor, 2000). To correct for
this the USNO\footnote{United States Naval
Observatory} B1.0 catalogue (Monet et al. 2003) was used to create a
catalogue of reference stars, which was then used by the starlink program \emph{astrom} to calculate a 
distortion--corrected calibration. For the April 2004 observations
this astrometric calibration was only applied to the \emph{r}--band data, as
it was felt that no further radio host--galaxy identifications would be obtained from considering
the significantly shallower \emph{i}--band data also. However, the \emph{i}--band images were
tweaked locally to each \emph{r}--band detection to ensure that the
images lined up. The calculated errors were $\sim0.3$\arcsec\ for all
CCDs in both fields.

\subsection{UKIRT observations and data reduction}

In contrast, UFTI consists of one 1024x1024 HgCdTe array with a
plate scale of $0.091\arcsec$/pixel. This results in a field of view
of 92$\arcsec$ which is significantly smaller than that of the WFC. This meant
that it could only be used to obtain images of, mainly, individual sources
rather than the complete--field observations done with the INT. 

\begin{table}
\centering
\caption{\protect \label{ukirt_obs} Summary of UKIRT observations. }
\begin{tabular}{c|c|c|c|c}
\hline
Field          & Date     & Target  & Exposure & Seeing (\arcsec)\\
&&Source&Time&\\
\hline
Hercules       & 25/07/04 & 53w054        & 54x60s        & 0.7    \\
	       & 28/07/04 & 53w084        & 36x60s        & 0.5    \\
	       &    ''    & 53w087        & 36x60s        & ''     \\ 
	       & 22/08/04 & 53w089        & 36x60s        & 0.9    \\
	       &    ''	  & 66w031        & 36x60s        & ''     \\
               & 12/09/04 & 66w009        & 36x60s        & 0.8    \\
 	       & 11/09/04 & 53w091        & 36x60s        & 0.7    \\
	       & 14/09/04 & 66w035        & 36x60s        & 1.1    \\
               & 20/09/04 & 66w036        & 36x60s        & 0.9    \\
Lynx           & 15/01/05 & 55w119        & 36x60s        & 0.9    \\
               & 21/01/05 & 55w128        & 36x60s        & 0.7    \\
               &    ''    & 55w132        & 36x60s        & ''     \\ 
               &    ''    & 55w120	  & 36x60s        & ''     \\
               &    ''    & 55w125	  & 27x60s        & ''     \\
               &    ''    & 55w126	  & 36x60s        & ''     \\
               &    ''    & 55w133        & 36x60s        & ''     \\
               & 23/01/05 & 55w135        & 36x60s        & 0.7     \\ 
               &    ''    & 55w138	  & 36x60s        & ''     \\
               &    ''    & 55w147        & 36x60s        & ''     \\
               & 24/01/05 & 55w155        & 36x60s        & 0.5     \\
               &    ''    & 55w136        & 36x60s        & ''     \\
               &    ''    & 55w128        & 36x60s        & ''     \\
               &    ''    & 55w132        & 36x60s        & ''     \\
               & 16/02/05 & 55w133        & 18x60s        & 0.9    \\
               & 17/02/05 & 55w121        & 36x60s        & 1.4    \\
               &    ''    & 55w123        & 36x60s        & ''     \\	 
               &    ''    & 55w156        & 36x60s        & ''     \\
               &    ''    & 55w143        & 36x60s        & ''     \\
\hline
\end{tabular}
\end{table}  

The UKIRT UFTI observations were done in a combination of service and
visitor mode, spread over the period July 2004 to January 2005. All
the observations were done using the \emph{K}--band filter;
details of the observations, along with the sources observed can be
found in Table \ref{ukirt_obs}. The sources selected for these
observations were those with a faint optical detection or no optical detection
at all. 

The sources were observed using a 9--point dither pattern, 
with offsets of 10$\arcsec$, and an exposure time of 60s per dither
position. In general this was repeated 4 times resulting in a total of 36
exposures. The exceptions to this were 53w054, where the observation had to be
re--started due to high humidity and 55w125, where software problems
meant the observation had to be halted after 27 exposures (3 repeats
of the dither pattern). Also, because
some observations were done in service, two sources which were not
detected after 36 exposures were able to be re--observed for a further 18 exposures
at a later date. The observations of 55w128 and 55w132 were also
repeated since the originals were taken at a very high airmass which
resulted in  
significant elongation of the objects in the field. 

Appropriate standard stars (FS125 for Lynx and FS27 for Hercules)
were observed multiple times throughout the night if multiple targets were also
observed, but only observed once on other nights. All nights were photometric.

The infra--red data reduction method is similar to that already
described for the optical data; again the IRAF software was
used to process the images, but each individual source observation was reduced
independently. The first step was the subtraction of the
appropriate dark frame from each image. Flat--fields were then
made for each source by median combining the first 9 observations
only, rejecting pixels according to the noise properties of the detector,
to minimise the effects of sky variability over the full length of the
exposure. The images were then divided by the flat--field which had
been normalised using its median pixel value. 

The final steps in the reduction process -- sky--subtraction, cosmic
ray removal and image
combining -- were done using the IRAF package \emph{dimsum}, by
P. Eisenhardt, M. Dickinson and S.A. Stanford, and, in
particular, the task \emph{reduce}. The
images were again registered using, on average, 10 stars. For the source
with 2 separate observations (55w133) the sky--subtraction was done
for the two nights separately, but all the 
images were then registered and combined together to produce a single final image. 

The astrometric calibrations for the images were done where possible
using the INT images as references. In cases with no INT overlap,
rough astrometry was derived from the telescope pointing position and
then the RA and DEC positions were improved using the one or two stars from the USNO
catalogue available in the small fields. 

\subsection{Aperture photometry and source identification}

This section outlines the steps taken to identify the host galaxies of
the sample sources and the subsequent magnitude measurements of these
objects.  

\subsubsection{Optical standard star calibration}

The standard star observations were reduced in the same way as the
science observations. The fields used were from the Landolt Faint
Equatorial Standards catalogue (Landolt, 1992), each containing an
average of 10 standard stars. The April 2004 standard field
was SA104 in \emph{r}--band only and SA107 in \emph{r} and
\emph{i}, whereas the January 2005 standard fields were SA104 and
SA98. The April 2003 observations were not photometric. Object counts
were measured using a 5\arcsec\ radius aperture for all standards using
the \emph{gaia} package; the only 
exception to this was for the January 2005 standards where high seeing
meant a larger aperture (15\arcsec\ radius) was needed, and stars with
near neighbours were ignored to minimise errors.   

The Landolt Faint Equatorial Standards are only available in 
the Johnson--Kron--Cousins photometric system, and are given in Vega magnitudes. Therefore the apparent magnitudes
for the Landolt standard stars needed to be transformed to the Sloan
photometric system. This transformation was done using the following two
relations from Smith et al., (2002):
\begin{eqnarray}
\label{trans1}
r' = V - 0.84(V-R) + 0.13 \\
r' - i' = 1.00(R-I) - 0.21 
\end{eqnarray}
In the above equations lowercase letters indicate the
Sloan photometric system and uppercase letters the Johnson
system. It should be noted that this also transforms the magnitudes to
the AB--magnitude system.   

Once the transformations had been applied, the calibration coefficients, zeropoint magnitude,
$m_{\rm zpt}$, and extinction co--efficient, $\kappa$, were determined for the
photometry. These are summarised in Table \ref{calib} and are in 
good agreement with previously published values for INT 
extinction. 

\begin{table}
\centering
\caption{\label{calib} The calibration co--efficients for the two observations}
\begin{tabular}{c|c|c|c}
\hline
Date            & Filter   & $\kappa$ (mag/airmass) & $m_{\rm zpt}$ \\
\hline
April 2004   & i & -0.03    & 24.24 $\pm$ 0.05 \\    
             & r & -0.07    & 24.68 $\pm$ 0.05 \\
\hline
January 2005 & i & -0.01     & 24.31 $\pm$ 0.05 \\
\hline
\end{tabular}
\end{table}

\subsubsection{Infra--red standard star calibration}

The UKIRT standard star observations were reduced in a similar way to
the science images, the one difference being that all 5 observations for each
standard were used to make the flat--field image. In contrast to the
INT Landolt standard star fields, which contained many stars, only one star was
used for Hercules (FS27) and one for Lynx (FS125). These were both
from the UKIRT Faint \emph{JHK} Standards catalogue and are given in
Vega magnitudes (Casali, 1992). The
aperture size used to measure the standards was 2.5\arcsec\ radius. 

As there is only one star per field and, in many cases, just one
standard star observation per night, the zeropoint magnitude and
extinction co--efficient could not both be determined. The extinction 
co--efficient, $\kappa$, was therefore taken to be
0.05 mags/airmass, the published value for UFTI (Leggett, 2005). The
zeropoint magnitudes for each observing night were then 
calculated using the published $m_{\rm app}$ for each standard; the
values found all lay in the range 22.35--22.40 and are in good
agreement with previous values given for UKIRT. For nights where more
than one source and hence more than one standard, were  
observed the mean value for $m_{\rm zpt}$ was used; the uncertainty of
$\pm$0.02 on this zeropoint was incorporated into the error estimates
of the source magnitudes as described in \S\ref{err_section}.  

\subsubsection{Identifications and magnitudes}

The source host--galaxy positions were found by overlaying the VLA
A--array radio contour maps with the optical and infra--red
data. Figures B1 and B2 show the radio/optical 
and, where appropriate, radio/infra--red  overlays resulting from the
UKIRT and INT April 2004 observations. Radio contour maps for sources with no host galaxy
identification are shown in Figure 3. The corresponding host galaxy
positions are given in Tables \ref{mags_herc} and \ref{mags_lynx}. 

The aperture photometry of all the sources was then done, again using the \emph{gaia} package. The
counts received for each source in \emph{r}, \emph{i}, and
\emph{K}--band (if available), were measured in 4 different
sized apertures -- $1.5\arcsec$,
$2.5\arcsec$, $4.0\arcsec$ and $8.0\arcsec$ radius -- and then, depending on the
extent of the source, the measurement from the most appropriate
aperture was selected and used thereafter. The aperture chosen for each source was the same 
in the three bands to enable colours to be accurately
determined. Sky--subtraction was achieved either using an annulus round the object or, in cases where 
this could not be done because of the proximity of other objects, using a sky--aperture placed nearby. 

The magnitudes were then calibrated using the appropriate values of
$\kappa$ and $m_{\rm zpt}$ determined for the optical and infra--red
standard stars; these can be found in Tables \ref{mags_herc} and \ref{mags_lynx} and are in
reasonable agreement with previously published results. The
\emph{K}--band magnitudes for sources observed with UKIRT but not
included in the complete sample are given in Table \ref{mag_nosamp}. 

\subsubsection{\protect \label{ap_cor} Aperture corrections}

The next step was to correct all the calculated source apparent magnitudes to a
metric aperture of 63.9~kpc diameter, thus allowing accurate
comparisons to be made between sources at all redshifts. The 63.9~kpc aperture, corresponding
to an aperture of \squig\ 8$\arcsec$  at $z=1$ has become
a standard metric size following previous work by Eales et
al. (1997) and others. 

At low redshift ($z<0.6$) this correction is carried out using the
curve of growth for elliptical galaxies tabulated by Sandage (1972);
this method assumes that the hosts of radio galaxies are 
all giant ellipticals and that they share the same intensity
profile. This assumption is a good approximation at low redshift, but
is not valid for higher redshift radio galaxies which can exhibit very
different structure. For radio galaxies located at $z>0.6$ therefore
the measured emission within an aperture of radius $r$ was assumed to
be proportional to $r^{\alpha}$ where $\alpha=0.35$ (Eales et al., 1997).

These aperture correction methods obviously depend on the redshifts for
the sources being known; only a small fraction of
the sample satisfied this condition. For the remaining objects 
redshifts were estimated iteratively using the K--z and r--z
relations, as described in \S\ref{z_est_1}. The calculated magnitude corrections for
the \emph{r}, \emph{i} and \emph{K}--band magnitudes can be found in
Tables \ref{mags_herc} and \ref{mags_lynx}. The aperture corrections
range from +0.15 to -0.52 magnitudes and are typically negative, with
an average correction of -0.3 mag to account for missing flux.     

\subsubsection{Magnitude error}
\label{err_section}
The magnitude error was determined by combining in quadrature four
potential sources of error: (i) the error on the received counts as
determined by Poisson error on the incoming photons, (ii) the error in the
determination of $m_{\rm zpt}$, (iii) the error due to the subtraction of the sky
background, found by taking the standard deviation of 10
apertures placed randomly on empty regions in the fields and (iv) the
error in the aperture correction which is taken as 50\% of the
correction value. If the source is 
bright (i) dominates; (iii) is most important for the infra--red
observations where the background is very high.  

\section{\protect\label{id_chap} Imaging results}
\protect\label{id_chap}

The April 2003 optical data resulted in an
identification fraction of 53\% and 63\% for the Lynx and Hercules fields
respectively. These numbers rose to 76\% and 87\% with the inclusion
of the optical data from April 2004. 80\% 
of the Hercules sources observed in the infra--red were identified,
compared with 57\% of the sources observed in the Lynx
field. Combining these figures gave an identification
fraction for the Lynx field of 83\%, and 90\% for
Hercules. 

In total, out of the complete sample, 4 radio sources in the Hercules
field and 7 radio sources in the Lynx field remain unidentified after
the \emph{r$^{\prime}$}, \emph{i$^{\prime}$} and \emph{K}--band observations. The observations reached
optical 3$\sigma$ limiting magnitudes of $r^{\prime}<25.17$ mag and
$i^{\prime}<23.76$ mag for Hercules, and $r^{\prime}<24.38$ mag and
$i^{\prime}<23.46$ mag for Lynx; the infra--red 3$\sigma$ limiting 
magnitudes were $K<19.85$ mag, $K<19.98$ mag and $K<20.16$ mag for the 36x60s, 54x60s and
72x60s observations respectively. Notes on individual sources can be found in Appendix \ref{notes_im}.

\begin{table}
\caption{\label{mag_nosamp} The
  host galaxy positions and \emph{K}--band magnitudes for the sources
  not included in the complete sample. The
  radius (in \arcsec) of the aperture used for photometry is given in
  brackets and 3$\sigma$ limits are given for undetected sources. The
  corresponding radio positions can be 
  found in Table \ref{a_table_no}.}
\begin{tabular}{l|c|c|c|c}
\hline
Name    &RA             & DEC        & K                       & K (63.9 kpc)   \\
&(J2000)& (J2000)&&\\
\hline
53w091  & 17 22 32.73 	& 50 06 01.9 & 18.40 $\pm$ 0.15 (2.5)  & 18.25 $\pm$ 0.17 \\ 
55w119  & 08 43 47.98 	& 44 50 41.4 & 19.54 $\pm$ 0.37 (2.5)  & 19.36 $\pm$ 0.38\\   	 
55w125  & 08 44 15.25 	& 44 11 16.7 & 17.44 $\pm$ 0.06 (2.5)  & 17.26 $\pm$ 0.11\\ 	 
55w126  &    --         &            &       $>$19.85          &        --       \\	 
\hline
\end{tabular}
\end{table}
\begin{table*}
\caption[The host galaxy positions and magnitudes for Hercules]{The
  host galaxy positions and magnitudes for the Hercules field. The
  radius (in \arcsec) of the aperture used for photometry is given in brackets. Sources which were unmeasurable
due to the presence of a nearby bright object are labelled with a * and
  3$\sigma$ limits are given for undetected sources\label{mags_herc}. The corresponding radio positions can be found in Table \ref{a_table_herc}. }
\begin{minipage}{180mm}
\begin{tabular}{l|c|c|c|c|c|c|c|c}
\hline
\multicolumn{9}{l}{Hercules}  \\
\hline
Name    &RA &DEC & r                     & r (63.9 kpc)    & i                     & i (63.9 kpc)    & K                     & K (63.9 kpc)     \\
&(J2000)& (J2000)& &  &  & & &\\
\hline	                 										     
53w052  & 17 18 34.07 & 49 58 50.2  & 21.31 $\pm$ 0.05 (4)  &21.19 $\pm$ 0.08 & 20.86 $\pm$ 0.07 (4)  & 20.74 $\pm$ 0.09&        --              &        --        \\  
53w054a & 17 18 47.30 & 49 45 49.0  & 23.74 $\pm$ 0.14 (2.5)&23.58 $\pm$ 0.16 & 23.62 $\pm$ 0.26 (2.5)& 23.46 $\pm$ 0.27& 18.32 $\pm$ 0.13 (2.5) & 18.17 $\pm$ 0.15 \\  
53w054b & 17 18 49.97 & 49 46 12.2  &       $>$25.17        &       --        &      $>$23.76         &        --       & 19.95 $\pm$ 0.59 (2.5) & 19.75 $\pm$ 0.60 \\  
53w057  & 17 19 07.29 & 49 45 44.8  & 24.69 $\pm$ 0.31 (2.5)&24.53 $\pm$ 0.32 &       $>$23.76        &        --       &        --              &        --        \\  
53w059  & 17 19 20.26 & 50 00 19.6  & 24.32 $\pm$ 0.22 (2.5)&24.17 $\pm$ 0.23 &       $>$23.76        &        --       &        --              &        --        \\  
53w061  & 17 19 27.34 & 49 43 59.7  & 21.13 $\pm$ 0.05 (4)  &21.13 $\pm$ 0.05 & 20.77 $\pm$ 0.07 (4)  & 20.77 $\pm$ 0.07&        --              &        --        \\  
53w062  & 17 19 32.07 & 49 59 06.8  & 21.91 $\pm$ 0.06 (2.5)&21.67 $\pm$ 0.14 & 21.04 $\pm$ 0.06 (2.5)& 20.80 $\pm$ 0.14&        --              &        --        \\  
53w065  & 17 19 40.07 & 49 57 40.8  & 23.00 $\pm$ 0.08 (2.5)&22.84 $\pm$ 0.11 & 23.31 $\pm$ 0.20 (2.5)& 23.14 $\pm$ 0.22&        --              &        --        \\  
53w066  &    --       &             &       $>$25.17        &       --        &       $>$23.76        &        --       &        --              &        --        \\	  
53w067  & 17 19 51.27 & 50 10 58.5  & 22.15 $\pm$ 0.06 (2.5)&21.94 $\pm$ 0.12 & 21.43 $\pm$ 0.06 (2.5)& 21.22 $\pm$ 0.12&        --              &        --        \\  
53w069  & 17 20 02.54 & 49 44 51.0  & 25.12 $\pm$ 0.46 (2.5)&24.97 $\pm$ 0.47 &       $>$23.76        &        --       &        --              &        --        \\	  
53w070  & 17 20 06.07 & 50 06 01.7  & 22.20 $\pm$ 0.06 (2.5)&22.05 $\pm$ 0.10 & 21.37 $\pm$ 0.06 (2.5)& 21.21 $\pm$ 0.10&        --              &        --        \\  
53w075  & 17 20 42.36 & 49 43 49.2  & 21.10 $\pm$ 0.05 (4)  &21.12 $\pm$ 0.05 & 20.67 $\pm$ 0.06 (4)  & 20.69 $\pm$ 0.06&        --              &        --        \\  
53w076  & 17 20 55.78 & 49 41 03.1  & 19.57 $\pm$ 0.05 (4)  &19.41 $\pm$ 0.10 & 18.91 $\pm$ 0.05 (4)  & 18.75 $\pm$ 0.10&        --              &        --        \\	  
53w077  & 17 21 01.32 & 49 48 34.1  & 21.71 $\pm$ 0.05 (4)  &21.69 $\pm$ 0.05 & 20.82 $\pm$ 0.07 (4)  & 20.80 $\pm$ 0.07&        --              &        --        \\  
53w078  & 17 21 18.17 & 50 03 34.9  & 18.28 $\pm$ 0.05 (8)  &18.29 $\pm$ 0.05 & 17.54 $\pm$ 0.05 (8)  & 17.54 $\pm$ 0.05&        --              &        --        \\  
53w079  & 17 21 22.62 & 50 10 31.2  & 20.62 $\pm$ 0.05 (4)  &20.54 $\pm$ 0.07 & 19.71 $\pm$ 0.05 (4)  & 19.62 $\pm$ 0.07&        --              &        --        \\  
53w080  & 17 21 37.46 & 49 55 36.9  & 18.22 $\pm$ 0.05 (8)  &18.37 $\pm$ 0.09 & 17.85 $\pm$ 0.05 (8)  & 18.00 $\pm$ 0.09&        --              &        --        \\  
53w081  & 17 21 37.81 & 49 57 56.9  & 23.99 $\pm$ 0.13 (1.5)&23.64 $\pm$ 0.19 & 23.36 $\pm$ 0.26 (1.5)& 23.01 $\pm$ 0.31&        --              &        --        \\  
53w082  & 17 21 37.64 & 50 08 27.4  & 25.01 $\pm$ 0.42 (2.5)&24.86 $\pm$ 0.43 &       $>$23.76        &        --       &        --              &        --        \\  
53w083  & 17 21 48.93 & 50 02 39.8  & 22.18 $\pm$ 0.06 (2.5)&21.94 $\pm$ 0.13 & 21.52 $\pm$ 0.06 (2.5)& 21.28 $\pm$ 0.13&        --              &        --        \\  
53w084  & 17 21 50.43 & 49 48 30.5  & 24.78 $\pm$ 0.34 (2.5)&24.61 $\pm$ 0.35 &       $>$23.76        &        --       & 19.46 $\pm$ 0.38 (2.5) & 19.29 $\pm$ 0.39 \\  
53w085  & 17 21 52.47 & 49 54 34.0  & 22.17 $\pm$ 0.06 (2.5)&22.01 $\pm$ 0.10 & 21.93 $\pm$ 0.07 (2.5)& 21.77 $\pm$ 0.10&        --              &        --        \\  
53w086a & 17 21 56.42 & 49 53 39.8  & 20.22 $\pm$ 0.05 (4)  &20.10 $\pm$ 0.08 & 19.44 $\pm$ 0.05 (4)  & 19.32 $\pm$ 0.08&        --              &        --        \\  
53w086b & 17 21 57.65 & 49 53 33.8  & 22.08 $\pm$ 0.06 (2.5)&21.69 $\pm$ 0.12 & 20.95 $\pm$ 0.06 (2.5)& 20.56 $\pm$ 0.12&        --              &        --        \\  
53w087  &   --        &             & $>$25.17              &       --        &       $>$23.76        &        --       &       $>$19.85          &        --        \\	  
53w088  &  --         &             & $>$25.17              &       --        &       $>$23.76        &        --       &        --              &        --        \\	  
53w089  & 17 22 01.02 & 50 06 51.7  & 24.27 $\pm$ 0.16 (1.5)&23.84 $\pm$ 0.22 &       $>$23.76        &        --       &       $>$19.85          &        --        \\ 
66w009a & 17 18 32.87 	& 49 55 53.9 &	23.11 $\pm$ 0.08 (1.5)&22.68 $\pm$ 0.22  & 22.36 $\pm$ 0.11 (1.5)& 21.93 $\pm$ 0.24& 16.94 $\pm$ 0.02 (1.5) & 16.51 $\pm$ 0.21 \\ 
66w009b & 17 18 33.80 	& 49 56 02.2 &	17.71 $\pm$ 0.05 (4)  &17.19 $\pm$ 0.26  & 17.19 $\pm$ 0.05 (4)  & 16.67 $\pm$ 0.26& 13.79 $\pm$ 0.01 (4)   & 13.26 $\pm$ 0.26 \\ 
66w014  & 17 18 53.49   & 49 52 39.3 &        *               &        --        &        *              &        --       &        --              &        --       \\ 
66w027  & 17 19 52.11 	& 50 02 12.7 &	18.33 $\pm$ 0.05 (8)  & 17.99 $\pm$ 0.19 & 17.81 $\pm$ 0.05 (8)  & 17.46 $\pm$ 0.19&        --              &        --       \\ 
66w031  & 17 20 06.87 	& 49 43 57.0 &	22.65 $\pm$ 0.07 (2.5)& 22.45 $\pm$ 0.12 & 22.43 $\pm$ 0.10 (2.5)& 22.23 $\pm$ 0.14& 17.96 $\pm$ 0.10 (2.5) & 17.76 $\pm$ 0.14\\ 
66w035  & 17 20 12.41 	& 49 57 08.7 &	23.47 $\pm$ 0.11 (2.5)& 23.31 $\pm$ 0.14 & 23.12 $\pm$ 0.17 (2.5)& 22.95 $\pm$ 0.19& 19.10 $\pm$ 0.29 (2.5) & 18.94 $\pm$ 0.30\\ 
66w036  & 17 20 21.46 	& 49 46 58.3 &	22.79 $\pm$ 0.07 (2.5)& 22.60 $\pm$ 0.12 & 21.82 $\pm$ 0.07 (2.5)& 21.63 $\pm$ 0.12& 17.45 $\pm$ 0.06 (2.5) & 17.26 $\pm$ 0.11\\ 
66w042  & 17 20 52.20 	& 49 42 49.2 &	21.21 $\pm$ 0.05 (4)  & 21.16 $\pm$ 0.06 & 21.01 $\pm$ 0.07 (4)  & 20.96 $\pm$ 0.08&        --              &        --       \\ 
66w047  & 17 21 05.48 	& 49 56 55.9 &	19.30 $\pm$ 0.05 (8)  & 19.38 $\pm$ 0.06 & 18.80 $\pm$ 0.06 (8)  & 18.87 $\pm$ 0.07&        --              &        --       \\ 
66w049  & 17 21 11.21 	& 49 58 32.9 &	22.59 $\pm$ 0.07 (2.5)& 22.41 $\pm$ 0.11 & 22.16 $\pm$ 0.08 (2.5)& 21.98 $\pm$ 0.12&        --              &        --       \\ 
66w058  &   --          &            &       $>$25.17         &        --        &       $>$23.76        &        --       &        --              &        --       \\ 
\hline
\end{tabular}
\end{minipage}
\end{table*}

\begin{table*}
\caption[The host galaxy positions and magnitudes for Lynx]{The
  host galaxy positions and magnitudes for the Lynx field. The
  radius (in \arcsec) of the aperture used for photometry is given in
  brackets and 3$\sigma$ limits are given for undetected sources\label{mags_lynx}. The corresponding radio positions can be
  found in Table \ref{a_table_lynx}. }  
\begin{minipage}{180mm}
\begin{tabular}{l|c|c|c|c|c|c|c|c}		
\hline
\multicolumn{9}{l}{Lynx}  \\
\hline
Name    &RA &DEC & r                     & r (63.9 kpc)    & i                     & i (63.9 kpc)    & K                     & K (63.9 kpc)     \\
&(J2000)& (J2000)& & &  &  & &\\
\hline
55w116  & 08 43 40.79 	& 44 39 25.5 & 22.03 $\pm$ 0.07 (2.5)&21.84 $\pm$ 0.12& 21.16 $\pm$ 0.10 (2.5)& 20.97 $\pm$ 0.14 &        --               &        --       \\	 
55w118  & 08 43 46.86 	& 44 35 49.7 & 21.29 $\pm$ 0.06 (4)  &21.24 $\pm$ 0.07& 20.89 $\pm$ 0.08 (4)  & 20.84 $\pm$ 0.08 &        --               &        --       \\	 
55w120  & 08 43 52.87 	& 44 24 29.1 &       $>$24.38        &      --        &       $>$23.46        &        --        & 18.12 $\pm$ 0.10 (2.5)  & 17.96 $\pm$ 0.13\\	 
55w121  & 08 44 04.01 	& 44 31 20.3 & 23.15 $\pm$ 0.14 (2.5)&22.98 $\pm$ 0.16&       $>$23.46        &        --        & 19.35 $\pm$ 0.35 (2.5)  & 19.17 $\pm$ 0.36\\	 
55w122  & 08 44 12.10 	& 44 31 17.5 & 20.74 $\pm$ 0.05 (4)  &20.66 $\pm$ 0.07& 20.56 $\pm$ 0.08 (4)  & 20.47 $\pm$ 0.09 &        --               &        --       \\	 
55w123  & 08 44 14.54 	& 44 35 00.2 & 22.90 $\pm$ 0.12 (2.5)&22.71 $\pm$ 0.15& 22.98 $\pm$ 0.36 (2.5)& 22.79 $\pm$ 0.37 & 17.30 $\pm$ 0.06 (2.5)  & 17.10 $\pm$ 0.11\\	 
55w124  & 08 44 14.93 	& 44 38 52.2 & 21.22 $\pm$ 0.06 (4)  &21.16 $\pm$ 0.07& 21.28 $\pm$ 0.10 (4)  & 21.22 $\pm$ 0.10 &        --               &        --       \\	 
55w127  & 08 44 27.15 	& 44 43 08.0 & 14.18 $\pm$ 0.05 (8)  &13.63 $\pm$ 0.28& 14.12 $\pm$ 0.07 (8)  & 13.57 $\pm$ 0.28 &        --               &        --       \\	 
55w128  &   --          &            &       $>$24.38        &       --       &       $>$23.46        &        --        &       $>$20.16          &        --       \\	 
55w131  & 08 44 35.51 	& 44 46 04.1 & 23.18 $\pm$ 0.15 (2.5)&23.02 $\pm$ 0.17& 21.79 $\pm$ 0.14 (2.5)& 21.62 $\pm$ 0.16 &        --               &        --       \\	 
55w132  &  --         	&            &       $>$24.38        &       --       &       $>$23.46        &        --        &       $>$20.16          &        --       \\ 	 
55w133  & 08 44 37.24   & 44 26 00.4 & 25.51 $\pm$ 1.03 (1.5)&25.15 $\pm$ 1.05&       $>$23.46        &        --        &       $>$19.98          &        --       \\	 
55w135  & 08 44 41.10 	& 44 21 37.7 &	     $>$24.38        &       --       &       $>$23.46        &        --        & 13.23 $\pm$ 0.02 (10)   & 13.12 $\pm$ 0.23\\	 
55w136  & 08 44 45.09 	& 44 32 27.3 & 23.80 $\pm$ 0.22 (1.5)&23.45 $\pm$ 0.28&      $>$23.46         &        --        & 19.17 $\pm$ 0.16 (1.5)  & 18.81 $\pm$ 0.24 \\  	 
55w137  &     --        &            &       $>$24.38        &       --       &      $>$23.46         &        --        &        --               &        --        \\         
55w138  & 08 44 54.45 	& 44 26 22.0 &       $>$24.38        &       --       &       $>$23.46        &        --        & 19.72 $\pm$ 0.25 (1.5)  & 19.34 $\pm$ 0.31 \\  	 
55w140  & 08 45 06.06 	& 44 40 41.2 & 20.72 $\pm$ 0.05 (4)  &20.75 $\pm$ 0.05& 20.94 $\pm$ 0.08 (4)  & 20.96 $\pm$ 0.08 &        --               &        --        \\	 
55w141  &     --        &            &       $>$24.38        &      --        &       $>$23.46        &        --        &        --               &         --        \\		 
55w143a & 08 45 05.62 	& 44 25 42.9 & 25.38 $\pm$ 0.92 (1.5)&25.03 $\pm$ 1.00&       $>$23.46        &        --        &       $>$19.85          &        --        \\	 
55w143b & 08 45 04.25   & 44 25 53.3 & 25.46 $\pm$ 0.98 (1.5)&25.11 $\pm$ 0.98&       $>$23.46        &        --        &       $>$19.85          &        --        \\	 
55w147  & 08 45 23.83 	& 44 50 24.6 & 23.07 $\pm$ 0.13 (2.5)&22.90 $\pm$ 0.16&       $>$23.46        &        --        & 17.68 $\pm$ 0.07 (2.5)  & 17.51 $\pm$ 0.11 \\         
55w149  & 08 45 27.17  	& 44 55 25.9 & 16.49 $\pm$ 0.05 (8)  &16.34 $\pm$ 0.09& 15.94 $\pm$ 0.07 (8)  & 15.80 $\pm$ 0.10 &        --               &        --        \\   
55w150  & 08 45 29.47  	& 44 50 37.4 & 20.82 $\pm$ 0.05 (4)  &20.70 $\pm$ 0.08& 20.12 $\pm$ 0.07 (4)  & 20.00 $\pm$ 0.09 &        --               &        --        \\  
55w154  & 08 45 41.30  	& 44 40 11.9 & 19.19 $\pm$ 0.05 (8)  &19.25 $\pm$ 0.06& 18.59 $\pm$ 0.08 (8)  & 18.64 $\pm$ 0.08 &        --               &        --        \\  
55w155  &    --         &            &       $>$24.38        &       --       &       $>$23.46        &        --        &       $>$19.85          &        --        \\  
55w156  & 08 45 50.92  	& 44 39 51.5 & 22.75 $\pm$ 0.10 (2.5)&22.56 $\pm$ 0.14& 23.03 $\pm$ 0.37 (2.5)&22.84 $\pm$ 0.38  & 17.27 $\pm$ 0.05 (2.5)  & 17.07 $\pm$ 0.11 \\  
55w157  & 08 46 04.44  	& 44 45 52.7 & 22.07 $\pm$ 0.07 (2.5)&21.77 $\pm$ 0.16& 21.45 $\pm$ 0.11 (2.5)& 21.15 $\pm$ 0.18 &        --               &        --        \\  
55w159a & 08 46 06.67  	& 44 51 27.5 & 23.54 $\pm$ 0.08 (2.5)&23.38 $\pm$ 0.22&       $>$23.46        &        --        &        --               &        --        \\  
55w159b & 08 46 06.66  	& 44 50 53.8 & 18.70 $\pm$ 0.05 (8)  &18.74 $\pm$ 0.05& 18.11 $\pm$ 0.07 (8)  & 18.15 $\pm$ 0.07 &                         &                  \\	   
55w160  & 08 46 08.57  	& 44 36 47.4 & 21.40 $\pm$ 0.06 (4)  &21.33 $\pm$ 0.07& 20.24 $\pm$ 0.07 (4)  & 20.17 $\pm$ 0.08 &        --               &        --        \\  
55w161  & 08 46 27.46  	& 44 29 57.1 & 20.07 $\pm$ 0.05 (4)  &19.94 $\pm$ 0.08& 19.46 $\pm$ 0.07 (4)  & 19.33 $\pm$ 0.10 &        --               &        --        \\  
55w165a & 08 46 34.78  	& 44 41 37.6 & 21.36 $\pm$ 0.06 (4)  &21.31 $\pm$ 0.06& 20.31 $\pm$ 0.07 (4)  & 20.26 $\pm$ 0.07 &        --               &        --        \\  
55w165b & 08 46 33.37  	& 44 41 24.4 & 21.65 $\pm$ 0.07 (4)  &21.62 $\pm$ 0.07& 20.89 $\pm$ 0.08 (4)  & 20.86 $\pm$ 0.08 &        --               &        --        \\  
55w166  & 08 46 36.02 	& 44 30 53.5 & 22.72 $\pm$ 0.10 (2.5)&22.54 $\pm$ 0.13& 21.99 $\pm$ 0.16 (2.5)& 21.81 $\pm$ 0.18 &        --               &        --        \\  
60w016  & 08 44 03.58  	& 44 38 10.2 & 22.74 $\pm$ 0.10 (2.5)&22.55 $\pm$ 0.14& 21.58 $\pm$ 0.12 (2.5)& 21.38 $\pm$ 0.15 &        --               &        --        \\  
60w024  & 08 44 17.83  	& 44 35 36.9 & 21.97 $\pm$ 0.08 (4)  &21.94 $\pm$ 0.08& 20.81 $\pm$ 0.08 (4)  & 20.78 $\pm$ 0.08 &        --               &        --        \\  
60w032  &     --        &            &       $>$24.38        &       --       &       $>$23.46        &        --        &        --               &        --        \\  
60w039  & 08 44 42.50  	& 44 45 32.5 & 17.23 $\pm$ 0.05 (8)  &17.08 $\pm$ 0.08& 16.78 $\pm$ 0.07 (8)  & 16.63 $\pm$ 0.10 &        --               &        --        \\  
60w055  & 08 45 14.00  	& 44 53 08.7 & 21.85 $\pm$ 0.06 (2.5)&21.63 $\pm$ 0.12& 20.79 $\pm$ 0.08 (2.5)& 20.57 $\pm$ 0.13 &        --               &        --        \\  
60w067  &    --         &            &       $>$24.38        &       --       &       $>$23.46        &        --        &        --               &        --        \\  
60w071  & 08 46 00.34   & 44 43 22.1 & 23.44 $\pm$ 0.18 (2.5)&23.28 $\pm$ 0.20&       $>$23.46        &        --        &        --               &        --        \\  
60w084  & 08 46 40.23   & 44 33 44.7 & 17.79 $\pm$ 0.05 (8)  &17.58 $\pm$ 0.11& 17.06 $\pm$ 0.07 (8)  & 16.85 $\pm$ 0.12 &                         &                  \\  
\hline
\end{tabular}
\end{minipage}
\end{table*}

\subsection{Colours and magnitude distribution}

The $(r-i)$ and $(r-K)$ colour--magnitude diagrams for both Lynx and
Hercules sources are shown in Figure \ref{k_plot}. The $(r-i)$ plot
suggests a slightly greater range in the colours of the sources in the Lynx
field compared to the Hercules field, but this is likely to be the
result of the relative shallowness of the Lynx \emph{i}--band data,
which would introduce a bias against the fainter bluer sources in this
field, rather than a result of the aforementioned radio source count
underrepresentation. The $(r-i)$ colours for both fields though show
that the majority of the sources are hosted by red galaxies as
expected.   

\begin{figure*}
\begin{minipage}{170mm}
\includegraphics[scale=0.4, angle=90]{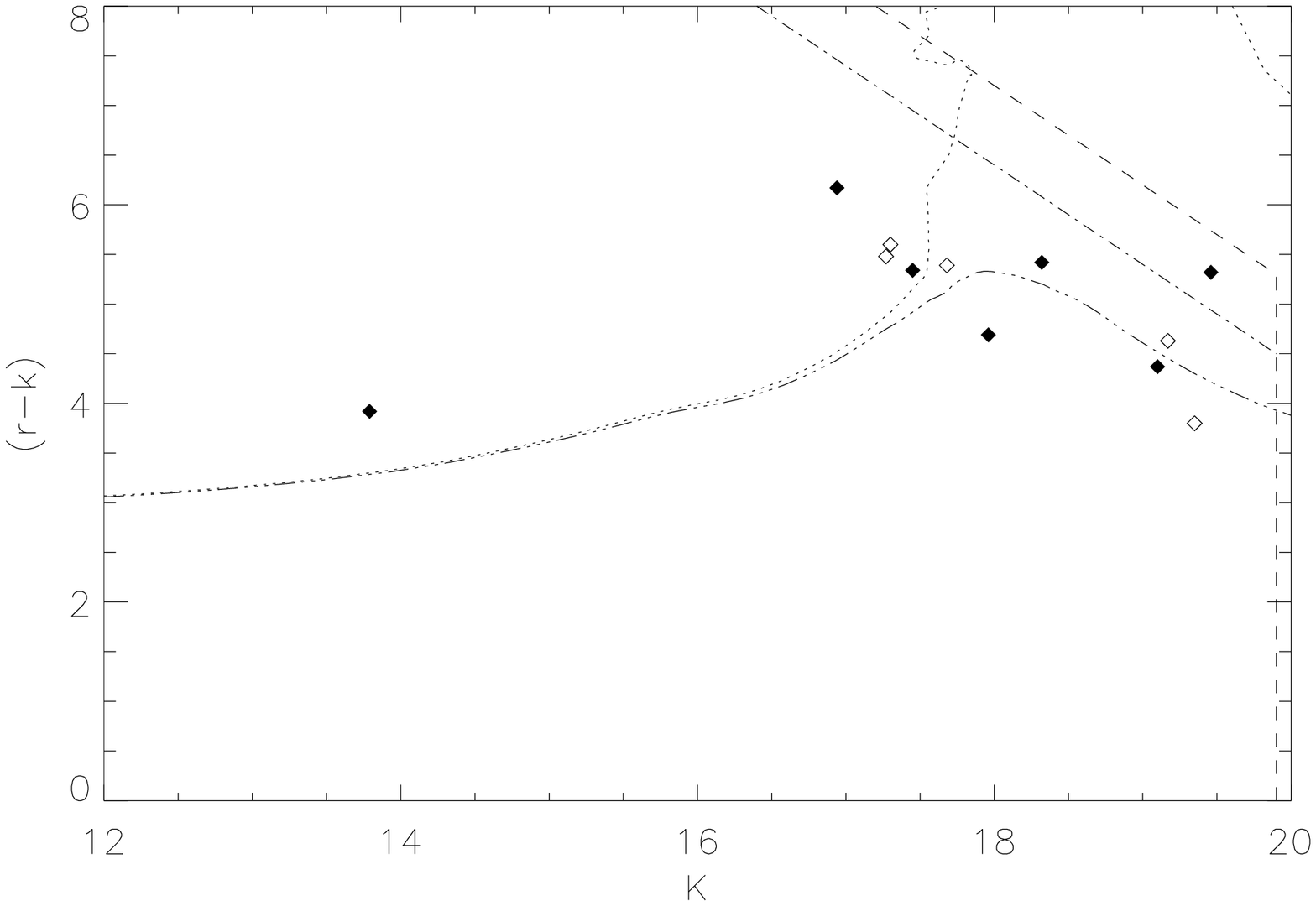} 
\includegraphics[scale=0.4, angle=90]{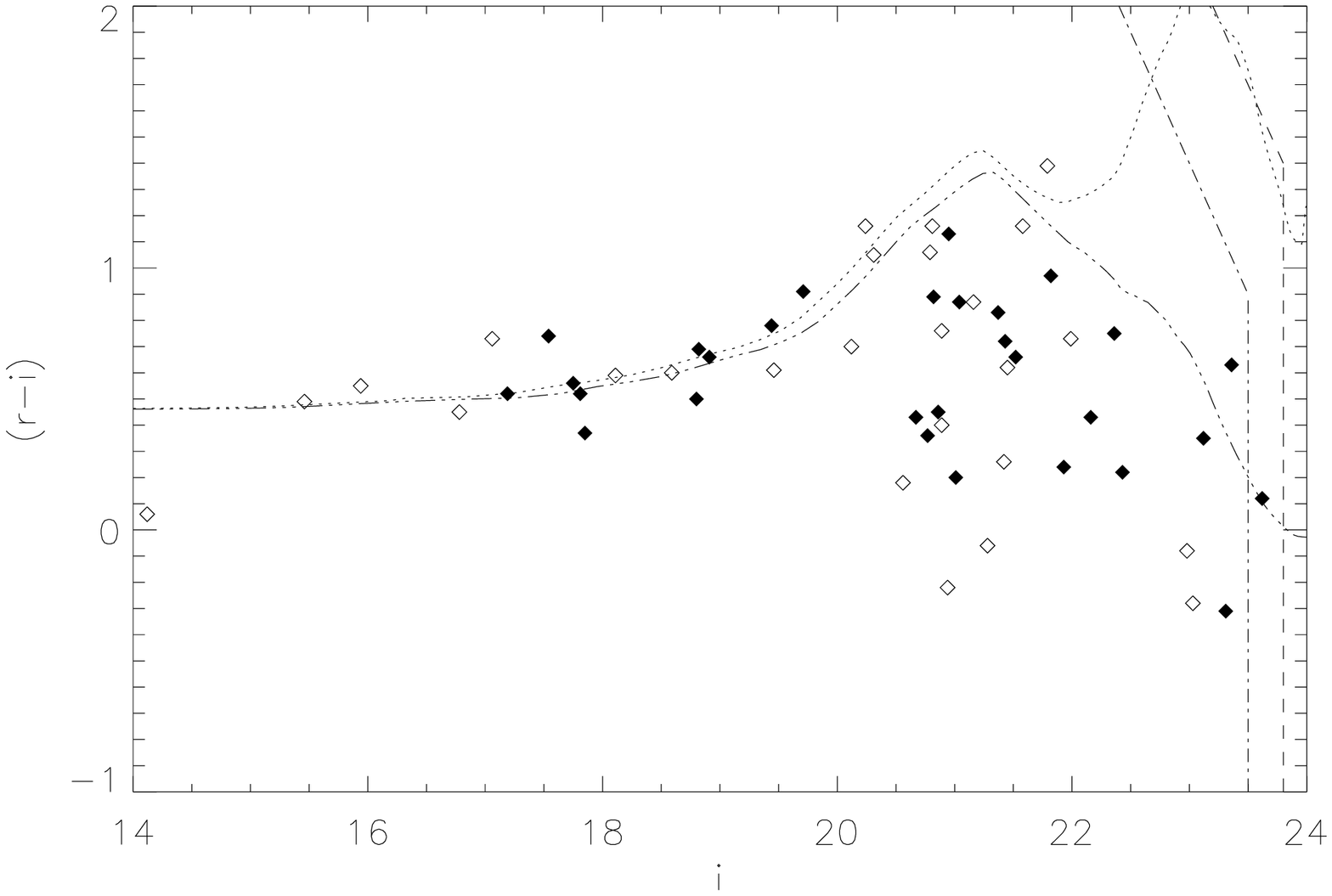}
\caption{\label{k_plot} Colour magnitude diagrams and completeness limits for the
Lynx (open diamonds; dot--dashed lines) and Hercules (filled diamonds;
dashed lines) results. Overplotted is the predicted behaviour of an
L$^{*}$ type galaxy for two models (Bruzual \& Charlot, 2003): passive
evolution following an instantaneous burst of star formation at $z=5$
(dotted line), and exponentially declining star formation, formed at
$z=10$, with an e--folding time of 1 Gyr (dot--dot--dot--dashed
lines).} 
\end{minipage}
\end{figure*}

\begin{figure*}
\begin{minipage}{170mm}
\includegraphics[scale=0.4, angle=90]{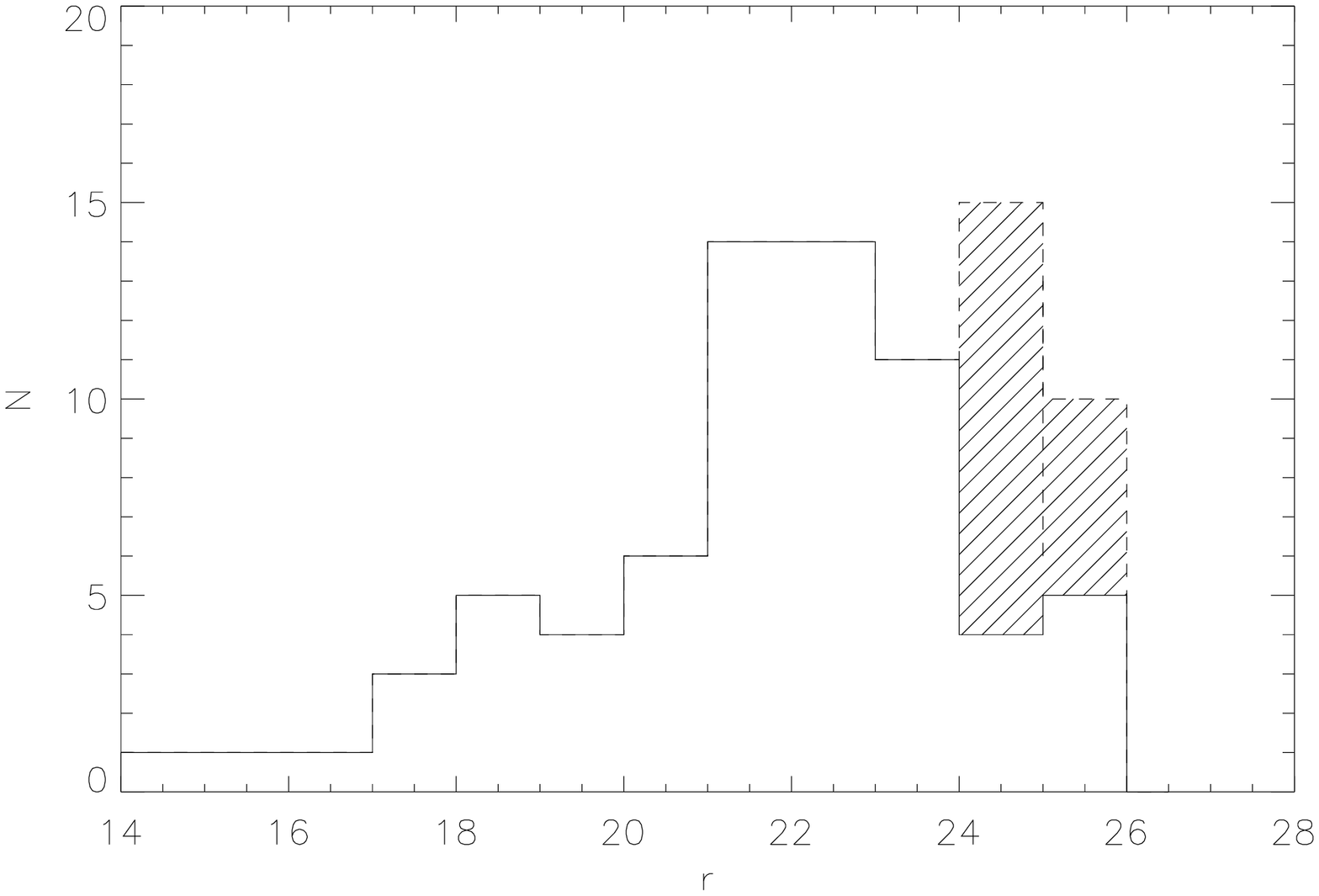}
\includegraphics[scale=0.4, angle=90]{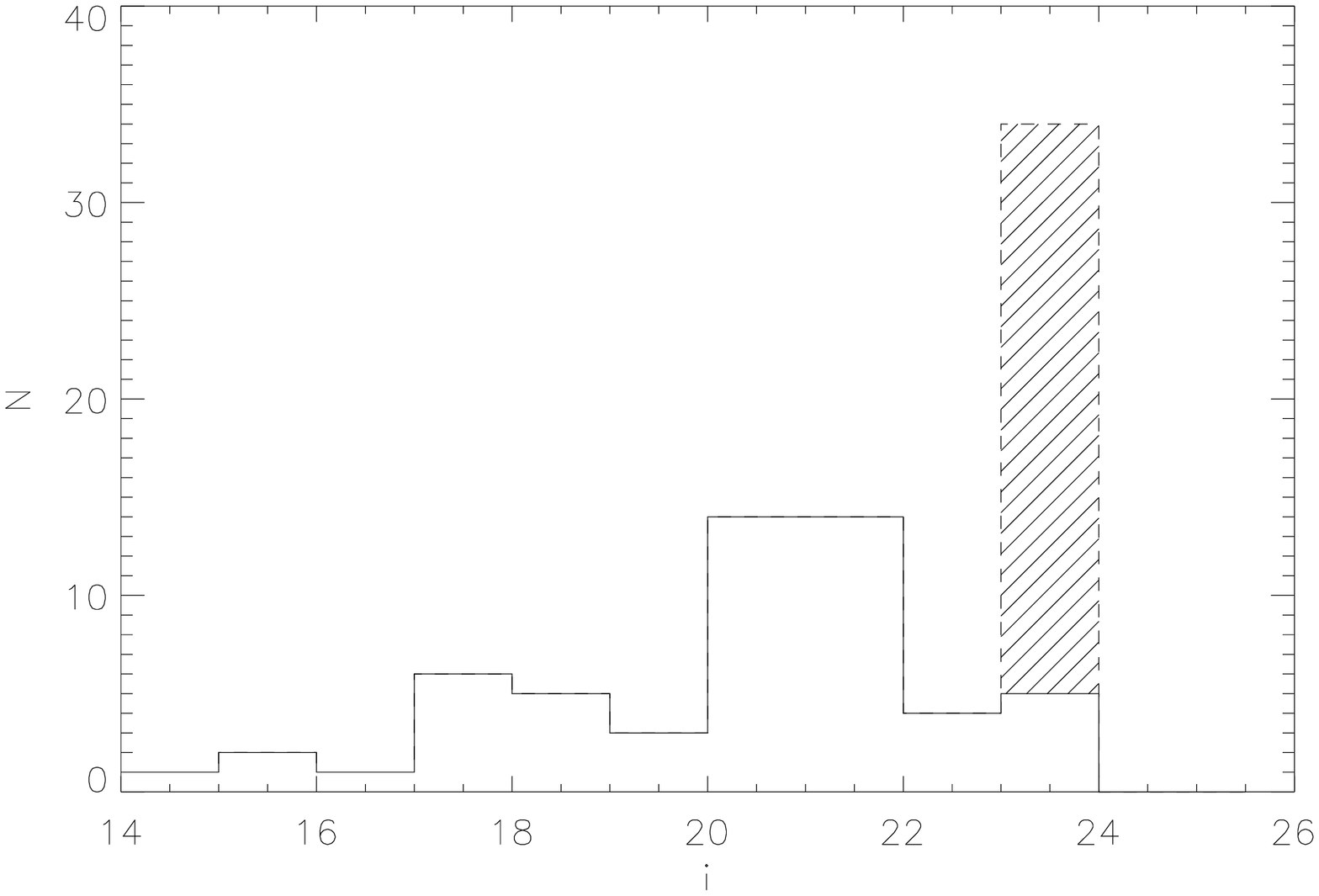}
\caption{\label{hist} Magnitude distribution for the combined Lynx and
Hercules complete sample in \emph{r$^{\prime}$} (left) and \emph{i$^{\prime}$} (right). The shaded
regions show the undetected sources at the magnitude limits.}
\end{minipage}
\end{figure*}

The Lynx field sources in the $(r-K)$ plot have a similar colour
distribution to the Hercules sources, but the small 
number of sources in this diagram make a comparison difficult. The
small numbers and poor population are 
 the result of the selection criteria used for the
\emph{K}--band observations; sources with either an undetected or
faint \emph{r}--band detection were those chosen.

Also shown on Figure \label{k_plot} are two likely models for an
L$^{*}$ type galaxy (assuming $M^{*}_{\rm i} = -22.0$ (Blanton et al.,
2001)), calculated using the GALAXEV code (Bruzual \& Charlot 2003): passive evolution
following an initial burst of star formation at $z=5$, and
exponentially declining star formation, with an e--folding time of 1
Gyr, beginning at $z=10$. The $(r-i)$ vs \emph{i} model lines track
the data well out to \emph{i}\squig\ 20 (corresponding to $z$ \squig~
0.6), beyond which the observed galaxies tend to be bluer than the
models. This tendency is not unexpected as the blue rest--frame
wavelengths probed at those redshifts mean that a smaller amount of
recent star formation or AGN activity will significantly bluen the
galaxy colours. In the $(r-K)$ plot, the observed galaxies are in good
agreement with the models, but again the small number of sources
included make drawing conclusions difficult. 

The magnitude distribution histograms (Figure \ref{hist}) are useful as they provide a
first look at the redshift distribution of the sample through the
magnitude--redshift relations for radio galaxies (these are described
in more detail in \S\ref{relns}), which indicate that the optically
faintest objects should lie at the largest distances. The
\emph{K}--band magnitude distribution is ignored here 
because the low number of K--magnitudes taken would not result in a
meaningful diagram. 

Both the colour--magnitude and magnitude distribution diagrams are
also in reasonable agreement with those shown in Figures 9 and 10
respectively of Waddington et al. (2000). 

\section{Spectroscopic TNG observations \protect\label{spec_chap}}
 \protect\label{spec_chap}
The redshifts for the radio sources identified in the Lynx and
Hercules fields are vital in determining their  cosmic
evolution. Previously published spectroscopic or photometric redshifts (Waddington et
al. 2000 and references therein; Bershady et al. 1994) already exist for 19 of the
Hercules field sources, and 3 of the Lynx field sources had
spectroscopic redshifts from the SDSS; the remaining sources had no previous redshift
information. This section covers the
spectroscopic observations made of a selection of the sample with the
multi--object spectrograph 
DOLORES on the 3.58 m Telescopio Nazionale Galileo (TNG), along with the
redshift estimation methods used for the remaining sources.

DOLORES, the Device Optimized for the LOw RESolution, consists of one
Loral back--illuminated and thinned 2048x2048 pixel CCD, with a scale of
0.275$\arcsec$/pixel, resulting in a field of view of
9.4\arcmin x 9.4\arcmin. For multi--object spectroscopy (MOS)
observations, rectangular masks with dimensions
6.0\arcmin x 7.7\arcmin,  are used. Vertical slits of constant width
(either 1.1$\arcsec$ or 1.6$\arcsec$) but varying length are drilled
in the masks according to the positions of the various sources for
which spectra are required. 

Since some uncertainties in the astrometry remained the 
1.6$\arcsec$ slit was used for the observations. Only 10 masks
were permitted per observing run, so it was 
decided to create 5 masks of varying position angle for each for the
two fields, thus covering as many sources as possible. \S\ref{MOS}
below describes the mask creation and source selection process. 

The DOLORES observations took place from 18th to 20th April
2004. All nights were photometric and the standard star Feige 34 was observed
(with a 5$\arcsec$ longslit) at regular intervals. Each mask observation consisted
of several long exposures 
to avoid saturation of the CCD, and to allow cosmic ray hits to be
identified in the final spectra. The good weather 
conditions also allowed two observations with
the 1.5$\arcsec$ longslit of three sources (two in Hercules and one in
Lynx), which were not included on the masks. All observations were
carried out using the LR-R grism which has a wavelength range of
4470--10360\AA\ and a resolution of 11.0\AA. Full details of the
observations can found in Table \ref{TNG_obs}. 

The masks for the Lynx field were, on average, observed for less time
than those for Hercules due to the early setting time of the Lynx field.

\subsection{MOS Mask Creation \label{MOS}}

The mask limitations meant that not all the radio sources could be
included in the observations, therefore a ranking system was
introduced to ensure that the optimum number of interesting, high--redshift, sources
were included. First, \emph{r}--band magnitudes were estimated for all the
sources using a rough $m_{\rm zpt}$ and ignoring the airmass and
extinction corrections. (These magnitudes could only be estimated as,
at the time these observations were being prepared, only the
non--photometric April 2003 WFC data were available.) 

The resulting magnitudes were then used to give an indication of the redshifts of the
sources through the r--z relation (Snellen et al. 1996). Those sources
with previously published redshifts were 
obviously excluded from the rankings. Table \ref{rank}
gives the full details of the ranking scheme.  

\begin{table}
\centering 
\caption{\label{rank} The ranking scheme for the radio sources. }
\begin{tabular}{c|c|c}
\hline
Redshift & \emph{r} magnitude & Rank \\
\hline
z$>$0.5 & 21-22       & 1    \\
      & $>$22         & 2    \\
\hline                           
z$<$0.5 & $<$21         & 3    \\ 
      & $<$15         & 4    \\
\hline
\end{tabular}
\end{table} 

The masks were then arranged such that the maximum number of first
rank sources would be observed. This was done using an IDL script
which allowed possible combinations of objects to be rotated to
determine the best fit parameters for each mask. Table \ref{TNG_obs}
gives the final determined parameters. In the 
Lynx field 56\% of all the sources were included, and of these 30\%
were first rank and 78\% were
second rank or above. The Hercules field was slightly better with 62\% of
the total number included; 80\% classed as second rank or
higher and 11\% first rank. 

\begin{table*}
\centering
\begin{minipage}{150mm}
\caption{\label{TNG_obs} Summary of the TNG observations. H and L
  refer to the Hercules and Lynx fields respectively.} 
\begin{tabular}{c|c|c|c|c|c|c}
\hline
Mask & Position Angle ($^{\circ}$) & Centre (J2000) & Slit Length
($\arcsec$) & Observation & Exposure & Seeing (\arcsec) \\
& & & &Date &Time & \\
\hline
L1 & 0    &8 45 53.78 +44 43 01.8 &  12 & 19/04/04 &  4x1500s  & 1.3 \\
L2 & 20   &8 45 24.20 +44 53 43.4 &  12 & 20/04/04 &  3x1800s  & 0.8 \\
L3 & 0    &8 44 55.98 +44 44 04.8 &  12 & 19/04/04 &  2x1800s  & 1.3 \\
L4 & -5   &8 44 36.14 +44 46 50.1 &  10 & 18/04/04 &  3x1200s  & 1.3 \\
L5 & -95  &8 44 00.00 +44 37 08.0 &  12 & 18/04/04 &  3x1500s  & 1.3 \\ 
\hline	   		
H1 & -40  &17 21 51.57 +49 50 31.3 &  11 & 18/04/04 & 4x1500s  & 1.3 \\
H2 & 10   &17 21 52.74 +50 09 11.4 &  12 & 19/04/04 & 4x1800s  & 1.3 \\
H3 & -15  &17 20 14.86 +49 47 20.3 &  12 & 18/04/04 & 3x1800s, 1x492s  & 1.3 \\ 
H4 & -45  &17 20 09.00 +50 01 33.6 &  12 & 20/04/04 & 4x1800s  & 0.8 \\ 
H5 & -90  &17 19 05.97 +49 45 59.4 &  12 & 19/04/04 & 3x1800s, 1x100s, 1x1200s & 1.3 \\ 
\hline
\end{tabular}
\end{minipage}
\end{table*}

\subsection{Data reduction}

The initial data reduction of the DOLORES observations was very similar
to the WFC reduction process; again the IRAF software package was
used throughout. First a master bias was constructed for each night by
averaging all the appropriate bias frames together; this was then subtracted from
the rest of the exposures. The science, flat--field and arc data were
then grouped according to mask. The calibration arc--lamp used was HeNeAr. 

The science and arc images for each mask were then combined and split
into individual two--dimensional spectra for each source. The normalised flat field for
each source was then applied and the strong background sky--lines were
removed. This process was not perfect and
some background residuals remained; these had to be taken into account when
attempting to identify spectral features.

One--dimensional spectra were then extracted for the sources and their
corresponding arcs. Lines in the arc spectrum were then identified and
the resulting calibration was applied to the science spectra. The
calibration was improved by adding sky--lines ([OI] at 6300.3\AA~and
5579\AA~and Na at 5896\AA), extracted from the un--background
subtracted science spectra, to the arc spectra to improve the
wavelength coverage.  

The spectra were then flux calibrated using observations of a standard
star, Feige 34. The standard star spectrum was also used to correct,
to a first approximation, for sky absorption features.

\subsection{Spectroscopic Results}

The spectroscopic observations included 41 sources in total; 17 in
the Hercules field and 24 in the Lynx field. The resulting spectra
yielded 3 and 11 definite redshifts in the Hercules and Lynx fields
respectively. A single line was detected in a further 4 spectra in
Lynx and 2 in Hercules; this mainly provided a redshift `best--guess' 
only, except where the line identification was obvious (e.g. broad
MgII). 65\% of the redshifts were from emission lines and the
remainder were from absorption lines only. These results are given more 
fully in Table \ref{z_results} and the spectra for sources with one or
more detected lines or absorption features, are shown in Figure
\ref{z_fig}. The list of sources observed with DOLORES, but for which
no lines were detected can be found in Table \ref{no_z}. Notes on
individual sources can be found in Appendix \ref{notes_spec}.

\begin{table}
\centering
\caption{The sources targeted in the DOLORES observations with no lines detected.}
\label{no_z}
\begin{tabular}{c|l}
\hline
Mask & Source \\
\hline
H1 & 53w084, 53w086a, 66w058 \\  
H2 & 53w082, 53w087, 53w088, 53w089 \\
H3 & 53w069 \\   
H4 & 66w035 \\
H5 & 53w054a, 53w057, 53w061 \\ 
L1 & 55w156, 60w071 \\
L2 & 55w147 \\ 
L3 & 55w138 \\ 
L4 & 55w127, 55w132, 60w032 \\
L5 & 55w118, 55w123 \\
\hline
\end{tabular}
\end{table}

\begin{table*}
\begin{minipage}{200mm}
\caption{\label{z_results} Spectroscopic redshifts and line
  information for the
Hercules and Lynx fields. }
\begin{tabular}{c|c|c|c|c|c|c|c|c}
\hline
Mask & Source & $\lambda$ & Line & Flux & $\Delta_{\rm fwhm}$ & W & z & Final z \\
     &        & (\AA)   &   & ($\times10^{-16}{\rm erg~s^{-1}cm^{-2}}$) &
$(kms^{-1})$ & (\AA)  &   &         \\             
\hline
\multicolumn{9}{l}{Hercules}  \\
\hline
H4&53w070 &6480.5 &  MgII &0.48 $\pm$ 0.07 &--&27 $\pm$ 5& 1.315 $\pm$ 0.001 & 1.315 $\pm$ 0.001 \\
H1&  53w086b & -- &4000\AA~break &--&--&--&0.73 $\pm$ 0.01& 0.73 $\pm$ 0.05\\
H4& 66w027 &7136.6& H$\alpha$ &16.9 $\pm$ 2.55& 1091 $\pm$ 228 &39
$\pm$ 4& 0.087 $\pm$ 0.001 & 0.086 $\pm$ 0.002 \\ 
&           &7136.5& NII &6.9 $\pm$ 2.55 &1074  $\pm$ 228  &38  $\pm$ 4 &0.084 $\pm$  0.001 &\\
&           &7311.7& [SII] &$\sim$3.94 &--&$\sim$6& 0.088 $\pm$ 0.001 &\\
&           &6401& NaD &--&--&-- & -- &\\
&           &5282& H$\beta$ & 10.77 $\pm$ 1.94&--&17 $\pm$ 3& -- &\\
H3&66w031&6751.8& [OII] &2.47 $\pm$ 0.25 & 987 $\pm$ 251 & 508 $\pm$ 
205& 0.812 $\pm$ 0.001 & 0.812 $\pm$ 0.001 \\ 
&           &8822& H$\beta$ & --  & -- & -- & \\
H3&  66w036 &8273.4& G--band &--&--&--& 0.924 $\pm$ 0.001 & 0.924 $\pm$ 0.001 \\
&           &7581.4& CaH &--&--& --   & 0.927 $\pm$ 0.012 &\\
&           &7650& CaK &--&--&--& --  &\\
\hline
\multicolumn{6}{l}{Lynx}  \\
\hline
L5&  55w116 &7281.5 & CaH & -- &--&--  & 0.854 $\pm$ 0.031 & 0.851 $\pm$ 0.007 \\
  &         &7343.4 & CaK & -- &--& -- & 0.851 $\pm$ 0.008 & \\

L5&  55w124 &6536.5 &MgII & 4.85 $\pm$ 0.49 & 5471 $\pm$ 894&39 $\pm$
4 & 1.335 $\pm$  0.003& 1.335 $\pm$ 0.003 \\   

L4&  55w128 &8159.3&[OII]&0.17 $\pm$ 0.03  &-- &--&1.189 $\pm$ 0.001& 1.189 $\pm$ 0.001\\

L4&  55w131 &7917.1 &[OII]&0.75 $\pm$ 0.10 & 409 $\pm$ 281 & -- &
1.124 $\pm$ 0.001& 1.124 $\pm$ 0.001\\ 

L3&  55w137 &5755.8 & [OIII] & 1.64 $\pm$ 0.22&--&4 $\pm$ 1& 0.150 $\pm$ 0.001& 0.151 $\pm$ 0.001 \\
  &         &5703.8 & [OIII] & 9.94 $\pm$ 2.56   &--& 26 $\pm$ 7  & \\
  &         &7567.1 & H$\alpha$ & 11.7 $\pm$ 2.63 & 2000 $\pm$ 266 &
37 $\pm$ 4 & 0.153 $\pm$ 0.001 & \\ 
  &         &7575.2 & [NII] & 13.1 $\pm$ 2.56 & 1975 $\pm$ 265 & 36 $\pm$ 4 &0.150 $\pm$ 0.001 & \\
  &         &7738.4 & [SII] & 2.77 $\pm$ 0.31 &--&--& 0.152 $\pm$ 0.001 & \\
  &         &6778.4 & NaD & --&--&--&0.150 $\pm$ 0.001 & \\
  &         &7248.8 & [OI] &0.57 $\pm$ 0.15 & --&-- &0.151 $\pm$ 0.001 & \\

L3&55w140 &7514.3 &MgII & 12.10 $\pm$ 1.40 & 6192 $\pm$ 2647 & 36
$\pm$ 4 &1.685 $\pm$ 0.012& 1.685 $\pm$ 0.012 \\ 
 
L2&  55w149 &6793.7 &NaD &--&--&--&0.152 $\pm$ 0.001& 0.151 $\pm$ 0.001\\
  &         &5950.6 &Mgb &--&--&--& 0.150 $\pm$ 0.001 & \\
  &         &5747.7 & [OIII] &--&--&--& -- & \\
  &         &5601.2 & H$\beta$ &--&--&--& -- & \\
 
L2&55w150 &7359.3 & [OIII] & 6.16 $\pm$ 0.64 & 727 $\pm$ 233 & 67
  $\pm$ 8& 0.470 $\pm$ 0.001 &0.470 $\pm$ 0.001 \\ 
  &         &7292.6 & [OIII] & 1.65 $\pm$ 0.21 & --& 22 $\pm$ 3 & 0.471 $\pm$ 0.001 & \\
  &         &7138.4 & H$\beta$ & 0.96 $\pm$ 0.16 &-- &21 $\pm$ 4& 0.469 $\pm$ 0.001 & \\
  &         &9660.4& H$\alpha$ &--&--&--& -- & \\
 
L1&  55w154 &6475.6 &H$\beta$ & --&--&--& 0.332 $\pm$ 0.001 & 0.330 $\pm$ 0.001 \\
  &         &6891.1 & Mgb &--&--&--&0.332 $\pm$ 0.001 & \\
  &         &5717.9 & G--band &--&--&--& 0.330 $\pm$ 0.002 &\\
  &         &5227.3 & CaH &--&--&--& 0.329 $\pm$  0.001 & \\
  &         &5277.5 & CaK &--&--&--& -- & \\
  &         &5456.5 & H$\delta$ &--&--&--& -- & \\

L1&  55w157 &6760.2& H$\gamma$ &--&--&--& 0.558 $\pm$ 0.001 & 0.559 $\pm$ 0.002 \\
  &         &6201.4& CaK &--&--&--& 0.563 $\pm$ 0.001 &\\
 
LS&  55w160 &6292.3 & CaH &--&--&--& 0.600 $\pm$ 0.002 & 0.600 $\pm$ 0.002 \\
  &         &6352.0 & CaK &--&--&--& -- & \\
 
L5&  60w016 &7237.0&CaH&--&--&--&0.840 $\pm$ 0.001 & 0.840 $\pm$ 0.001 \\
  &         &7332.7&CaK &--&--&--&-- & \\
 
L5&  60w024 &6609.9&[OII] &0.41 $\pm$ 0.06& 814 $\pm$ 374 & 13 $\pm$ 2
&0.774 $\pm$ 0.001& 0.774 $\pm$ 0.001 \\

  &         &6974.2& CaH  & -- & -- & -- & -- & \\ 

  &         &7032.2& CaK  & -- & -- & -- & -- & \\ 

L3/4&60w039&7562.2&H$\alpha$& 51.0 $\pm$ 6.18 & 939 $\pm$ 223 & 47
    $\pm$ 5 &0.152 $\pm$ 0.001& 0.151 $\pm$ 0.001\\ 
    &       &7585.2& [NII] & 11.4 $\pm$ 6.12 & 925 $\pm$ 223 &46 $\pm$ 5 & 0.149 $\pm$ 0.001 & \\
    &       &7738.7& [SII] & 8.32 $\pm$ 1.33 &--&-- &0.152 $\pm$ 0.002 & \\ 
    &       &6786.8& NaD &--&--&--& 0.151 $\pm$ 0.001 & \\
    &       &5597.6& H$\beta$ &--&--&--& 0.152 $\pm$ 0.001 & \\

L2& 60w055  &6405.0& [OII] &1.40 $\pm$ 0.14 & 529 $\pm$ 255 & 45 $\pm$
  5 & 0.718 $\pm$ 0.001 & 0.718 $\pm$ 0.005\\ 
  &         &6754.4& CaH   &--&--&--& 0.717 $\pm$ 0.001&\\
  &         &6818.8 &CaK &--&--&--& -- & \\
  &         &7397.9& G--band &--&--&--& --& \\
  &         &7437.8& H$\gamma$ &--&--&--& -- & \\
\hline
\end{tabular}
\end{minipage}
\end{table*}

\section{\protect\label{sb_chap} Cleaning the sample: identifying quasars and starburst galaxies}
\protect\label{sb_chap}
Not all the radio sources detected in the two fields will be radio
galaxies; some contamination of the sample by quasars and starburst
galaxies is inevitable. The quasars need to be identified as the photometric 
redshift estimation methods are not valid for them, whilst the
starburst galaxies need to identified and removed. 

\subsection{\protect\label{id_gal} Identifying starburst galaxies}
\protect\label{id_gal}
The radio emission from starburst galaxies is mainly due to
synchrotron emission from supernovae instead of, as in radio
galaxies, accretion onto a supermassive black hole.
The radio luminosity function of Best et
al. (2005) shows
that, in general, the radio power of these galaxies is lower than that
of AGN and that it is only at
these very low radio powers ($\sim10^{23}$WHz$^{-1}$) that their
number density dominates; this suggests that only low--power sources need
to be considered here. There were 16 radio sources in the sample which
have $P_{\rm rad}\leq~10^{24}~{\rm WHz^{-1}}$ and were therefore possible starburst galaxy candidates. 

The main way to distinguish between radio and starburst galaxies is
through examination of the emission line ratios of their
spectra. Kauffmann et al. (2003) classify a source as an
AGN if 
\begin{equation}
\label{kauff_eqn}
\log(f{\rm[OIII]}/f({\rm H\beta}))>\frac{0.61}{\log(f{\rm [NII]}/f({\rm H\alpha)})-0.05}+1.3 . 
\end{equation}
where $f \rm [OIII]$, $f(\rm H\beta)$, $f \rm [NII]$ and $f(\rm H\alpha)$ are the fluxes of the
respective emission lines; [OIII] (5007\AA), [NII] (6583\AA),
$H_{\beta}$ (4861\AA) and $H_{\alpha}$ (6563\AA). This classification
method obviously requires a source to have   
spectroscopic data with the right lines detected. Of the
16 candidates, 11 have DOLORES spectra but none have all four of the
necessary emission lines. However, 3 candidates (66w027, 55w137 and
60w039) have both [NII] and H$\alpha$ detected and one, 55w150, has
both [OIII] and H$\beta$, hence an indication of
their classifications is possible; the results of this are outlined
below. The fluxes for these lines can be 
found in Table \ref{z_results}. A further two candidates, 55w135 and 60w084, were included in
the spectroscopic observations of the SDSS. The resulting spectrum for 55w135
clearly shows it to be a starburst galaxy since $f({\rm H\alpha}) \gg f{\rm[NII]}$ and
$f({\rm H\beta})\gg f{\rm [OIII]}$ whereas the spectrum for 60w084 suggests that it is
an AGN since $f({\rm H\alpha})\sim f \rm [NII]$ and $f{\rm [OIII]}
\gg f(\rm H\beta)$. 

\begin{itemize}
\item {\bf 66w027} (z=0.086, P$_{\rm 1.4 GHz}$=10$^{22.00}$WHz$^{-1}$) --
  $\log(f{\rm [NII]}/f(\rm H\alpha))=-0.38$ for this source which 
suggests that it is a starburst galaxy but is not sufficient to
be unambiguous. However, the H$\beta$ line was also detected whilst
the [OIII] line was not, implying that $\log(f{\rm[OIII]}/f(\rm H\beta)) < 0.0$ and
that this source is a starburst galaxy. 
\item {\bf 55w137} (z=0.151, P$_{\rm 1.4 GHz}$=10$^{22.97}$WHz$^{-1}$) --
 $\log(f{\rm [NII]}/f(\rm H\alpha))=0.05$ for this source which 
places it firmly in the AGN region. This
is further supported by the indication of extended radio emission
visible in the radio image. 
\item {\bf 60w039} (z=0.151, P$_{\rm 1.4 GHz}$=10$^{22.58}$WHz$^{-1}$) --
  $\log(f{\rm [NII]}/f(\rm H\alpha))=-0.65$ for this source which 
places it firmly in the starburst region. 
\item {\bf 55w150} (z=0.470, P$_{\rm 1.4 GHz}$=10$^{23.86}$WHz$^{-1}$) --
 $\log(f{\rm[OIII]}/f(\rm H\beta)) =0.78$ for this source which, coupled with the high radio power,  
strongly suggests that it is an AGN. 
\end{itemize}

The five remaining starburst candidates with spectra can be classified
by comparing their $f(\rm [OII])$ and radio fluxes. Best et
al. (2002) derive a rough relationship between these two
fluxes for starburst galaxies by equating the theoretical correlation
between mean star--formation rate and [OII] luminosity
(Barbaro \& Poggianti, 1997) with that for radio (Condon \& Yin, 1990) giving 

\begin{equation}
\label{oii_sf}
\frac{S_{\rm 1.4 GHz}}{\mu Jy} = 11.0 \frac{f{\rm [OII]}}{10^{-16}{\rm ergs^{-1}cm^{-2}}} (1+z)^{-0.8} 
\end{equation}

\noindent where $S_{\rm 1.4 GHz}$ is the radio flux of the object in question at a
frequency of 1.4 GHz and $z$ is the
redshift. It should be noted that dust extinction can cause the
measurement of the star--formation, from the [OIII] flux, to be underestimated. 

The radio--[OII] flux relationship for AGN is an extrapolation of the
results of Willott et al. (1999) for powerful radio galaxies to the
sub--mJy levels of these sources. This gives, after converting to
1.4 GHz (assuming a spectral index of 0.8), 

\begin{equation}
\label{oii_rad}
\frac{S_{\rm 1.4 GHz}}{\mu Jy} =
4.7\times10^{4}\Big(\frac{f{\rm [OII]}}{10^{-16}{\rm ergs^{-1}cm^{-2}}}\Big)^{1.45} 
\end{equation}

\noindent It should be noted however, that AGN show considerable scatter around this line.

\begin{figure}
   \includegraphics[scale=0.65,angle=90]{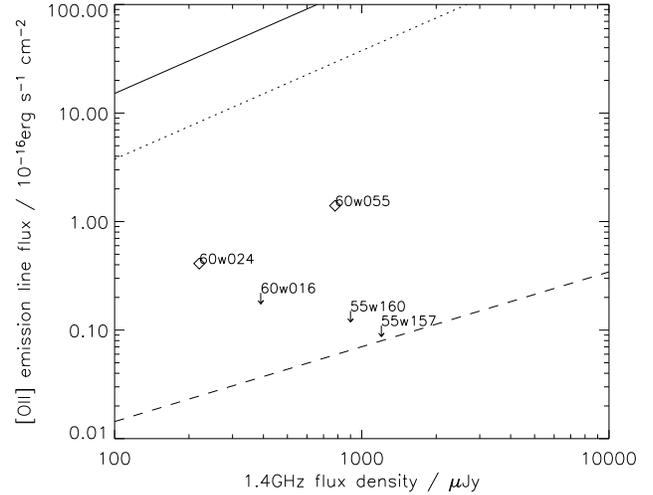} 
   \caption{Radio--[OII] flux relationships for starburst galaxies
   (solid line) and radio galaxies (dashed line). The dotted line
   shows the effect of two magnitudes of dust extinction in the
   measurement of the star formation rate. Overplotted are the
   starburst candidate sources for classification.} 
   \label{oii_plot}
\end{figure}

Equations \ref{oii_sf} and \ref{oii_rad} are shown in Figure
\ref{oii_plot}. Overplotted are the [OII] and radio fluxes for 60w055
and 60w024. The remaining 3 sources (60w016, 55w160 and 55w157) lack
an [OII] detection so an upper limit for their [OII] flux is plotted
instead. It is clear from inspection of  this diagram that all these
sources lie well below the starburst region, suggesting that they are
AGN. This is in accordance with their radio powers, all of which are
$>10^{23.3}{\rm WHz^{-1}}$. 

This leaves 5 candidates still to be classified. Two (66w009b and 55w127) have radio powers that 
are $\leq10^{23.01}$WHz$^{-1}$ which strongly suggests that these are also 
starburst galaxies. The final three candidate sources (55w118, 55w122
and 55w161) all have $P\gtrsim10^{23}{\rm WHz^{-1}}$ which is 
comparable with the powers of candidates already identified as
AGN. Therefore, on the balance of probability, these three objects are
classified as AGN also.

In summary, therefore, there are five starburst galaxies which need to be removed
from the sample: 66w027, 66w009b, 55w127, 55w135 and 60w039.

\subsection{Identifying quasars} 
\label{q_id}
Possible quasars in the sample were identified in three ways; firstly on the basis of their
point--like appearance (determined by measuring the FWHM using \emph{gaia}) in the
\emph{r}--band images, secondly by looking for broad lines
($\gg$1000$\rm kms^{-1}$) in the
available spectra and thirdly by examining their optical $(r-i)$
colours. The quasar candidates selected in this way are described below.

\begin{itemize}

\item {\bf 53w061} -- This source has a pointlike appearance (FWHM of
1.53\arcsec in \emph{r}--band, c.f. seeing of 1.5\arcsec) and a blue $(r-i)$
colour of 0.36. It is also classed as a likely quasar (Q?) by Kron et
al. (1985) on the basis of its colour. It was included in the DOLORES
observations but no continuum was detected. It is therefore classed as
a quasar here.

\item {\bf 53w065} -- This source has a blue $(r-i)$ colour of -0.31 but
Waddington et al. (2001) classify it as a galaxy on the
basis of the narrow lines seen in its spectrum.
 
\item {\bf 53w075} -- This source is classified as a quasar by Kron et
  al. (1985) on the basis of its spectrum; this is supported by its
  pointlike appearance.

\item {\bf 53w080} -- This source is classified as a quasar by Kron et
  al. (1985) on the basis of its spectrum; this is supported by its
  pointlike appearance. 

\item {\bf 55w121} -- This source is classified as a quasar by Kron et
  al. (1985) on the basis of its colour. However, whilst its
  non--detection in \emph{i} does suggest a blue colour, this is not
  supported by its, not very blue, $(r-K)$ of 3.8, and it does not
  appear to be pointlike. It is therefore classed as a galaxy here.  

\item {\bf 55w124} -- This source is classified as a quasar on the basis of
its very broad MgII line and blue $(r-i)$ colour of -0.06.

\item {\bf 55w140} -- This source is classified as a quasar on the basis of
its very broad MgII line and very blue $(r-i)$ colour of -0.21.

\end{itemize}

In summary, therefore, the objects classed as quasars in the sample
are 53w061, 53w075, 53w080, 55w124 and 55w140.

\section{ Redshift estimation and results}
\label{z_est_1}
Redshifts have now been determined for 44\% of sources in the Lynx
field and 62\% of sources in the Hercules 
field, either through the observations described above or with
previously published results. For the remainder, estimated
redshifts were calculated instead using the two different magnitude--redshift relations, K--z and
r--z,  outlined in \S\ref{relns} below. The K--z
relation, the more accurate of the two, was used for all sources
detected in the UKIRT observations; the remaining sources were
estimated with the r--z. 

\subsection{\protect \label{relns} The K--z and r--z
  magnitude--redshift relationships}

The K--z relation is a tight correlation between the \emph{K}--band
magnitudes and redshifts of radio source host galaxies; it exists
because the radio host galaxy population is made of up of passively
evolving, massive elliptical galaxies. 

At high redshift the relation is slightly different for the radio 
surveys of different flux density limits (Willott et al. 2003),  
which therefore implies that it is dependent on the
flux density of the radio source. The relation used here is that
found for the 7C survey as the source fluxes for that sample are
the best comparison to those considered here. The relation itself is
given by Willott et al. (2003) as,  
\begin{equation}
\label{kz_eqn}
K = 17.37 + 4.53\log{z} - 0.31(\log{z})^{2} .
\end{equation}

The majority of the sources with a host galaxy detection but without
redshifts, were also not included 
in the UKIRT observations so, therefore, the K--z relation outlined
above cannot be used for them. Instead, since all these sources have
an \emph{r}--band magnitude, the r--z relation  was used. This
relation describes a correlation between the redshift
and \emph{r}--band magnitude of the host galaxies of Gigahertz Peaked 
Spectrum (GPS) radio sources (Snellen et al., 1996).

In general, redshifts estimated using the r--z relation are not as
reliable as those found using the K--z, especially above $z\sim0.6$
where the \emph{r}--band samples shortward of the 4000\AA~break. At
these bluer wavelengths, factors such as star formation or, in more
powerful radio galaxies, the `alignment effect' (where the optical
structures of the galaxy align with the radio jets (e.g. McCarthy et
al., 1987)) can result in considerable scatter in the measured
magnitudes. GPS sources are less affected by this scatter (c.f. Figure
6 of Snellen et al. 1996) so are ideal for the redshift estimation
here. The GPS relation is 
\begin{equation}
\label{rz}
r_{\rm g} = 22.7 + 7.4\log{z} . 
\end{equation}
and is valid up to $z\sim1$; at higher redshifts it is
unreliable due to the scarcity of measurements and the blue rest--frame wavelength
range it is sampling. 

The relation given in Equation \ref{rz} above assumes that the
\emph{r}--band magnitude used was observed using the Thuan--Gunn filter
system. It therefore had to be transformed to the Sloan filter system,
via the Johnson filter system, before it could be used. In the
following description capital letters indicate the Johnson system,
lower--case letters indicate the Sloan system and \emph{g}
subscripts indicate the Thuan--Gunn system. To convert from Johnson to
Thuan--Gunn, Jorgensen et al. (1994) give:
\begin{equation}
r_{\rm g} = R + 0.111(g_{\rm g}-r_{\rm g}) + 0.317 ,
\end{equation}
whilst to convert from Johnson to Sloan, Equation \ref{trans1} from
Smith et al. (2002) gives:
\begin{equation}
r' = R + 0.16(V-R) + 0.13 .
\end{equation}

\noindent Making the assumption that $(V-R) \simeq (g_{\rm g}-r_{\rm
  g}) \simeq 1$ for radio galaxy hosts, 
then $r=r_{\rm g}-0.14$ and the r--z relation becomes
\begin{equation}
r' = 22.56 + 7.4\log{z} .
\end{equation}

The magnitude--redshift relations need aperture corrected magnitudes
to give reasonable results, but the estimated redshift is needed to
perform the aperture correction (as described in \S\ref{ap_cor}). To
solve this problem an IDL script was written to iterate the redshift 
estimates and subsequent aperture corrections until they converged on
a final value. 

Table \ref{z_table} summarises all the redshift information for the
sources in the two fields. The quoted redshift for each source is
ideally a spectroscopic or previously published one, if neither of
those is available then the redshift is one from a K--z estimation and
finally, for sources with no \emph{K}--magnitude or spectrum, the redshift
given is an r--z estimate.

\begin{table}
\caption{\label{z_table} The redshifts found for the sources in the
  complete sample. (1a) and (1b)
indicates a previously published value, (a -- spectroscopic 
  (Waddington et al. 2000 and references therein; Bershady et al. 1994), b--
photometric (Waddington et al. 2001)), (1c) indicates the
redshift came from the SDSS, (2) indicates a DOLORES 
spectroscopic value, (3) is a K--z estimate,  
(4) is a r--z estimate.}
\centering
\begin{tabular}{c|c|cc||c|c|cc}
\hline
\multicolumn{3}{l}{Hercules} &&\multicolumn{4}{l}{Lynx} \\
\hline
  Name    & z      & Origin &&Name    & z & Origin &       \\
\hline		     
  53w052  & 0.46 & 1a &&55w116  & 0.851& 2   \\
  53w054a & 1.51 & 3  &&55w118  & 0.66 & 4   \\ 
  53w054b & 3.50 & 3  &&55w120  & 1.35 & 3   \\
  53w057  & 1.85 & 4  &&55w121  & 2.57 & 3   \\
  53w059  & 1.65 & 4  &&55w122  & 0.55 & 4   \\         
  53w061  & 2.88 & 1b &&55w123  & 0.87 & 3   \\
  53w062  & 0.61 & 1a &&55w124  & 1.335& 2   \\
  53w065  & 1.185& 1a &&55w127  & 0.06 & 4   \\
  53w066  & 1.82 & 1b &&55w128  & 1.189& 2   \\
  53w067  & 0.759& 1a &&55w131  & 1.124& 2   \\ 
  53w069  & 1.432& 1a &&55w132  &$>$4.4& 3   \\
  53w070  & 1.315& 2  &&55w133  & 2.24 & 4   \\
  53w075  & 2.150& 1a &&55w135  & 0.090& 1c  \\
  53w076  & 0.390& 1a &&55w136  & 2.12 & 3   \\
  53w077  & 0.80 & 1a &&55w137  & 0.151& 2   \\
  53w078  & 0.27 & 1a &&55w138  & 2.81 & 3   \\
  53w079  & 0.548& 1a &&55w140  & 1.685& 2   \\
  53w080  & 0.546& 1a &&55w141  &$>$1.8& 4   \\
  53w081  & 2.060& 1a &&55w143a & 2.15 & 4   \\
  53w082  & 2.04 & 4  &&55w143b & 2.21 & 4   \\
  53w083  & 0.628& 1a &&55w147  & 1.07 & 3   \\
  53w084  & 2.73 & 3  &&55w149  & 0.151& 2   \\
  53w085  & 1.35 & 1a &&55w150  & 0.470& 2   \\
  53w086a & 0.46 & 4  &&55w154  & 0.330& 2   \\
  53w086b & 0.73 & 2  &&55w155  &$>$3.7& 3   \\
  53w087  &$>$3.7& 3  &&55w156  & 0.86 & 3   \\
  53w088  & 1.773& 1a &&55w157  & 0.557& 2   \\
  53w089  & 0.635& 1a &&55w159a & 1.29 & 4   \\
  66w009a & 0.65 & 3  &&55w159b & 0.311& 1c  \\ 
  66w009b & 0.156& 1a &&55w160  & 0.600& 2   \\  
  66w014  &  --  &--  &&55w161  & 0.44 & 4   \\  
  66w027  & 0.086& 2  &&55w165a & 0.68 & 4   \\  
  66w031  & 0.812& 2  &&55w165b & 0.75 & 4   \\                         
  66w035  & 2.26 & 3  &&55w166  & 0.99 & 4   \\  
  66w036  & 0.924& 2  &&60w016  & 0.840& 2   \\  
  66w042  & 0.65 & 4  &&60w024  & 0.773& 2   \\  
  66w047  & 0.37 & 4  &&60w032  &$>$1.8& 4   \\  
  66w049  & 0.95 & 4  &&60w039  & 0.151& 2   \\  
  66w058  &$>$2.3& 4  &&60w055  & 0.718& 2   \\  
          &      &    &&60w067  &$>$1.8& 4   \\  
	  &      &    &&60w071  & 1.25 & 4   \\  
	  &	 &    &&60w084  & 0.127& 1c  \\  
\hline
\end{tabular}
\end{table}   

\subsection{Redshift comparison}

\begin{figure}
\centering
\includegraphics[scale=0.35, angle=90]{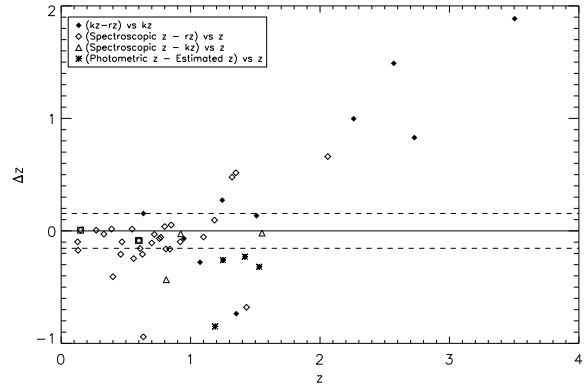}
\caption{\label{kz_comp} A comparison of the different methods used to
  obtain redshifts. Filled diamonds indicate the comparison between
  the two redshift estimation methods whilst empty diamonds and triangles
  indicate the comparison between the spectroscopically determined
  redshifts and the r--z and K--z estimates respectively. Square
  outlines indicate redshifts that came from 
  single--line spectra. The dotted lines are the 1$\sigma$ values of
  the spectroscopic $\Delta$z, $\pm$0.15.}
\end{figure}

A valuable test of the redshift estimation comes from comparing
the estimates with the DOLORES MOS and the previously published
(Waddington et al. 2000; 2001) spectroscopic and photometric redshifts. The results of this
comparison for both r--z and K--z redshift estimates are shown in
Figure \ref{kz_comp}; in general the
agreement between the spectroscopic and estimated results is very
good with a 1$\sigma$ $\Delta$z of 0.15. The agreement is also
reasonably good for the photometric redshifts. 

Additionally, several of the sources in the sample have both r and K--band
magnitudes, thus providing a useful means of comparing the two methods
of redshift estimation. The solid diamonds plotted in Figure
\ref{kz_comp} show the difference in 
the two estimates ((K--z)- (r--z)) plotted against the K--z value. The
two relations give similar redshifts up to z$_{\rm kz}$\squig~1.5, but, for the
redshifts higher than this, the K--z value is much greater than the
r--z further suggesting the lower accuracy of the r--z relation at
these values. 

\subsection{\protect\label{dist_chap} The redshift distribution of the sample}
\begin{figure}
\subfigure[][\label{z_hist1}]{\includegraphics[scale=0.35, angle=90]{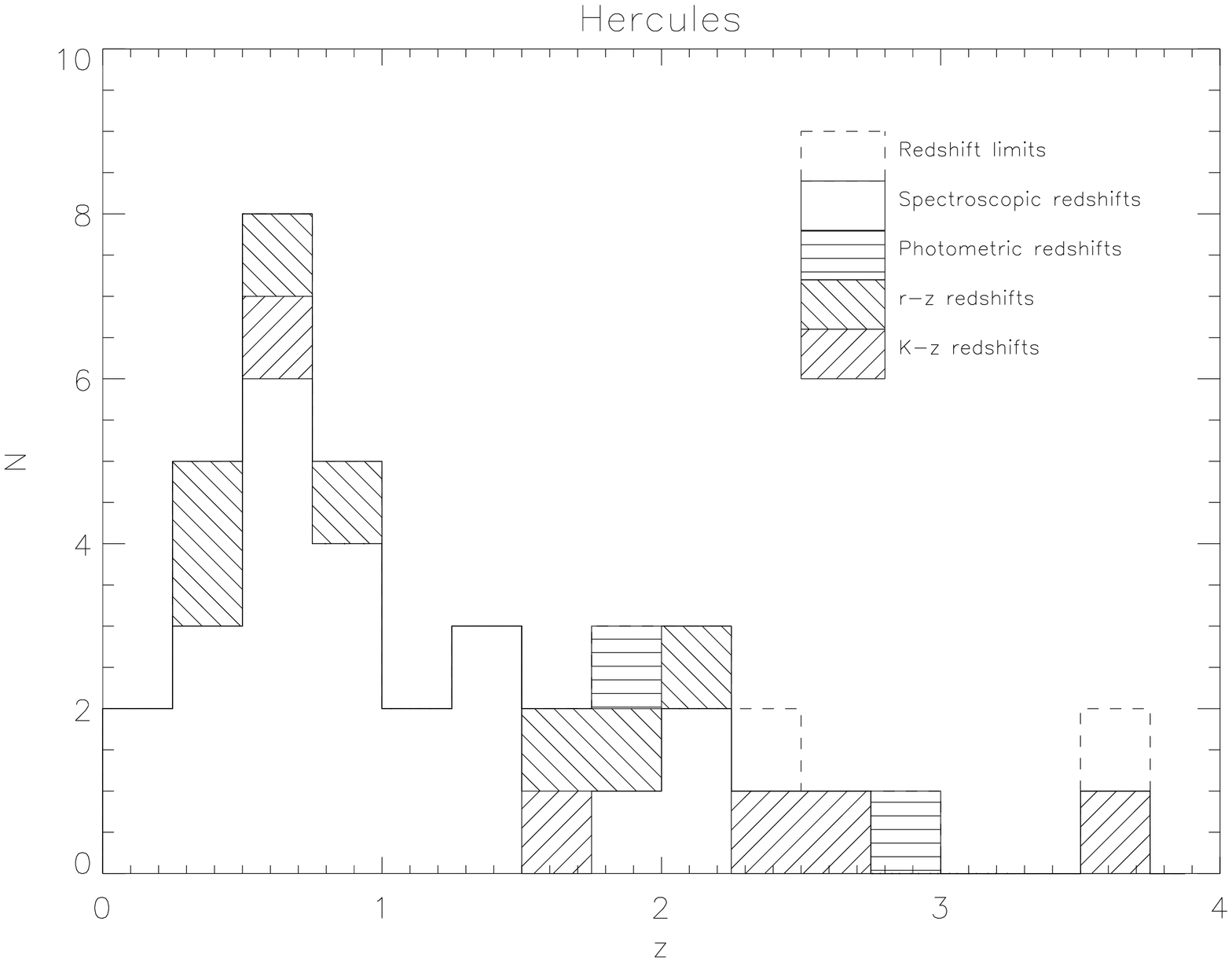} }
\subfigure[][\label{z_hist2}]{\includegraphics[scale=0.35, angle=90]{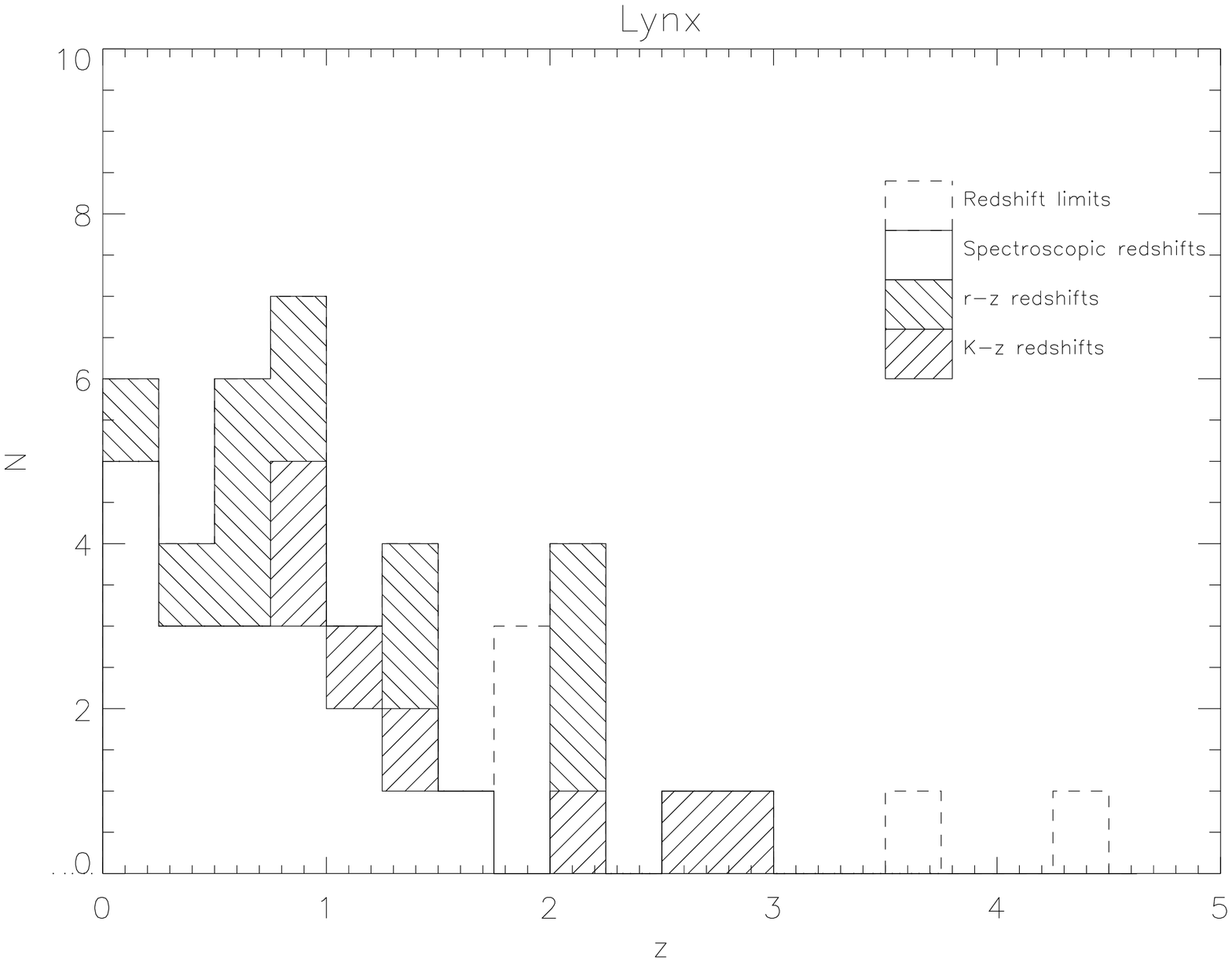} }
\caption{The redshift distribution for sources
  in  the Hercules (a) and Lynx (b) fields subdivided to show the contribution from the
  different redshift methods.} 
\end{figure}

\protect\label{dist_chap} 
Now that redshift information has been obtained or estimated for a
large proportion of the objects in the two fields, redshift histograms
can be constructed to compare the radio source distributions for the
two fields; these
are shown, split into the different redshift methods, in Figures
\ref{z_hist1} and \ref{z_hist2}. Both  histograms, 
peak at redshifts before $z=1.0$.

These histograms also illustrate the
lack of definite redshifts for the Lynx field, especially at the high end,
compared to the Hercules field. Whilst many of the Lynx sources
were included in the DOLORES MOS observations, lines in the resulting spectra
tended to be detected in the brighter, lower redshift objects. The
high--z end of both fields, however, is populated mainly by sources
with less accurate redshift estimates.  

\section{Conclusions}
\label{concl}
In summary, the data presented here form a complete sample of 81 radio
sources above the limiting flux density of S$_{\rm 1.4 GHz}>0.5$ mJy. Of
these, only 12 remain unidentified after the optical and infra--red
observations. Redshifts either previously existed or have been
determined for 49\% of the sample; the remaining objects with a host
galaxy detection, have redshift estimates. 

The radio, optical, infra--red and MOS observations outlined here
define a multi--wavelength sample which will form the basis for
investigations into the high--redshift space density and subsequent
cosmic evolution, if any, of FRI radio galaxies.  Paper II will
describe the high--resolution radio observations and subsequent
morphologically classification of the sources in the sample, along
with the subsequent space density measurements of the detected FRIs.  

\section*{Acknowledgements}
EER acknowledges a research studentship from the UK Particle Physics
and Astronomy Research Council. PNB
would like to thank the Royal Society for generous financial support
through its University Research Fellowship scheme. The United Kingdom Infra--red Telescope is operated
by the Joint Astronomy Centre on behalf of the UK Particle Physics and Astronomy
Research Council. We acknowledge the UKIRT Service Programme for some
of the infra--red imaging. The Isaac Newton Telescope is operated by
the Isaac Newton Group, and the Telescopio Nazionale Galileo is
operated by the Centro Galileo Galilei of the Consorzio Nazionale per
l'Astronomia e l'Astrofisica, both in the Spanish Observatorio del
Roque de los Muchachos of the Instituto de Astrofisica de
Canarias. The National Radio Astronomy Observatory is a facility of
the National Science Foundation operated under cooperative agreement
by Associated Universities Inc.  

{}

\bsp

\appendix

\section{Imaging and spectroscopy notes on individual sources}

\subsection{Imaging source notes}
\label{notes_im}
\setlength{\parindent}{0pt}

{\bf 53w054a and 53w054b} -- Windhorst et al. (1984) classified
these two sources as radio galaxy lobes but, in agreement with
Waddington et al. (2000), optical identifications
were found for both sources indicating they are two separate radio
galaxies. However, the Waddington et al. identification of 53w054b seems
incorrect -- a faint source is detected in the \emph{K}--band which is
more closely associated with the radio position. 

{\bf 53w087 and 55w132} -- Neither of these sources were detected in
the optical or infra--red imaging and, as a result, are quoted as
being at $z \gtrsim$4 in Table \ref{z_table}. However, inspection of their radio maps shows that both
appear to be extended, diffuse--like sources, which suggests that they
are more likely to be dusty star--bursting galaxies, or post--mergers,
at $z$ = 1--4. Their estimated redshift limits may, therefore, be
unreliable. 

{\bf 66w042} -- Whilst the centre of this source does not appear to align
with the indicated optical galaxy, the identification is valid as the
faint radio core is offset to the west. 

{\bf 55w133, and 55w143a/b} -- The \emph{r}--band magnitudes measured for
these 3 sources are all around the 1$\sigma$ level. They therefore
should be treated as unreliable. 

{\bf 55w137 and 60w032} -- The slightly different telescope
pointing used in the 2004 INT observations, compared to that used in
2003, meant that these sources were not present on the 2004 images. 

{\bf 55w159b} -- The magnitude measured for this source should be
treated as less reliable due to the presence of two other objects in
close proximity to it.

\subsection{Spectroscopy source notes}
\label{notes_spec}

{\bf 53w070} --  The single broad line is identified as MgII, giving a redshift of 1.32. 

{\bf 53w086b} -- The 4000\AA~break can be seen in this spectrum but no
lines are convincingly detected. The redshift of 0.73 is consistent
with the \emph{r}--band magnitude estimate of 0.81. 

{\bf 53w089} -- Waddington et al. (2000) detect [OII] and
[OIII] for this source, giving a redshift of 0.635 but nothing is
detected in these (shallower) observations.  

{\bf 66w031} -- The identification of single strong line as [OII] at a
redshift of 0.81 is consistent with the redshift estimated from the
\emph{r}--band magnitude (0.97).

{\bf 66w036} -- No line data are quoted for the G--band absorption
line for this source as the negative extent of the flux density
suggests that it is contaminated by noise.

{\bf 55w124} -- The single broad line is identified as MgII, giving a
redshift of 1.34.

{\bf 55w128} -- The weakly detected single line is identified as
[OII], giving a redshift of 1.19, but this is uncertain.

{\bf 55w131} -- The single line is identified as [OII] giving a redshift of
1.12, which is consistent with the redshift of 1.15 estimated using the r--z relation.

{\bf 55w140b} -- The redshift of 1.68 is from a single, broad line, identified as
MgII. 

{\bf 60w039} -- The two observations of this source on masks L3 and L4
have been combined together with the \emph{scomb} task in
\emph{iraf}. 

\setlength{\parindent}{18pt}

\section{The radio/optical images}
\label{rad_im}

\begin{figure}
\caption{{\it See attached .jpg files.} The identifications and VLA A--array contour maps for
  the Hercules field; from left 1.4 GHz radio
  contour map, \emph{r} with overlay if present, \emph{i} and \emph{K} with overlay
  if there is no \emph{r} detection. \label{overlays} Where
  necessary, host galaxy positions are marked with crosshairs and some images have been gaussian
  smoothed for clarity. Radio contours start at 24$\mu$Jy/beam and are
  separated by factors of $\sqrt{2}$. The primary beam correction has
  not been applied to the radio maps so that uniform images can be
  presented.} 
\caption{{\it See attached .jpg files.} Similarly, the identifications for the Lynx field. From left, 1.4 GHz radio
  contour map, \emph{r} with overlay if present, \emph{i} and \emph{K} with overlay
  if there is no \emph{r} detection. Radio contours start at 24$\mu$Jy/beam and are
  separated by factors of $\sqrt{2}$. The primary beam correction has
  not been applied to the radio maps so that uniform images can be
  presented. \protect\label{overlays2}}
\caption{{\it See attached .jpg file.} \label{no_overlay} The un--primary beam corrected radio contour maps for the
  sources in both fields, in the complete sample, 
  with no optical or infra--red identification. Radio contours start
  at 24$\mu$Jy/beam and are separated by factors of $\sqrt{2}$. }
\caption{{\it See attached .jpg file.} The radio images and infra--red identifications, if present, for the
  sources not included in the complete sample. Radio contours start at 24$\mu$Jy/beam and are
  separated by factors of $\sqrt{2}$. The primary beam correction has
  not been applied to the radio maps so that uniform images can be
  presented. \label{over_nosamp}}
\end{figure}

\section{The spectra}

Here Figure \ref{z_fig} shows the spectra resulting from the DOLORES
MOS observations. 

\begin{figure*}
\centering
\begin{minipage}{15cm}
\includegraphics[scale=0.60, angle=90]{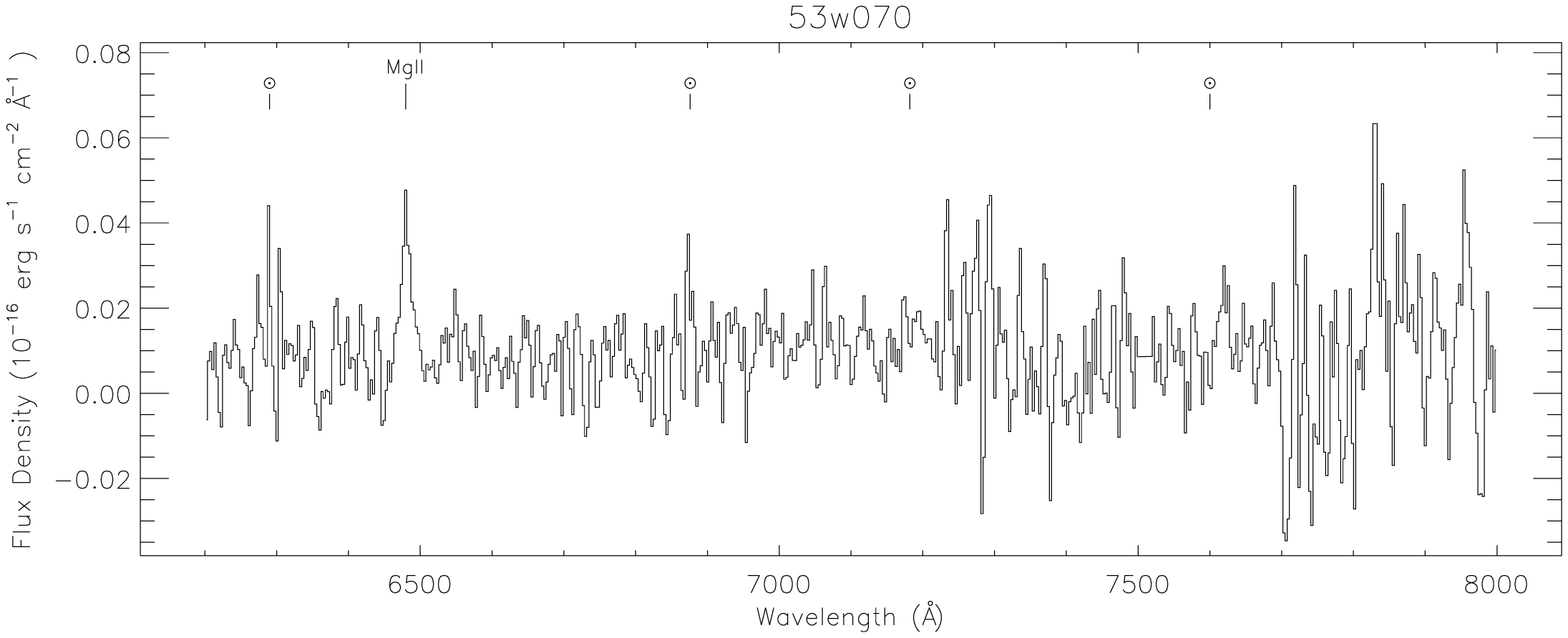}
\includegraphics[scale=0.60, angle=90]{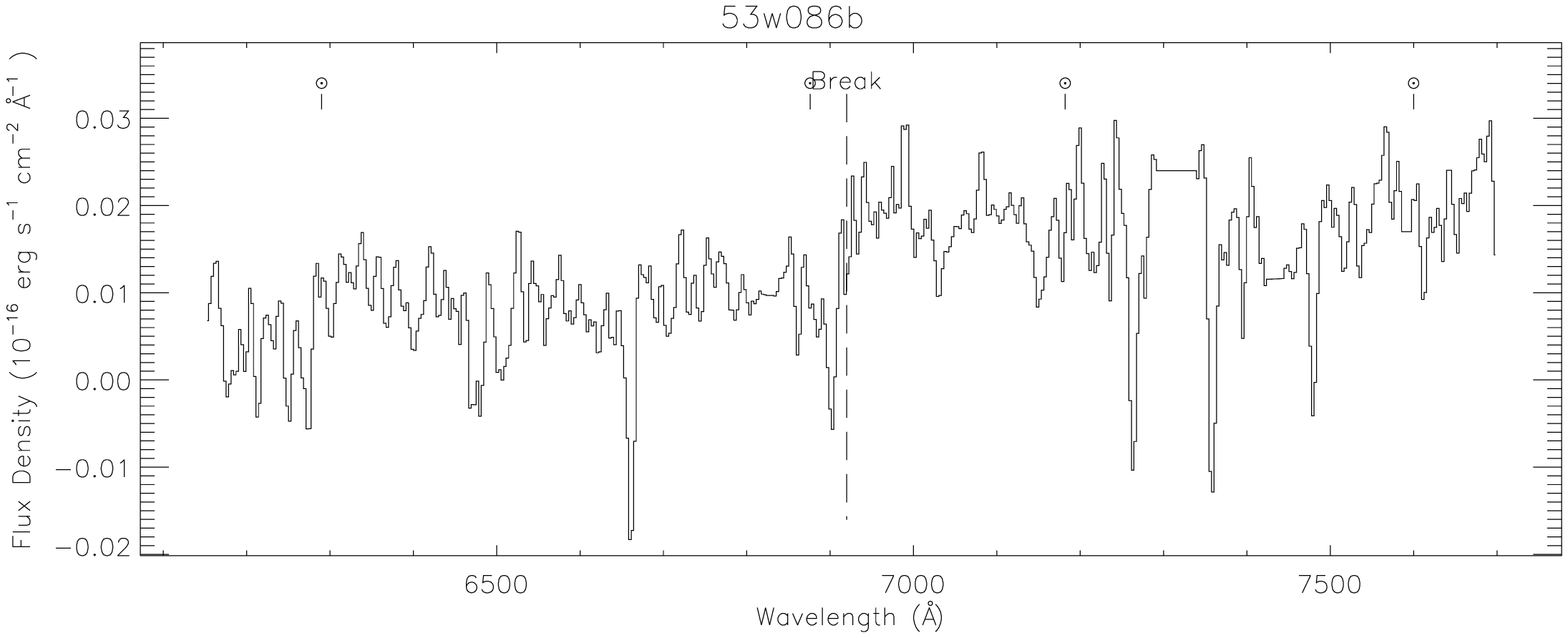}
\includegraphics[scale=0.60, angle=90]{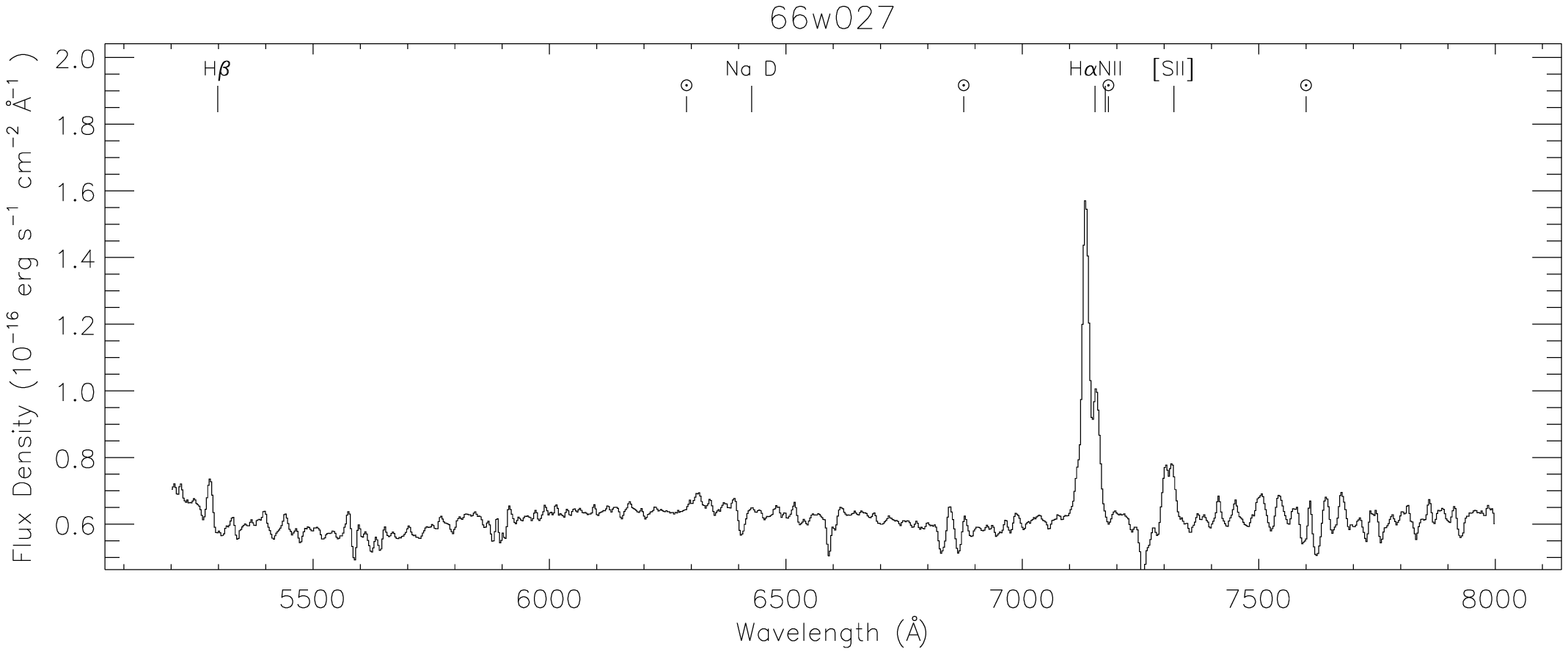}
\caption{The spectra resulting from the MOS observations. Residual sky
  features are marked with a $\odot$ \protect \label{z_fig} and dashed
  lines indicate the position of the 4000\AA break. 2D spectra are
  also shown in cases where the line detection is weak.} 
\end{minipage}
\end{figure*}

\begin{figure*}
\begin{minipage}{15cm}
\includegraphics[scale=0.60, angle=90]{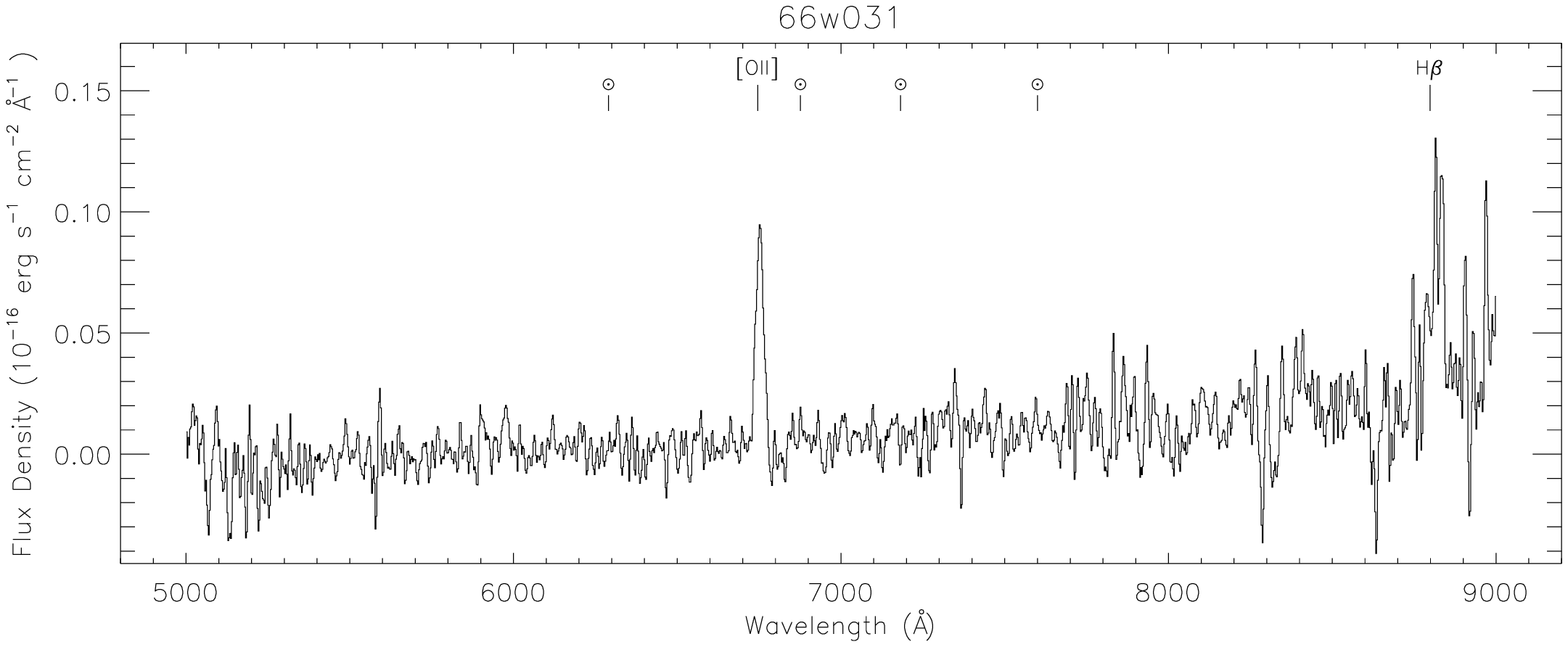}
\includegraphics[scale=0.60, angle=90]{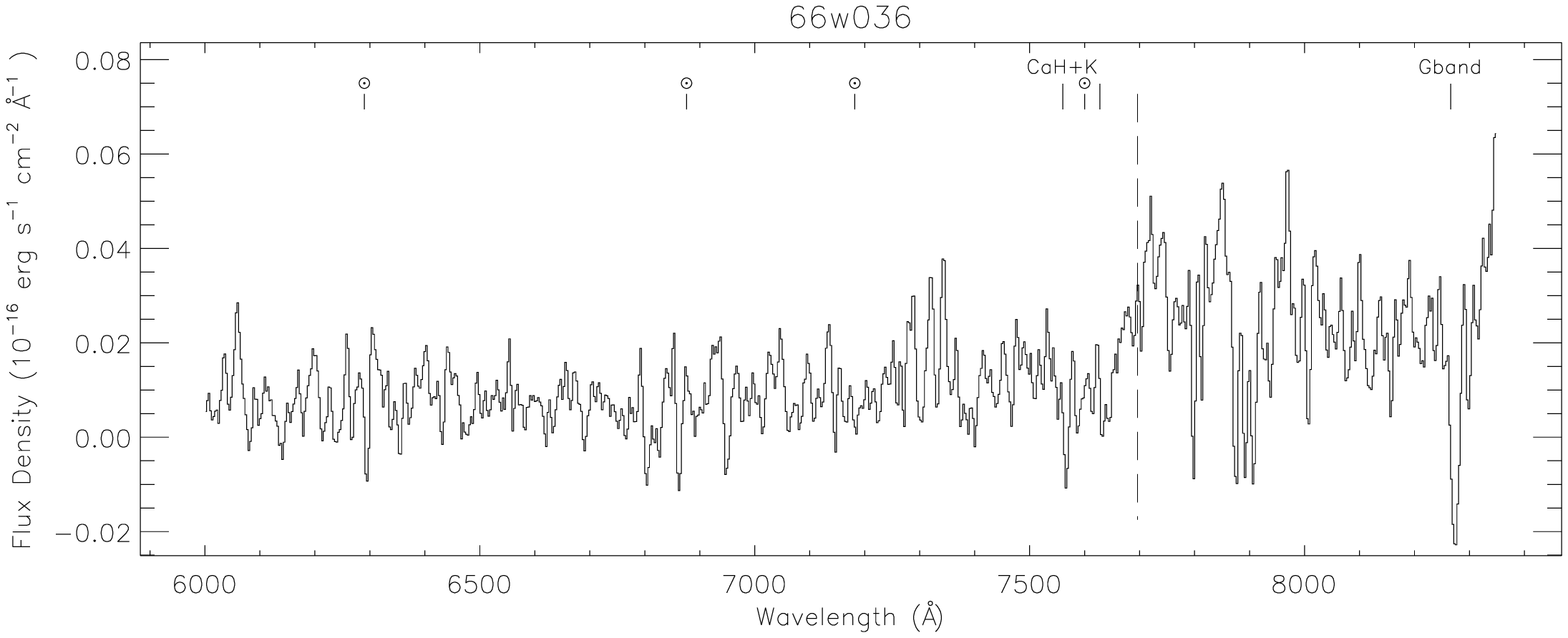}
\includegraphics[scale=0.60, angle=90]{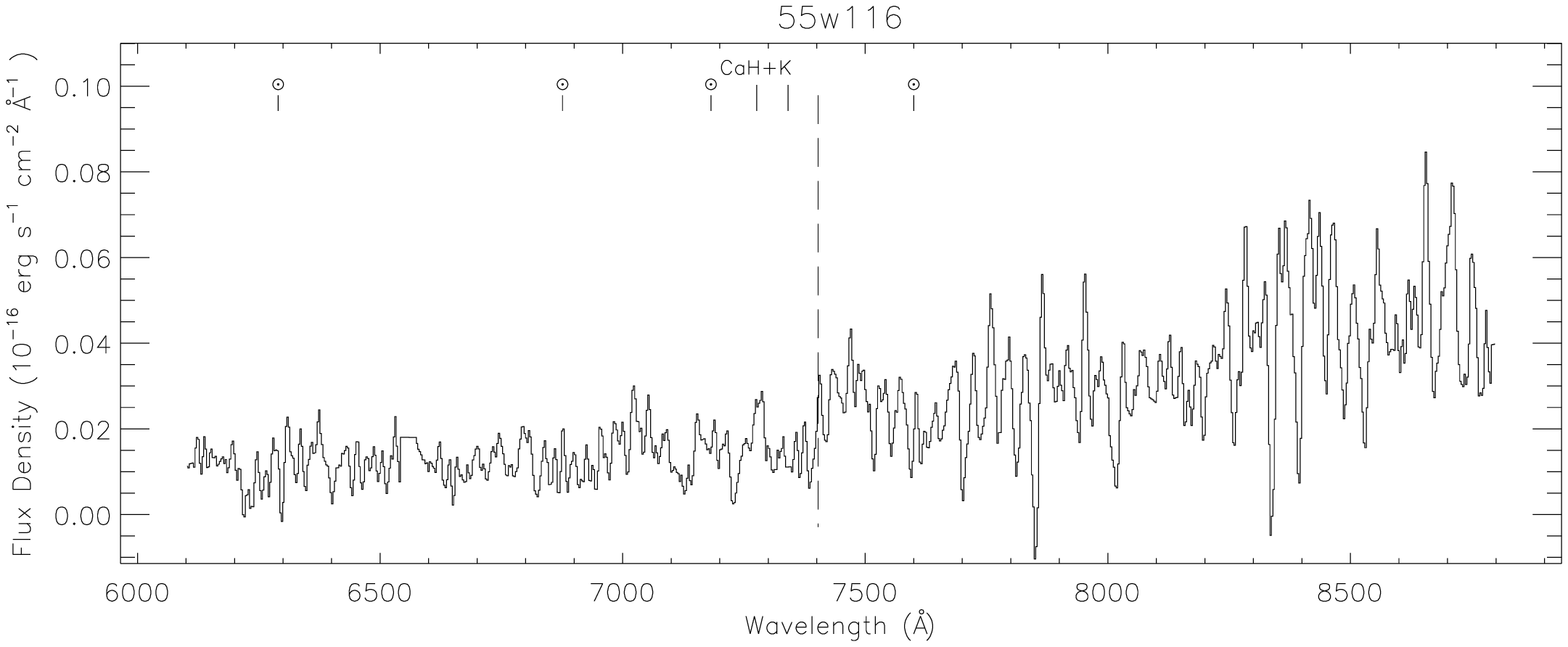}
\includegraphics[scale=0.60, angle=90]{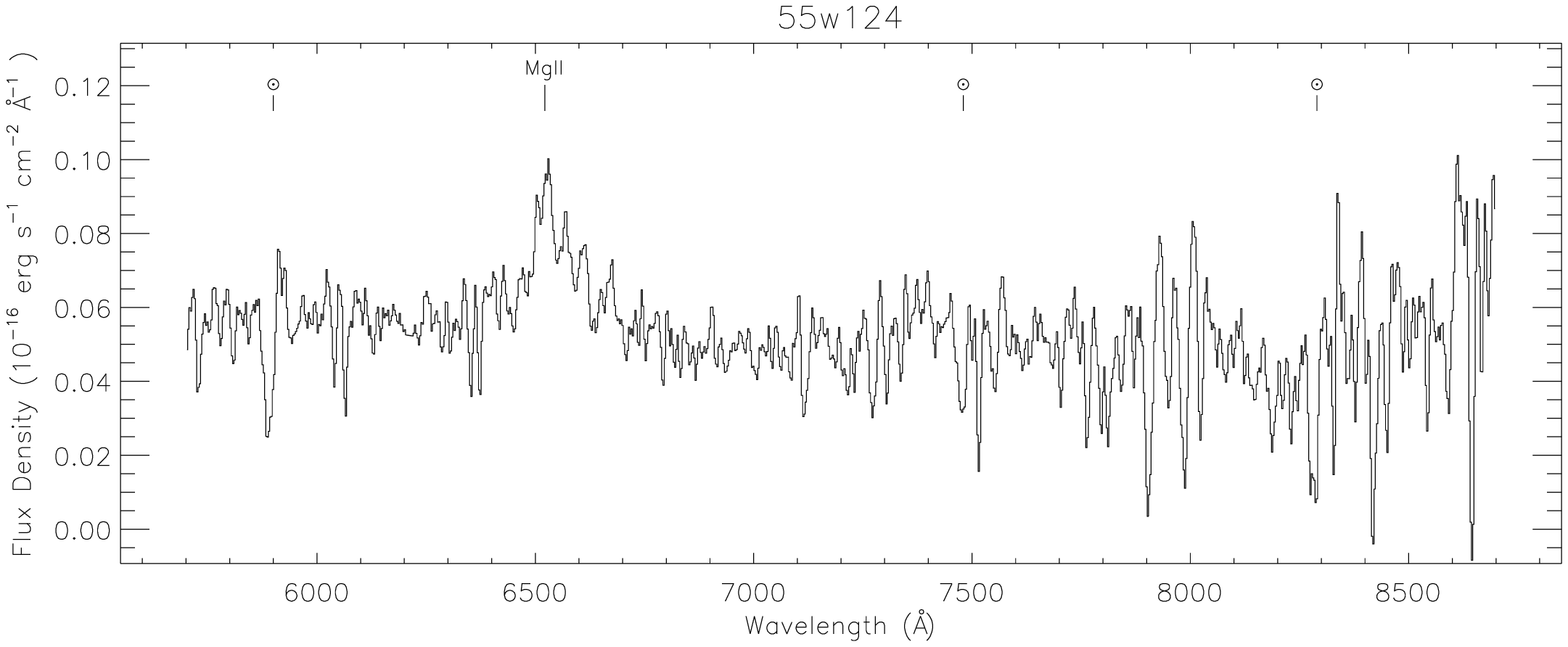}
\contcaption{}
\end{minipage}
\end{figure*}

\begin{figure*}
\begin{minipage}{15cm}
\includegraphics[scale=0.60, angle=90]{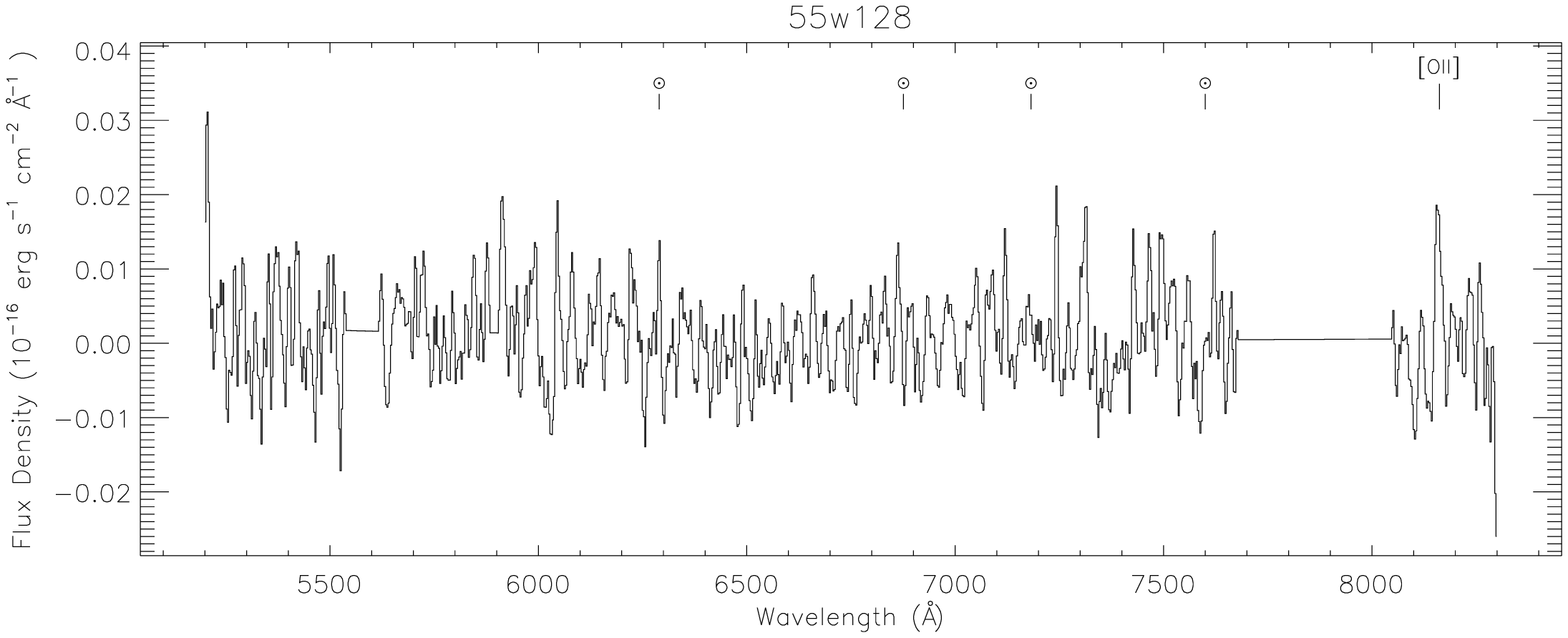}
\includegraphics[scale=0.60, angle=90]{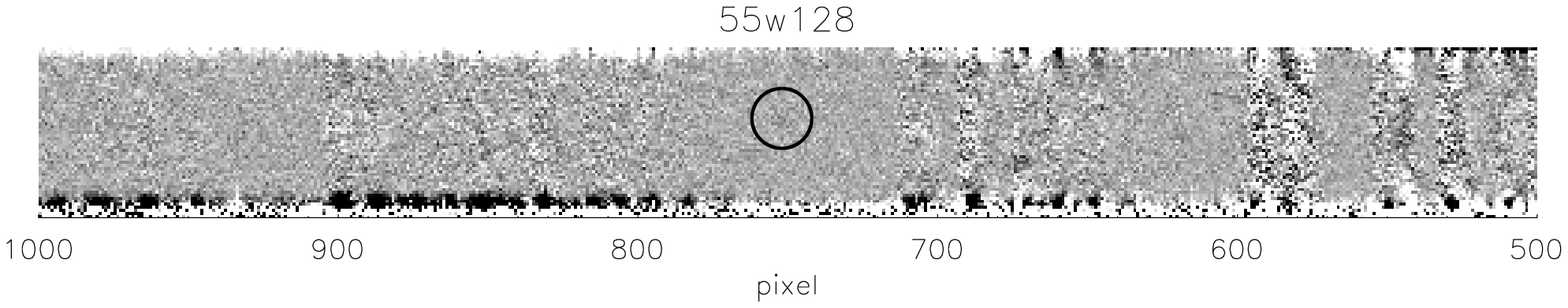}
\includegraphics[scale=0.60, angle=90]{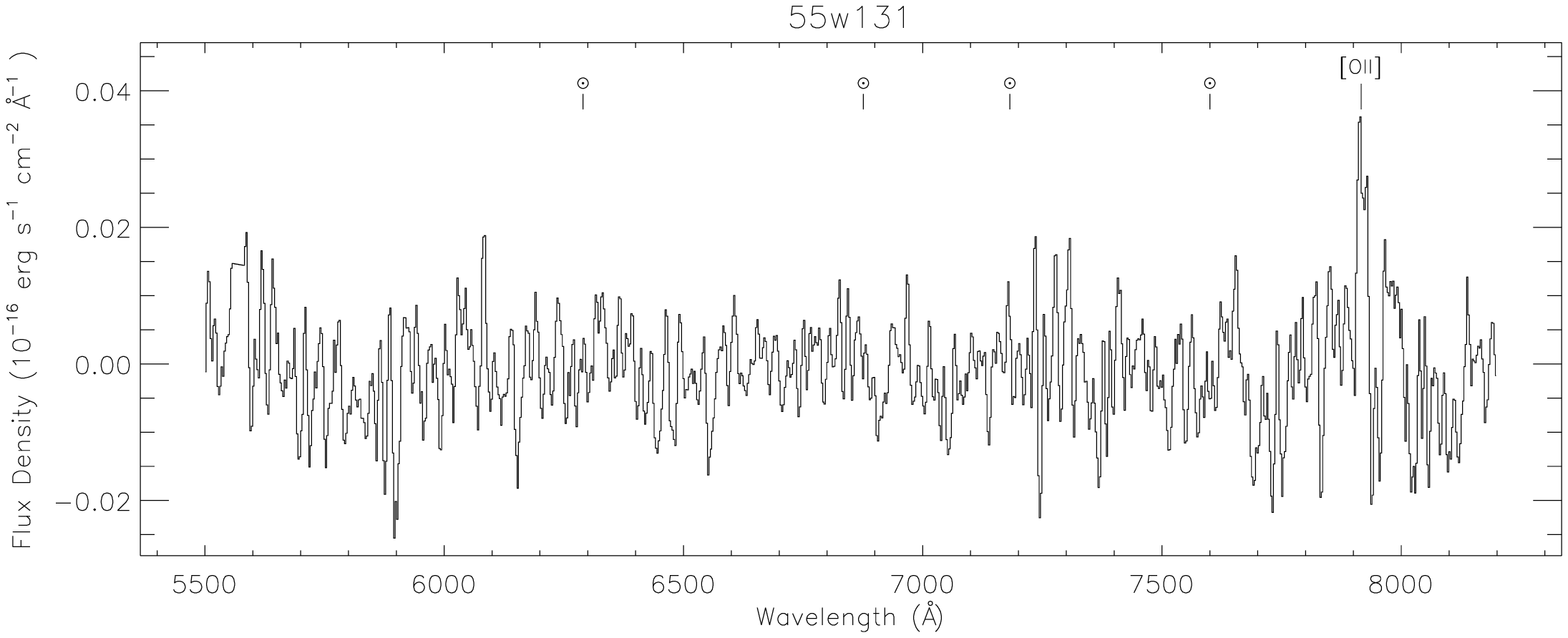}
\includegraphics[scale=0.60, angle=90]{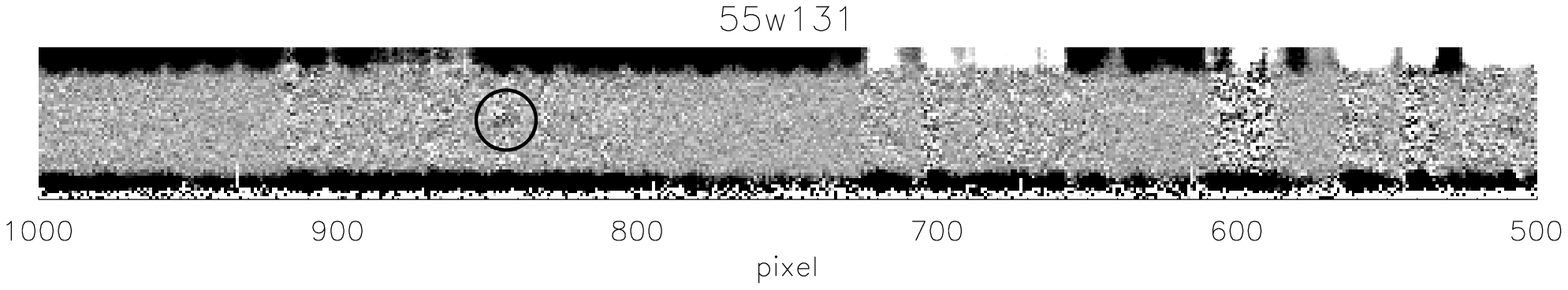}
\contcaption{}
\end{minipage}
\end{figure*}

\begin{figure*}
\begin{minipage}{15cm}
\includegraphics[scale=0.60, angle=90]{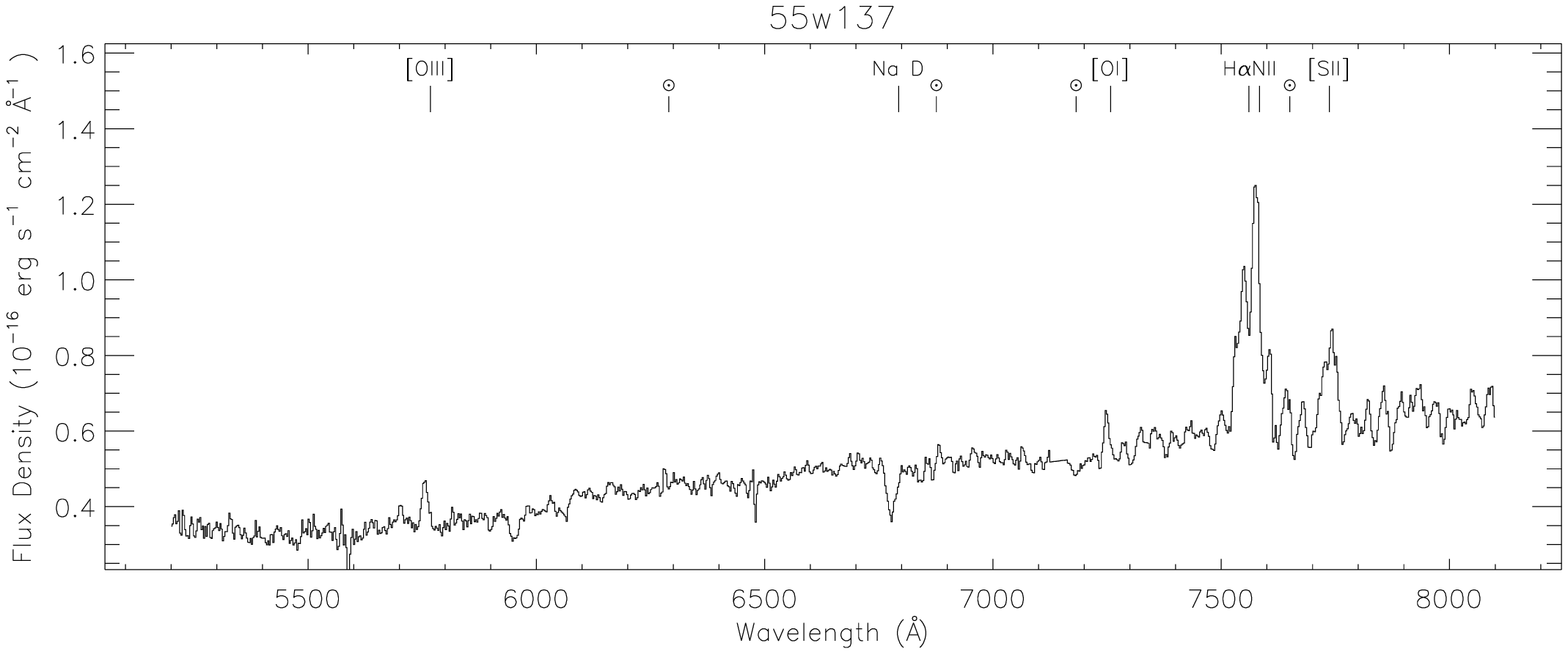}
\includegraphics[scale=0.60, angle=90]{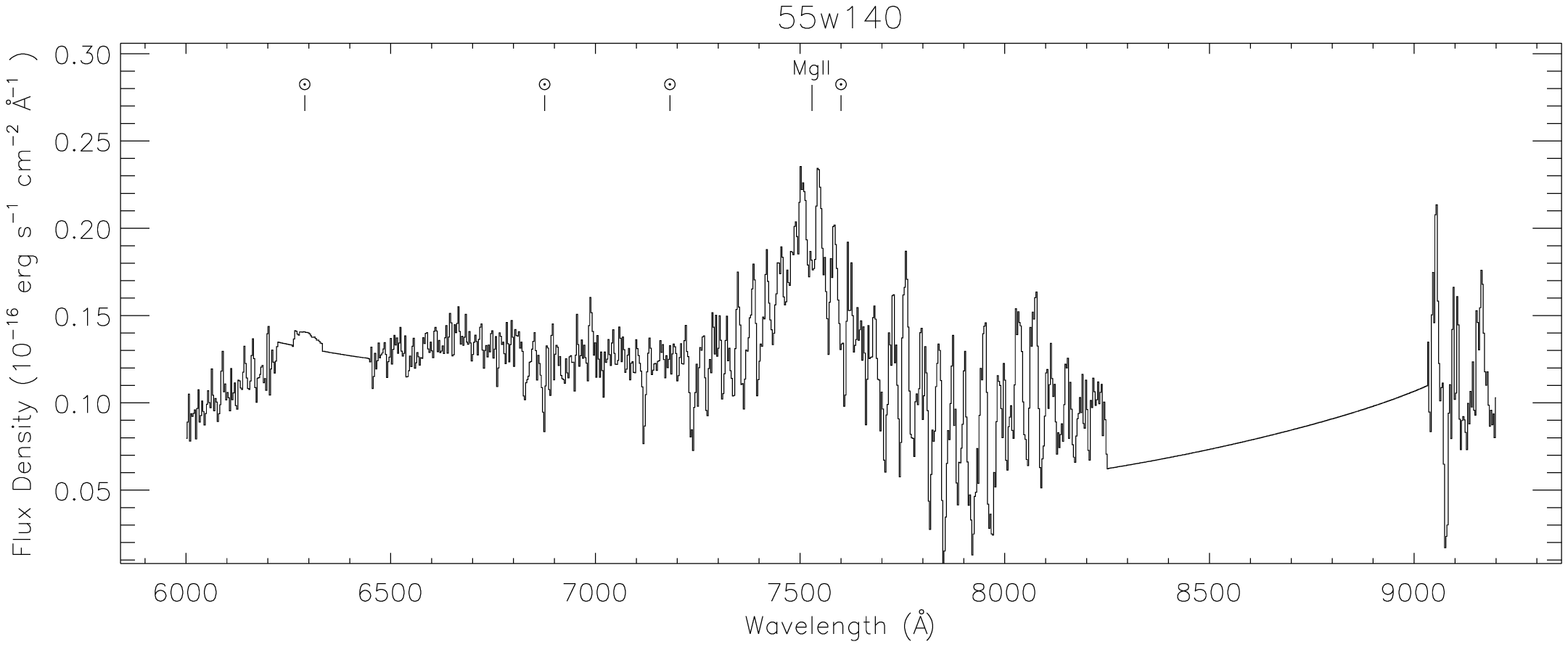}
\includegraphics[scale=0.60, angle=90]{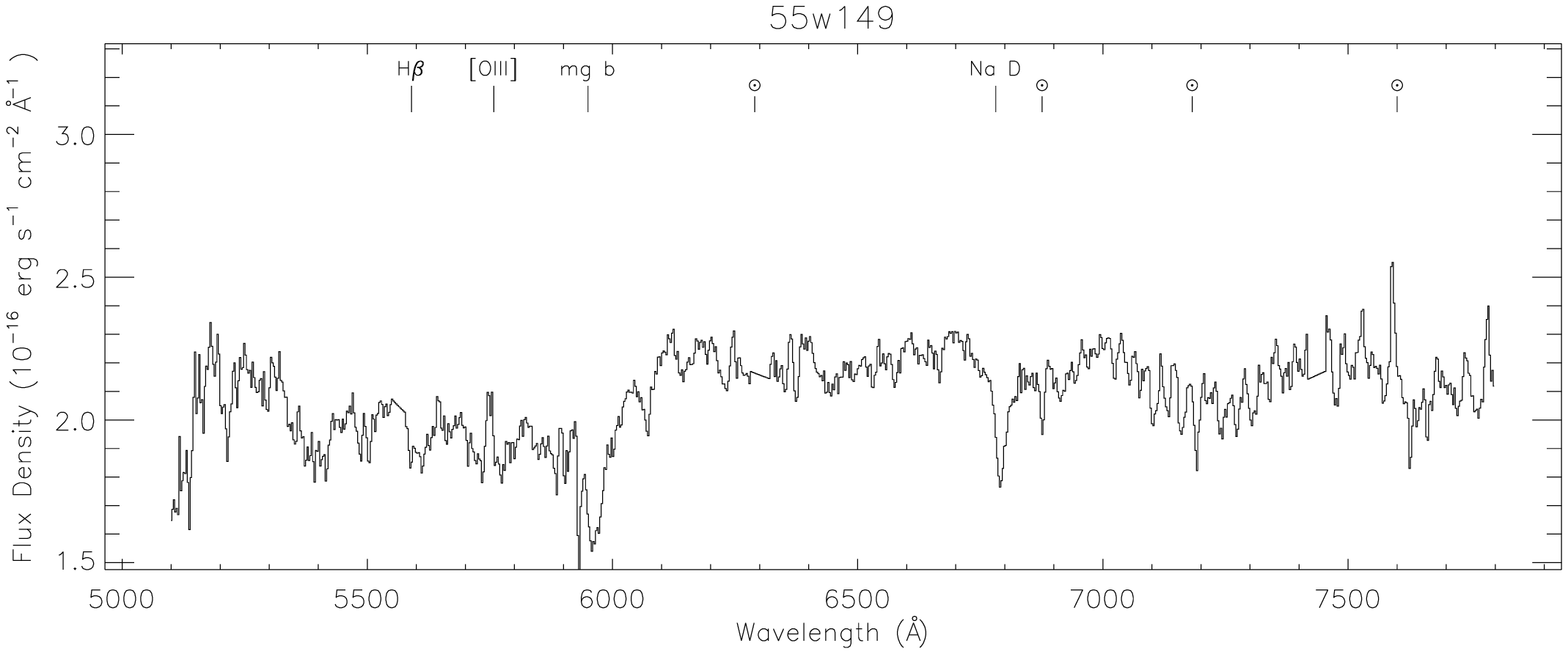}
\includegraphics[scale=0.60, angle=90]{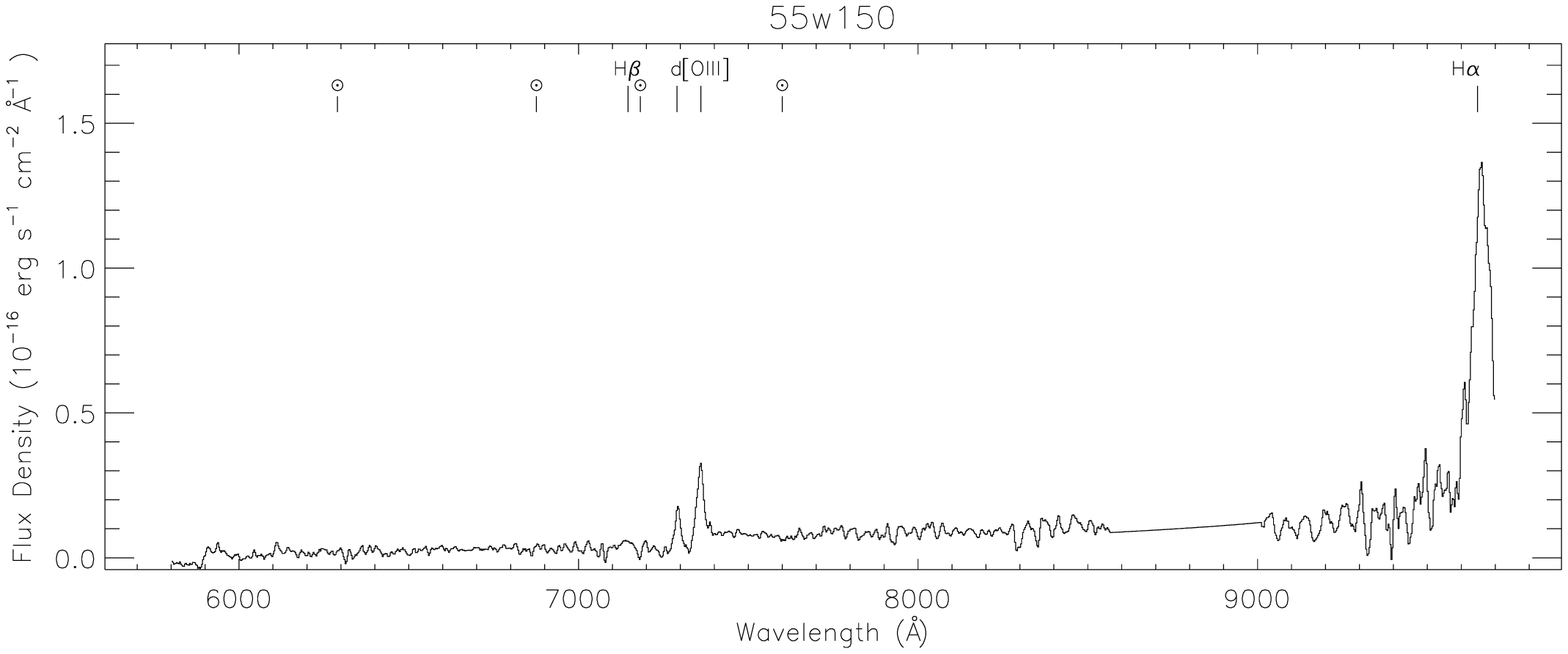}
\contcaption{}
\end{minipage}
\end{figure*}

\begin{figure*}
\begin{minipage}{15cm}
\includegraphics[scale=0.60, angle=90]{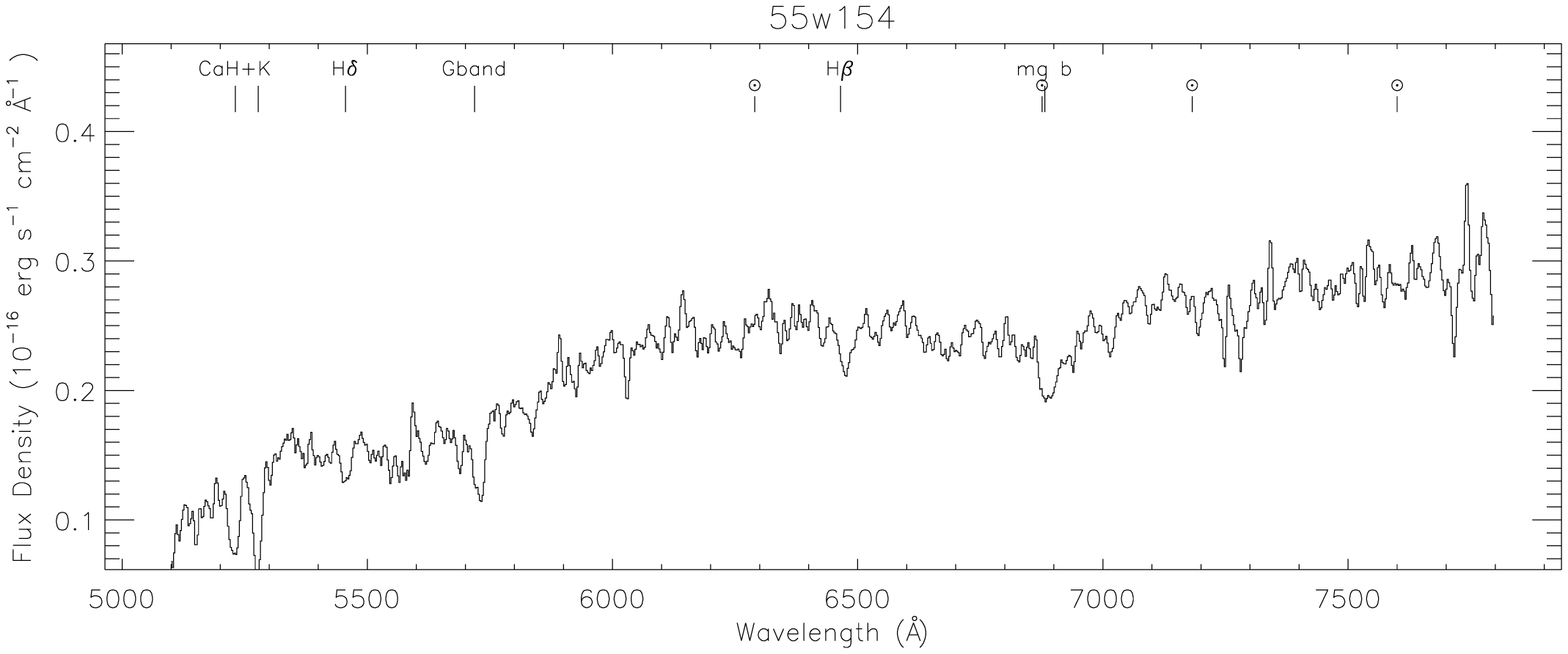}
\includegraphics[scale=0.60, angle=90]{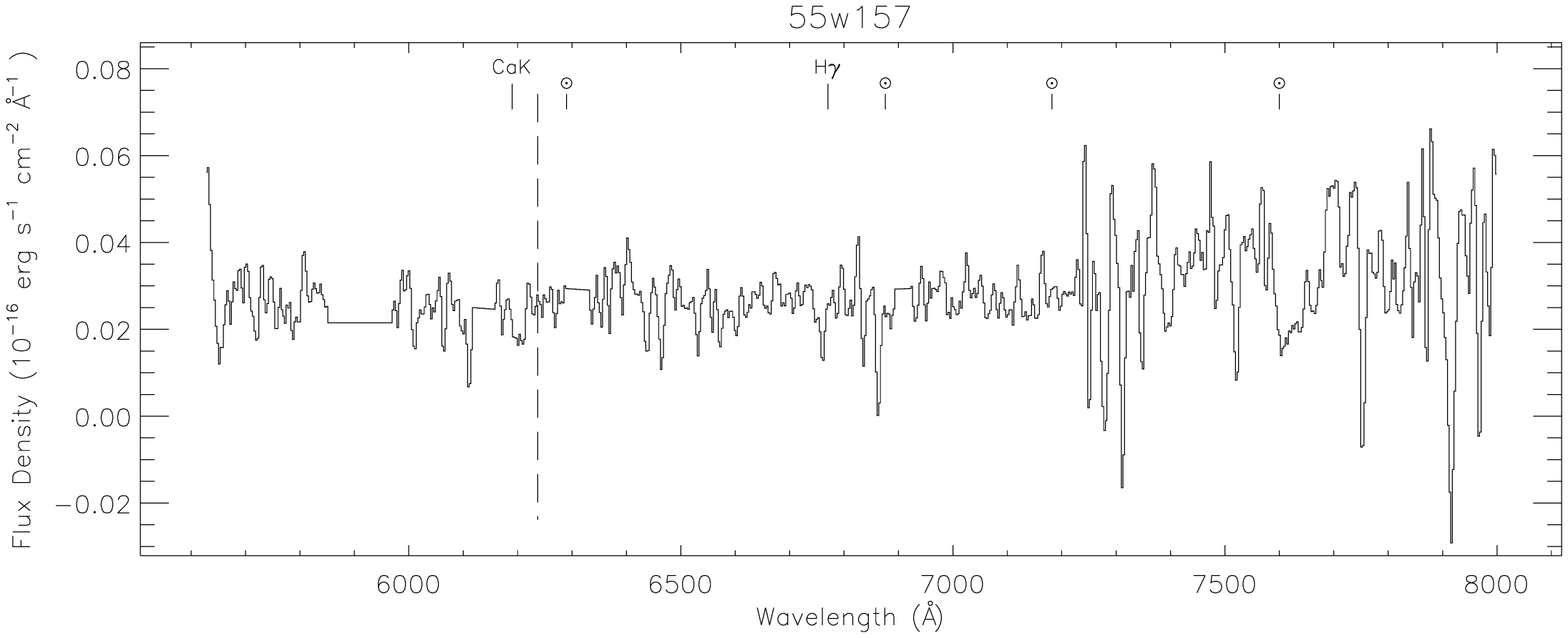}
\includegraphics[scale=0.60, angle=90]{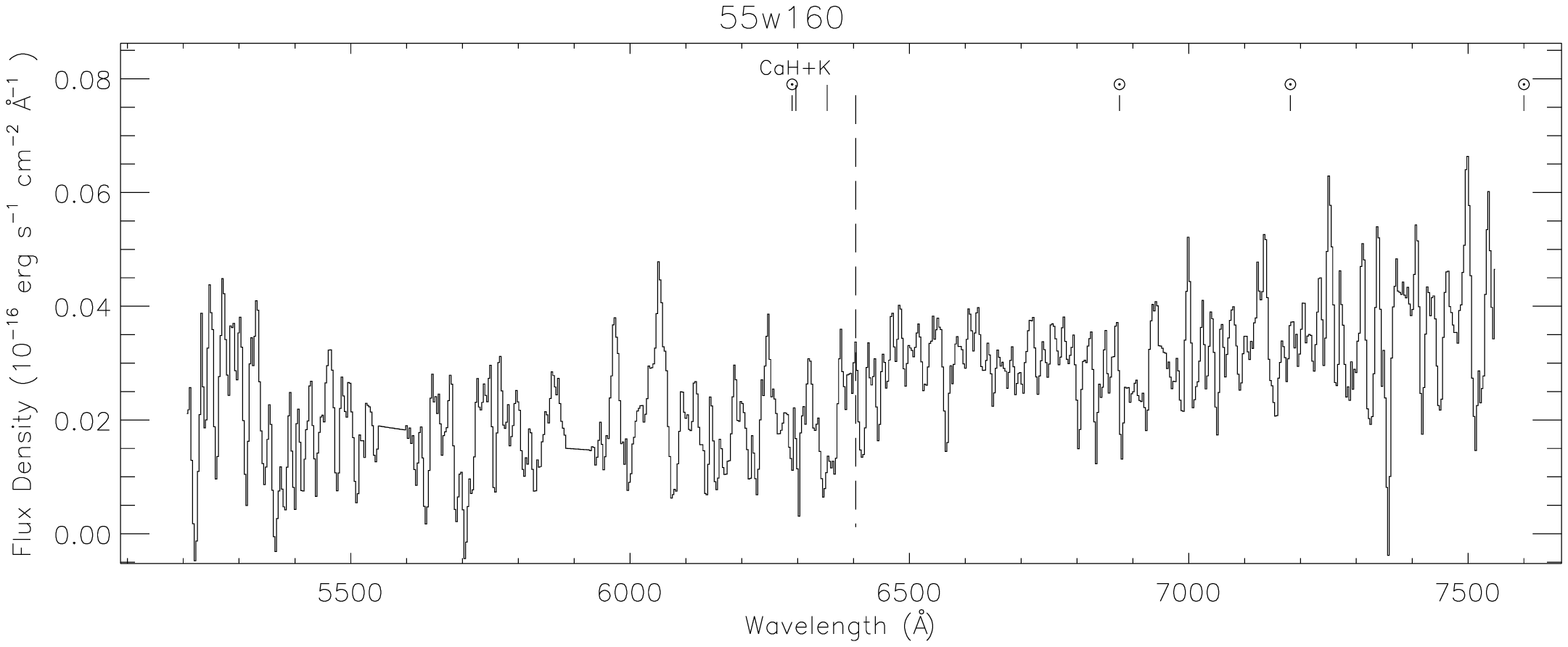}
\includegraphics[scale=0.60, angle=90]{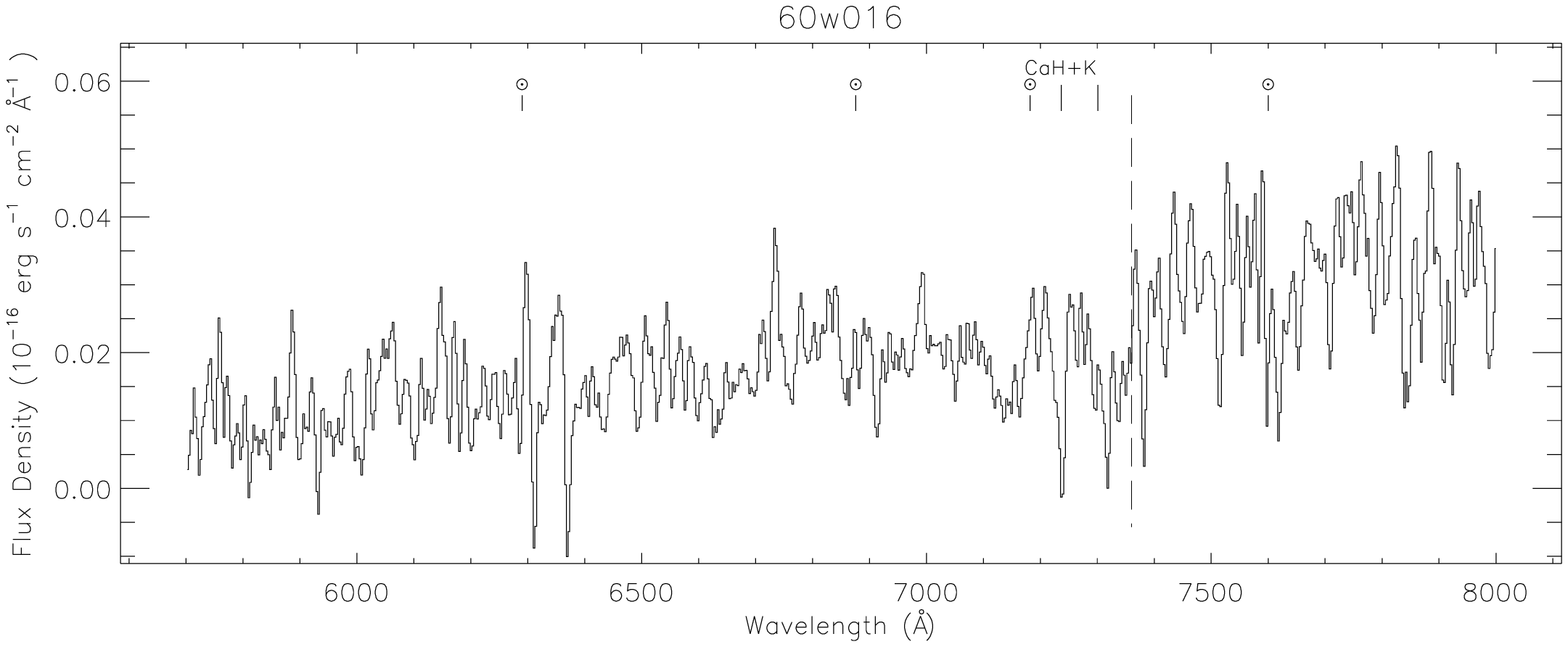}
\contcaption{}
\end{minipage}
\end{figure*}

\begin{figure*}
\begin{minipage}{15cm}
\includegraphics[scale=0.60, angle=90]{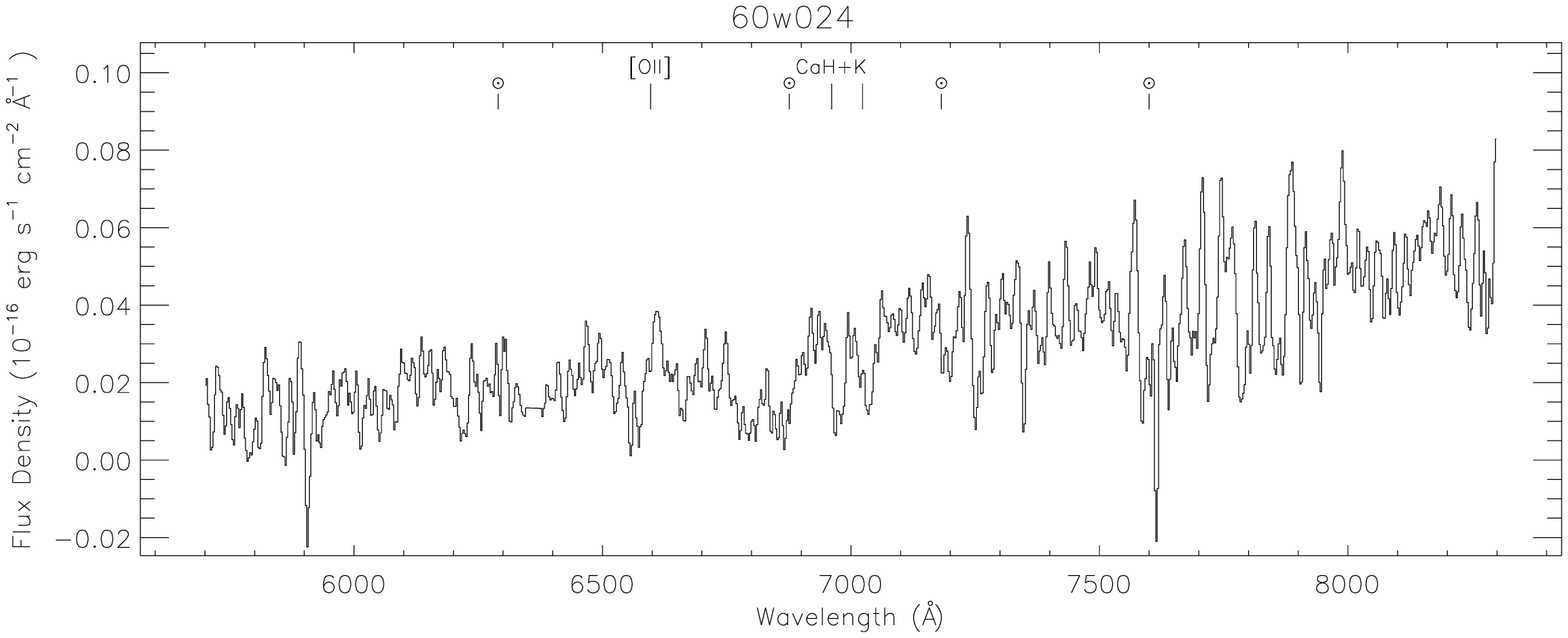}
\includegraphics[scale=0.60, angle=90]{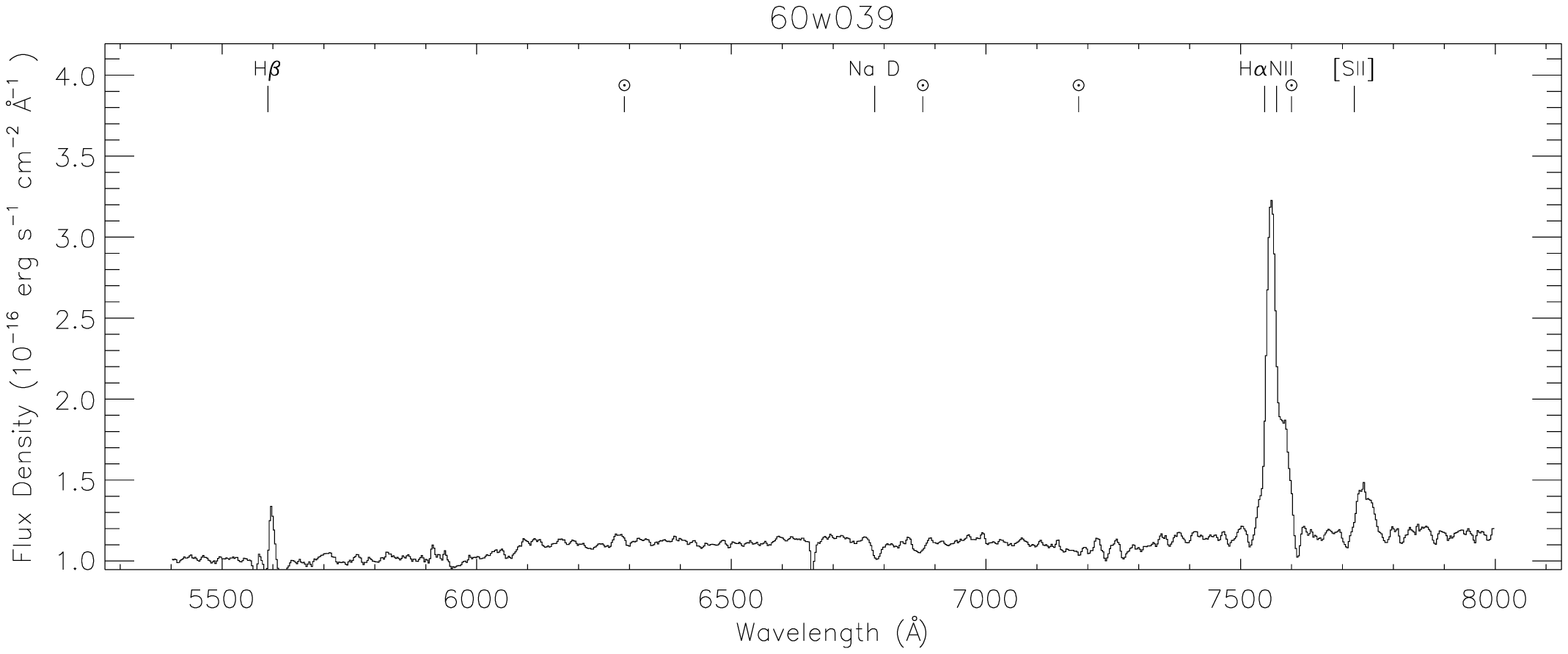}
\includegraphics[scale=0.60, angle=90]{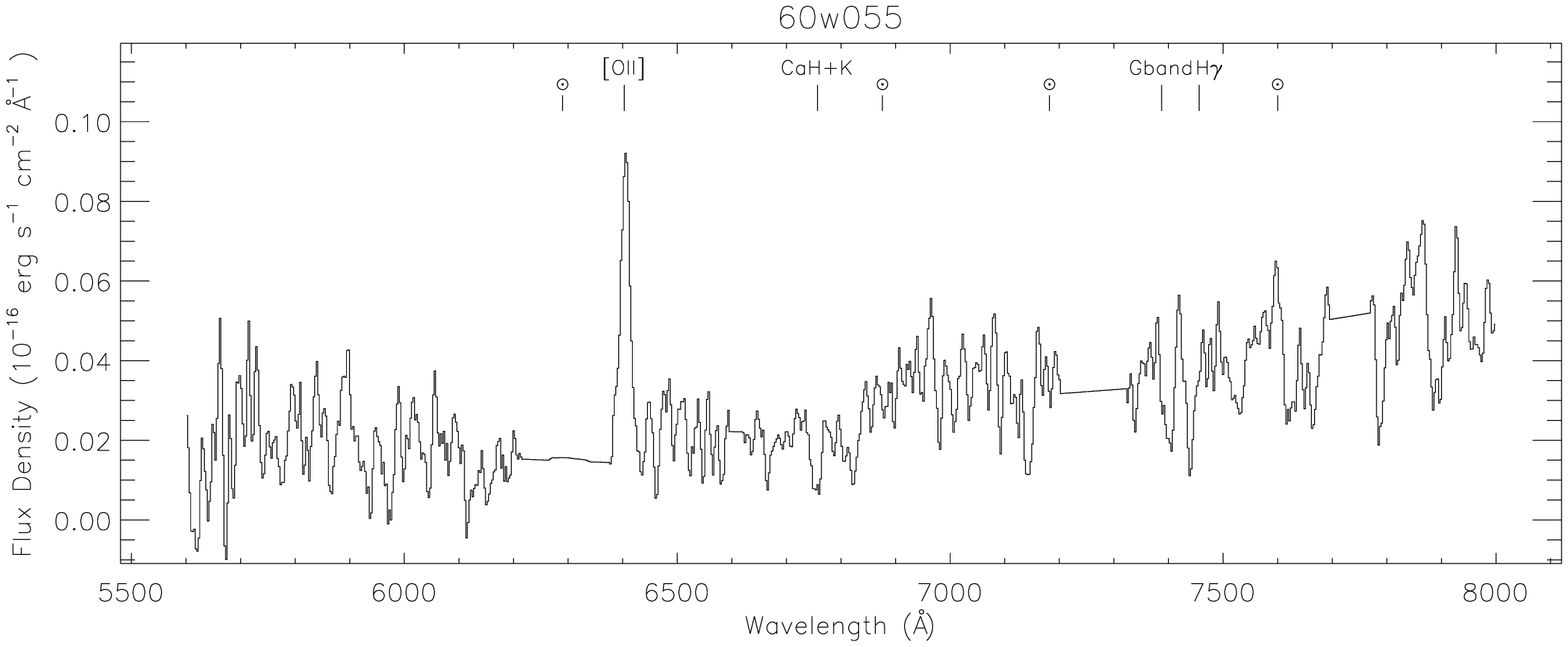}
\contcaption{}
\end{minipage}
\end{figure*}

\label{lastpage}

\end{document}